\documentclass[a4paper, oneside, 12pt, openright]{memoir}

\usepackage{amsmath}
\usepackage{amsfonts}
\usepackage{amssymb}
\usepackage{setspace}
\usepackage{bm}
\usepackage{indentfirst}
\usepackage[numbers, sort&compress]{natbib}
\usepackage{graphicx}
\usepackage{palatino}
\usepackage{oxfordthesis}
\usepackage{cancel}

\thetitle{Characteristics of Plasma Turbulence in the Mega Amp Spherical Tokamak}
\theauthor{Young-chul Kim}
\degreedate{October 2012}
\degree{Doctor of Philosophy}
\college{Merton College}

\usepackage{memhfixc}

\newcommand{\lp}{\left(}
\newcommand{\rp}{\right)}
\newcommand{\lab}{\left<}
\newcommand{\rab}{\right>}
\newcommand{\lang}{\left\langle}
\newcommand{\rang}{\right\rangle}
\newcommand{\lsb}{\left[}
\newcommand{\rsb}{\right]}
\newcommand{\lcb}{\left\{}
\newcommand{\rcb}{\right\}}
\newcommand{\labs}{\left|}
\newcommand{\rabs}{\right|}

\newcommand{\mydotfill}[2]{\noindent #1 \dotfill #2 \newline}

\newcommand{\grad}{\nabla}
\newcommand{\vct}[1]{\bm{#1}}
\newcommand{\Dx}{\Delta x}
\newcommand{\Dy}{\Delta y}
\newcommand{\DR}{\Delta R}
\newcommand{\DZ}{\Delta Z}
\newcommand{\Dt}{\Delta t}

\newcommand{\tor}{\phi}
\newcommand{\pol}{\theta}

\newcommand{\Btor}{B_\tor}
\newcommand{\Bpol}{B_\pol}

\newcommand{\rhoi}{\rho_i}
\newcommand{\rhoe}{\rho_e}
\newcommand{\rhoie}{\rho_{i, e}}
\newcommand{\rhostar}{\rho_*}
\newcommand{\kper}{k_\perp}
\newcommand{\kpar}{k_\parallel}
\newcommand{\kx}{k_x}
\newcommand{\ky}{k_y}
\newcommand{\lx}{\ell_x}
\newcommand{\ly}{\ell_y}
\newcommand{\lpar}{\ell_\parallel}
\newcommand{\lper}{\ell_\perp}
\newcommand{\lZ}{\ell_Z}
\newcommand{\lR}{\ell_R}
\newcommand{\LT}{L_T}
\newcommand{\LTi}{L_{T_i}}
\newcommand{\LTe}{L_{T_e}}
\newcommand{\LTie}{L_{T_{i, e}}}
\newcommand{\RLTi}{R/L_{T_i}}
\newcommand{\RLTic}{R/L_{T_i,c}}
\newcommand{\Ln}{L_n}
\newcommand{\LU}{L_U}
\newcommand{\LUz}{L_{U_z}}

\newcommand{\Lstar}{L_*}
\newcommand{\RLTe}{R/L_{T_e}}
\newcommand{\RLn}{R/\Ln}

\newcommand{\vth}{\ensuremath{ v_{\mathrm{th}} } }

\newcommand{\vti}{\ensuremath{v_{{\mathrm{th}}i}}}
\newcommand{\vte}{\ensuremath{v_{{\mathrm{th}}e}}}
\newcommand{\vtie}{\ensuremath{v_{{\mathrm{th}}i, e}}}
\newcommand{\Upar}{\ensuremath{U_\parallel}}
\newcommand{\Uper}{\ensuremath{U_\perp}}
\newcommand{\Utor}{\ensuremath{U_\tor}}
\newcommand{\vD}{\ensuremath{\vct{v_D}}}
\newcommand{\vbes}{\ensuremath{v_y^{\mathrm{BES}}}}
\newcommand{\delvper}{\delta u_\perp}

\newcommand{\Omegae}{\ensuremath{\Omega_e}}
\newcommand{\Omegai}{\ensuremath{\Omega_i}}
\newcommand{\taupeakcc}{\tau_{\mathrm{peak}}^{\mathrm{cc}}}
\newcommand{\taupeakenv}{\tau_{\mathrm{peak}}^{\mathrm{env}}}
\newcommand{\tauprop}{\tau_\mathrm{prop}}

\newcommand{\tc}{\tau_\mathrm{c}}
\newcommand{\tnl}{\tau_\mathrm{nl}}
\newcommand{\tnlnz}{\tau_\mathrm{nl}^\mathrm{NZ}}
\newcommand{\tst}{\tau_\mathrm{st}}
\newcommand{\tshear}{\tau_\mathrm{sh}}
\newcommand{\tM}{\tau_\mathrm{M}}
\newcommand{\tstar}{\tau_\ast}
\newcommand{\tstari}{\tau_{\ast i}}
\newcommand{\tstare}{\tau_{\ast e}}
\newcommand{\tstarie}{\tau_{\ast i, e}}
\newcommand{\tstarn}{\tau_{\ast n}}

\newcommand{\gE}{\gamma_E}
\newcommand{\gEbar}{\bar\gamma_E}
\newcommand{\nust}{\nu_{*i}}

\newcommand{\dn}{\delta n}
\newcommand{\dI}{\delta I}
\newcommand{\betabes}{\beta_\mathrm{BES}}
\newcommand{\Ntotal}{N_\mathrm{total}}
\newcommand{\Navg}{N_\mathrm{avg}}
\newcommand{\Ncorr}{N_\mathcal{C}}
\newcommand{\Dtsam}{\Dt_\mathrm{sam}}
\newcommand{\tres}{t_\mathrm{res}}
\newcommand{\corr}{\mathcal{C}}
\newcommand{\cov}{\mathcal{C}^\mathrm{v}}
\newcommand{\corrsub}{\mathcal{C}_\mathrm{sub}}
\newcommand{\hsbias}{\hat\sigma_\mathrm{bias}}

\newcommand{\hsrand}{\hat\sigma_\mathrm{rand}}

\newcommand{\hsmeanfit}{\hat\sigma_\mathrm{mean}^\mathrm{fit}}
\newcommand{\hsrandfit}{\hat\sigma_\mathrm{rand}^\mathrm{fit}}
\newcommand{\Qturb}{\bar Q_{\mathrm turb}}

\renewcommand{\eqref}[1]{Eq. (\ref{#1})}

\newcommand{\myfig}[4][4.5in]{
\begin{figure}[t]
\centering
\includegraphics[width=#1]{#2}%
\caption[#3]{#4\label{fig:#2}}%
\end{figure}
}

\newcommand{\mydoublesidefig}[8]{
\begin{figure}[t]
\begin{minipage}[t]{0.45\linewidth}
\centering
\includegraphics[width=#1]{#3}
\caption[#4]{#7}\label{fig:#3}
\end{minipage}
\hspace{0.5in}
\begin{minipage}[t]{0.45\linewidth}
\centering
\includegraphics[width=#2]{#5}
\caption[#6]{#8}\label{fig:#5}
\end{minipage}
\end{figure}
}

\newcommand{\figref}[1]{Figure \ref{fig:#1}}
\newcommand{\tableref}[1]{Table \ref{#1}}
\newcommand{\appendixref}[1]{Appendix \ref{#1}}
\newcommand{\chref}[1]{Chapter \ref{#1}}
\newcommand{\secref}[1]{Section \ref{#1}}

\newcommand{\mypart}[1]{\part{\hspace{2pt}#1}}

\usepackage{tikz}
\newcommand*\circled[1]{\tikz[baseline=(char.base)]{\node[shape=circle,draw,inner sep=0.5pt] (char) {#1};}}

\usepackage{color}

\usepackage[
pdftitle={\oxfthetitle},
pdfauthor={\oxftheauthor},
pdfsubject={Thesis for the Degree of \oxfdegree, \oxfdegreedate},
pdfborder=0,
bookmarks=true,
bookmarksnumbered=true,
bookmarksopen=true,
bookmarksopenlevel=1,
plainpages=false,
pdfpagelabels=true
]{hyperref}

\begin{document}

\titlepage

\frontmatter

\begin{dedication}
To Da Eun and Zane
\end{dedication}

\begin{abstract}
Turbulence is a major factor limiting the achievement of better tokamak performance as it enhances the transport of particles, momentum and heat which hinders the foremost objective of tokamaks.  Hence, understanding and possibly being able to control turbulence in tokamaks is of paramount importance, not to mention our intellectual curiosity of it.  We take the first step by making measurements of turbulence using the 2D ($8$ radial $\times$ $4$ poloidal channels) beam emission spectroscopy (BES) system on the Mega Amp Spherical Tokamak (MAST).  Measured raw data are statistically processed, generating spatio-temporal correlation functions to obtain the physical characteristics of the turbulence such as spatial and temporal correlation lengths as well as its motion. The reliability of statistical techniques employed in this work is examined by generating and utilizing synthetic 2D BES data. The apparent poloidal velocity of fluctuating density patterns is estimated using the cross-correlation time delay method. The experimental results indicate that the poloidal motion of fluctuating density patterns in the lab frame arises because the patterns are advected by the strong toroidal plasma flows while the patterns are aligned with the background magnetic fields which are not parallel to the flows.  Furthermore, various time scales associated with the turbulence are calculated using statistically estimated spatial correlation lengths and correlation times of turbulence. We find that turbulence correlation time, the drift time associated with ion temperature or density gradients, the ion streaming time along the magnetic field line and the magnetic drift time are comparable and possibly scale together suggesting that the turbulence, determined by the local equilibrium, is critically balanced.  Finally, we argue that we have produced a critical manifold in the experimentally obtained local equilibrium parameter space separating dominant turbulent transport from a non-turbulent or weakly turbulent state. It shows that the inverse ion-temperature-gradient scale length is correlated inversely with $q/\varepsilon$ (safety factor/inverse aspect ratio) and positively with the plasma rotational shear. Practically, this means that we can attain the stiffer ion-temperature-gradient, thus hotter plasma core, without increasing the rotational shear.

\end{abstract}

\tableofcontents

\listoffigures

\listoftables

\chapter{List of publications}
\subsection{First authored publications (4 papers)}
\noindent
Y.-c. Ghim, A. A. Schekochihin, A. R. Field, I. G. Abel, M. Barnes, G. Colyer, S. C. Cowley, F. I. Parra, D. Dunai, S. Zoletnik and the MAST team. \textbf{Experimental signatures of critically balanced turbulence in MAST.} \textit{submitted to Phys. Rev. Lett. [arXiv:1208.5970], 2012.}
\newline\newline\noindent
Y.-c. Ghim, A. R. Field, A. A. Schekochihin, E. G. Highcock, C. Michael and the MAST team. \textbf{Local dependence of ion temperature gradient on magnetic configuration, rotational shear and turbulent heat flux in MAST.} \textit{submitted to Phys. Rev. Lett. [arXiv:1211.2883], 2012.}
\newline\newline\noindent
Y.-c. Ghim, A. R. Field, D. Dunai, S. Zoletnik, L.Bardoczi, A. A. Schekochihin and the MAST team. \textbf{Measurement and physical interpretation of the mean motion of turbulent density patterns detected by the BES system on MAST.} \textit{Plasma Phys. Control. Fusion, 54(095012), 2012.}
\newline\newline\noindent
Y.-c. Ghim, A. R. Field, S. Zoletnik and D. Dunai. \textbf{Calculation of spatial response of 2D beam emission spectroscopy diagnostic on MAST.} \textit{Rev. Sci. Instrum., 81(10D713), 2010.}
\newline\newline\noindent
\textbf{Contributions for the listed publications and the thesis}
\newline\noindent
YCG has performed data analyses and created turbulence database. ARF, DD and SZ designed and installed the BES diagnostic on MAST and took the data. AAS, IGA, MB, GC, SCC, FIP and EH provided theoretical insights. CM provided EFIT data.
\subsection{Contributed publications (3 papers)}
\noindent
A. R. Field, D. Dunai, R. Gaffka, Y.-c. Ghim, I. Kiss, B. Meszaros, T. Krizsanoczi, S. Shivaev and S. Zoletnik. \textbf{Beam emission spectroscopy turbulence imaging system for the MAST spherical tokamak.} \textit{Rev. Sci. Instrum., 83(013508), 2012}
\newline\newline\noindent
A. R. Field, C. Michael, R. J. Akers, J. Candy, G. Colyer, W. Guttenfelder, Y.-c. Ghim, C. M. Roach and S. Saarelma. \textbf{Plasma rotation and transport in MAST spherical tokamak.} \textit{Nucl. Fusion, 51(063006), 2011.}
\newline\newline\noindent
B. Lloyd, \textit{et al.} \textbf{Overview of physics results from MAST} \textit{Nucl. Fusion, 51(094013), 2011.}

\begin{acknowledgements}
Since the start of my formal education in 1985 at the age of 6, it has been a long journey to come to this end, or should I say the ``beginning''?  I once dreamed to be a sprinter but gave up as I was not fast enough, a Taekwondo player but gave up not being strong enough, a soldier but gave up being myopic, a medical doctor but gave up abhorring biology, and a prosecutor but gave up not being lawful enough. The truth is, I did not pursue those dreams with my own excuses because I could not give up on endeavouring to understand the laws of the Nature, and as there are still almost nothing that I understand about it I will likely to keep this track for a while and hopefully for the rest of my life. Once, I had an illusion that I understood a little bit about the Nature, but now I, fortunately, realize that it was a mere delusion. There has been many great teachers to help me recognize this.  Without them, I would be still in a dark room deceiving myself. I give my greatest thanks to all my teachers.
\newline\indent
The most important teachers helping me to bring this work into reality are Dr. Alexander Schekochihin and Dr. Anthony Field.  Having a theoretician and an experimentalist as supervisors made me worried in the beginning of my D.Phil course, but it did not take me long to realize that this was a unique opportunity because I could learn both sides simultaneously.  Discussions on experimental data with Alex taught me how to link experiments with theories, while Anthony has trained me how to obtain and interpret the data.
\newline\indent
When I started my course I had no knowledge of statistical analyses, and I owe Prof. Troy Carter, Dr. Daniel Dunai, Dr. Clive Michael, Dr. Martin Valov\v{i}c and Dr. Sandor Zoletnik for teaching me how to apply them on data. As usual of me being ignorant, I've had many questions on plasma physics. But, I have had many friends who are willing to teach me not only in the office but also over the drinks: Ian Abel, Michael Barnes, Greg Colyer, Michael Fox, Edmund Highcock, Sarah Newton, Felix Parra and Alessandro Zocco. 
\newline\indent
During the course, I have had privilege to discuss plasma physics with leading scientists: Dr. Jack Connor, Prof. Steve Cowley, Prof. Bill Dorland, Prof. Greg Hammet, Dr. George McKee and Prof. Brian Taylor. I appreciate them for allocating their precious time to speak with me and the Leverhulme Trust International Network for Magnetised Plasma Turbulence for providing financial support on many international travels.
\newline\indent
From time to time, I have wanted to talk in Korean.  I thank my Korean friends whom I have met in Oxford: Hyun-joong Im, Yong-chool Jung, Hyun-tae Kim, Jin-hyok Kim, Shin-kwon Kim, Yong-soo Kim, Jin Park, Dong-meong Shin.  Also importantly, I thank Dr. JaeChun Seol and Dr. Myeon Kwon from National Fusion Research Institute in Korea for providing me a connection with Culham Centre for Fusion Energy so that I could carry on fusion research at CCFE. Of course, I need to thank the Kwanjeong Educational Foundation for its financial support while I have been in Oxford.
\newline\indent
There are two people who have sacrificed the most for me: my wife, Da Eun Yu, and my son, Zane Kim. Da Eun gave up her career to support my work, and she has done so without making me feel bad. She has endured all the emotional pains I have given her and helped me come this far successfully. It may have been the case that Zane would not be diagnosed to have autism spectra if we were living in Korea giving him more chances to interact with his relatives.  Or, if I were more devoted to him and spending more time with him, he could have speaking words by now. I greatly thank them, and at the same time I give them my sincere apologies. I promise them that I will be much better husband and dad so that I have nothing to apologize them any more for the rest of my life and their lives. Finally, I thank my family back in Korea for their encouragement.
\newline\indent
\begin{flushright}
Young-chul Ghim(Kim) \\
October, 2012 in Oxford, UK
\end{flushright}
\end{acknowledgements}

\clearpage

\mainmatter

\mypart{Introduction}

\chapter{Fusion and turbulence in tokamaks}
\begin{flushright}
We do \textit{not inherit} the planet from our parents, we \textit{borrow} it from our children.\\
-- Native American\\
\end{flushright}
Imagining our world without any electrical power just for a few seconds, we come to a non-negotiable conclusion: we require electrical power to sustain and flourish in our lives.  This power has been generated from natural resources, such as coal, oil, natural gas, and they inevitably produce carbon dioxide which may be argued as a cause of global warming.\footnote{IPCC, Synthesis Report in IPCC AR4 SYR 2007}  Moreover, the limited amount of such resources is another vital concern \cite{smil_mit_2005}.  Thus, it will be undoubtedly beneficial if we can generate electrical power without further enhancing global warming and without depleting the limited amount of resources so that we can return the planet to our children without corrupting it further.  Fusion power satisfies these criteria.  

\section{Fusion in tokamaks}
Energy can be generated by fusing two nuclei of deuterium and tritium, which produces a harmless helium nucleus and a neutron with a total energy release of $17.6 MeV$ \cite{freidberg_cambridge_2007}.  To fuse the two nuclei they must overcome the repulsive electrostatic force which can be achieved by heating them to (large thermal energy) around $\sim10-20\:keV$ at which reasonable performance of a fusion reaction can be achieved \cite{wesson_clarendon_2004}.  Note that temperatures are expressed in the energy unit of $eV$ throughout this work. At this high temperature, particles are ionized, i.e., become plasma.\footnote{Being ionized is not a sufficient condition to be in the state of plasma.  However, ionized gases in this work satisfy plasma criteria \cite{chen_springer_2006}: i) Debye length is smaller than the system size; ii) number of particles in a Debye sphere is large; and iii) typical plasma frequency is larger than collision frequency with neutrals.}  
\subsection{Tokamak concept}
Given that such hot plasmas with a sufficiently large density are created, the performance of a fusion power plant depends on how long they can be confined within a finite spatial domain.  One way to confine the plasma is using the Lorentz force, a basic concept of ``magnetic confinement'':  
\begin{equation}\label{eq:single_particle_motion}
m\frac{d\vct{v}}{dt}=Ze\lp\vct{E}+\frac{1}{c}\vct{v}\times\vct{B}\rp,
\end{equation}
where $m$, $\vct{v}$ and $Ze$ are the mass, the velocity and the charge of the particle, respectively, and $c$ is the speed of light.  $\vct{E}$ and $\vct{B}$ are the electric and magnetic fields, respectively.  By this law, a single charged particle in a strong magnetic field in the absence of an electric field is constrained to move along the magnetic field line with a helical trajectory, i.e., plasmas are confined in the perpendicular plane with respect to the magnetic field, as shown in \figref{helical2}(a). 
\myfig[4.5in]{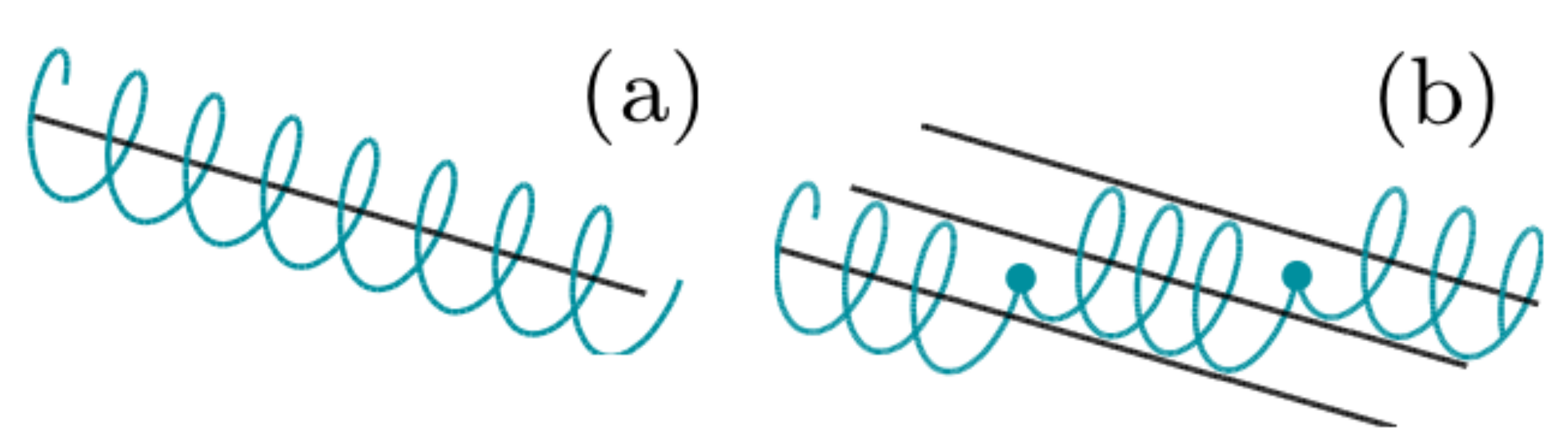}{Helical motion of a charged particle along a magnetic field line}{(a) Helical motion of a single charged particle along a magnetic field line in the absence of an electric field; (b) Cross-field motion of a single charged particle at the marked position due to cumulative effect of Coulomb collisions with other particles.  Figure is taken from Ref. \cite{highcock_phd_2012}}
\newline\indent
If the confined plasmas were collisionless in infinitely long parallel straight magnetic field lines, then a perfect confinement would be achieved.  But, such field lines cannot be generated in practice, and furthermore, plasmas should not be collisionless if the goal is to extract power from fusion reactions, i.e., particles need to collide with each other occasionally.
\newline\indent
Even though infinitely long straight magnetic field lines are not feasible, effectively ``infinitely long'' lines are certainly possible: a closed field line is infinitely long in the sense that starting and ending points are indistinguishable.  This is the basis for a concept of the ``TOKAMAK'' a transliteration of a Russian acronym a toroidal chamber with axial magnetic fields. The geometrical definition of a torus, on which ``infinitely long'' magnetic field lines lie, is\footnote{From Wikipedia.} 
\begin{quote}
a surface of revolution generated by a circle in three dimensional space about an axis coplanar with the circle
\end{quote}
 as shown in \figref{torus}.  A tokamak basically creates nested tori, where each torus is referred to as a flux surface, such that no magnetic field lines are connected between the two tori within the Last Closed Flux Surface (LCFS) unless there exist radial magnetic perturbations.  This means that magnetic fields can be described in two dimensional space, i.e., in poloidal and toroidal\footnote{Toroidal fields are generated by external coils, known as the toroidal field coils, whereas poloidal fields are generated by plasma currents flowing in the toroidal direction.} directions depicted with red and blue arrows, respectively, in \figref{torus}(a).  Although the poloidal cross-section of a torus is circular in its mathematical definition, that of a flux surface in a tokamak does not have to be a circle; in fact, it is usually `\textit{D}' shaped in practice.  In addition, the flux surfaces are not usually concentric due to the Shafranov shift \cite{shafranov_rpp_1966}, which means that the centres of the nested flux surfaces do not coincide.  A circular-shaped flux surface in a tokamak, as an example, is illustrated in \figref{toroidalcage}.
\mydoublesidefig{3.0in}{2.0in}{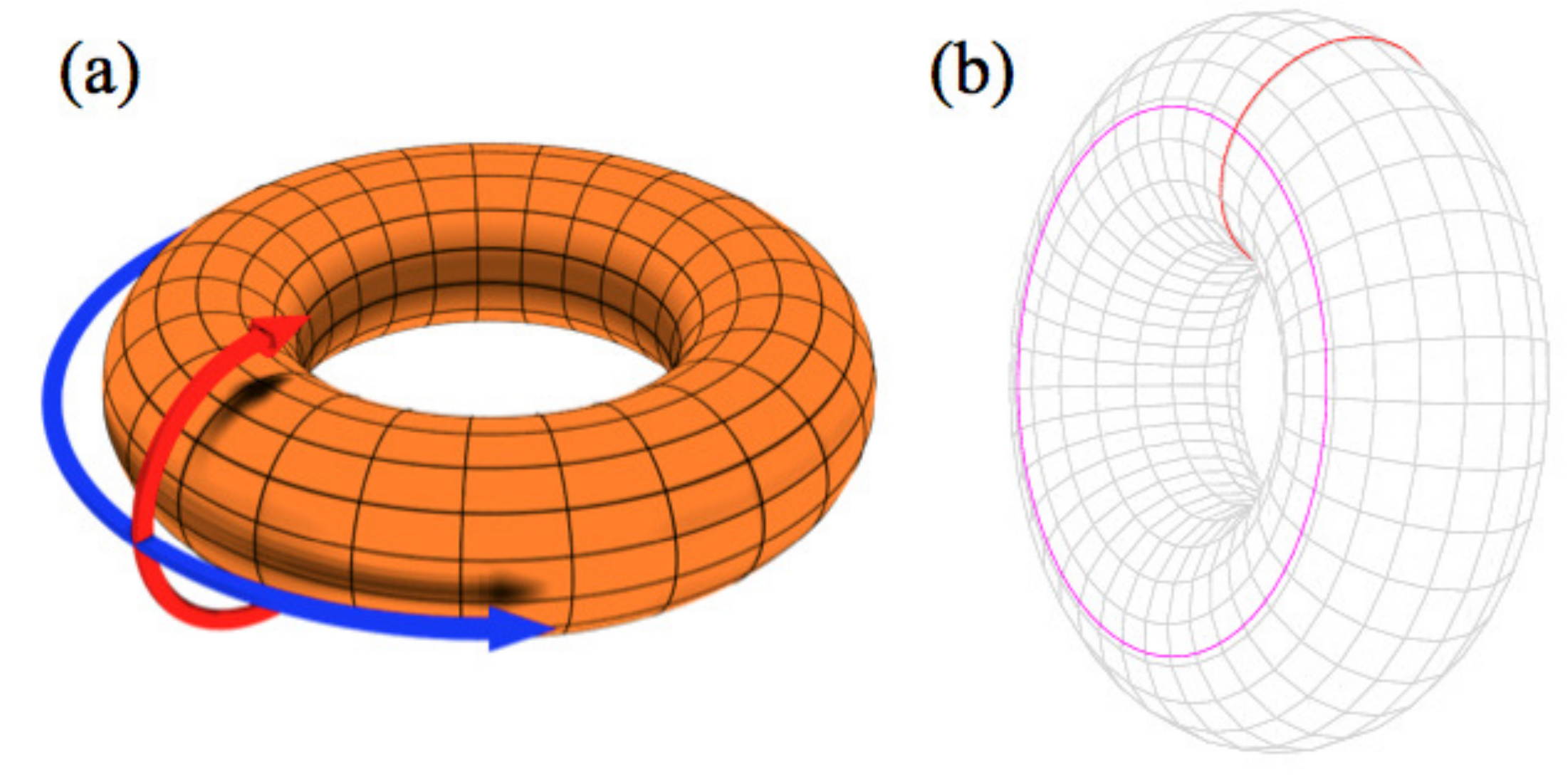}{Geometrical structure of a torus}{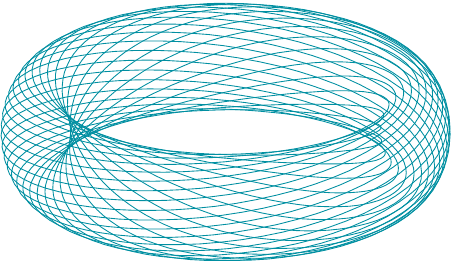}{Basic configuration of magnetic field lines in a tokamak}{Geometrical structure of a torus showing (a) poloidal (red arrow; short-way around) and toroidal (blue arrow; long-way around) directions; (b) a torus is a product of two circles: a red circle in a poloidal plane and a pink circle in a toroidal plane.  Figures are taken from Wikipedia.}{Basic configuration of magnetic field lines in a tokamak: toroidal cage with closed field lines.  Figure is taken from Ref. \cite{highcock_phd_2012}}
\newline\indent
A coordinate system of a tokamak is illustrated in \figref{coordinatesystem}.  The axis about which poloidal cross-sections are revolved is defined as $Z$-axis, and its value is the height from the midplane (a plane containing the magnetic axes).  Note that the centres of different flux surfaces are, to a good degree of precision, coplanar and perpendicular to the $Z$-axis, i.e., the Shafranov shift is mostly in the $R$-direction where $R$ denotes the major radius whose value is the distance from the $Z$-axis.  The minor radius $r$ is the distance from the magnetic axis, and $a_\psi$ is the half diameter of a flux surface $\psi$ at the magnetic axis height, i.e., at the midplane.  Here, $\psi$ is the flux surface label.  We use $a$ (without the subscript $\psi$) to denote the $a_\psi$ of the LCFS, and $R_0$ for $R$ at the point where $a_\psi\rightarrow 0$, which are measures of total plasma size.  Note that $r$ and $a_\psi$ can be different unless there is no Shafranov shift with circular poloidal cross-section.  $\tor$ and $\pol$ denote the toroidal and poloidal angles, respectively.  The origin of $\tor$ is not defined as it is not necessary for it being a symmetric direction; while $\pol$ is measured from the outboard midplane.  Outboard (inboard) is the region where $\vct{R}\cdot\vct{r}>0$ ($\vct{R}\cdot\vct{r}<0$). Note that in some chapters of this work we use a Cartesian coordinate system with a local approximation, and it is explicitly stated when we do so.
\myfig[5.0in]{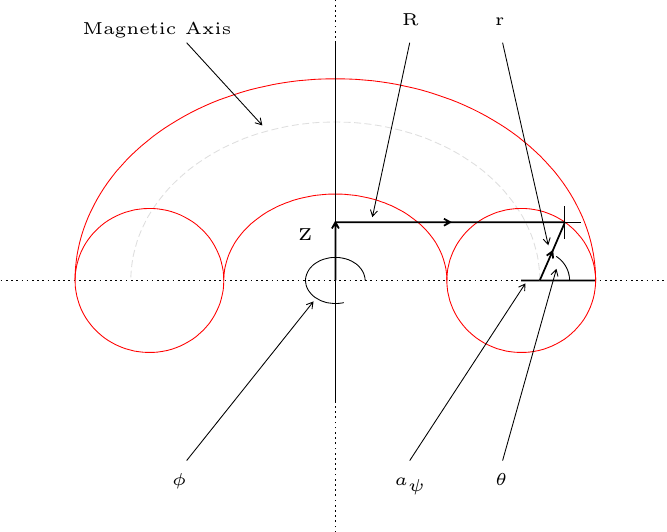}{A coordinate system of a tokamak}{A coordinate system of a tokamak showing the major radius $R$, minor radius $r$, height $Z$, toroidal $\tor$ and poloidal $\pol$ angles.  $a_\psi$ is a measure of plasma size at a flux surface $\psi$. Figure is taken from Ref. \cite{highcock_phd_2012}}
\newline\indent
Collisions in a plasma are not like ordinary instantaneous collisions in the sense that they are due to the long range Coulomb force with stochastic interaction between charged particles (binary) known as Coulomb collisions \cite{braginskii_rpp_1965}.  The characteristic Coulomb collision time of a plasma is usually considered as the time a particle takes to have a change of angle of order unity in the direction of velocity. \figref{helical2}(b) depicts the effect of Coulomb collisions where the marked position corresponds to the \textit{cumulative effect} of many Coulomb collisions.  Although collisions are necessary for fusion power generation, it is obvious that they will degrade the confinement time.  However, such a reduced confinement time due to Coulomb collisions is not what makes generating economical fusion power arduous since the associated particle diffusion is estimated to be at least an order of magnitude smaller than that experimentally observed \cite{freidberg_cambridge_2007}.  The observed transport is, in fact, anomalous exceeding both classical\footnote{Step size of random walk is the Larmor radius due to Coulomb collisions.} and neoclassical\footnote{Step size of random walk is the width of banana orbit of trapped particles due to the toroidal geometry.} transport \cite{helender_cambridge_2002} by more than an order of magnitude, and it is believed to be associated with plasmas being turbulent in a tokamak \cite{carreras_ieee_1997}, which has been the motivation of this work on plasma turbulence.
\newline\indent
As the final practical goal of the fusion community is to light up the whole world with economical fusion power plants, we must be able to ignite plasmas, i.e., generate self-sustained burning plasmas.  Known as the Lawson criterion \cite{lawson_psb_1957}, the ignition condition can be written as a triple product of density $n$, temperature $T$ and energy confinement time $\tau_E$ of plasmas:
\begin{equation}\label{eq:lawson_criterion}
nT\tau_E > 3\times10^{21}\:m^{-3}\:keV\:s,
\end{equation}
meaning that have as high a density as possible, be as hot as possible and retain particles as long as possible. The International Thermonuclear Experimental Reactor (ITER\footnote{www.iter.org}) endeavors to satisfy the condition with $n=10^{20}\:m^{-3}$, $T=10\:keV$ and $\tau_E\ge3\:s$.

\subsection{Mega Amp Spherical Tokamak (MAST)}
The spherical tokamak is a type of tokamak whose main difference from a ``conventional'' tokamak is the aspect ratio $\varepsilon_0^{-1}=R_0/a$.  A spherical tokamak has a tighter aspect ratio, $\varepsilon_0^{-1}\approx1.5$ than a conventional one $\varepsilon_0^{-1}\approx3$ as illustrated in \figref{tokamak_compare}.  Once it was predicted that a small aspect ratio tokamak could achieve higher beta $\beta$ operation \cite{troyon_ppcf_1984, peng_nf_1986} with a subsequent proposal of a new arrangement of magnetic coils \cite{peng_ornl_1984} in 1984, the Small Tight Aspect Ratio Tokamak (START) with $\varepsilon_0^{-1}=1.3$ was built at Culham Center for Fusion Energy (CCFE), known as Culham Laboratory then, and the first plasma was reported in 1992 \cite{sykes_nf_1992}.  Here, $\beta$ is the ratio of the plasma pressure to the magnetic field energy density which can be interpreted as the higher the $\beta$, the more economical it is because generating magnetic fields and building the coils are costly.  Successful results from START came out in 1997 that volume averaged (and central) $\beta$ reached $\sim11.5\:\%$ (and $\sim50.0\:\%$) and that $\beta$ could be increased further using higher input power \cite{sykes_pop_1997}.  The highest volume averaged $\beta$ achieved, then, was $12.6\:\%$ held by DIII-D\footnote{DIII-D has $R_0\sim1.7\:m$ and the central magnetic field $\sim2.0\:T$; whereas START had $R_0=0.3\:m$ with the central magnetic field $\sim0.5\:T$.} in San Diego, US \cite{strait_pop_1994}.  Subsequently, START had many experiments with the averaged $\beta\sim30\:\%$ and the highest (perhaps the highest record even to the current date) value of $40\:\%$ \cite{sykes_nf_1999}.
\myfig[4.5in]{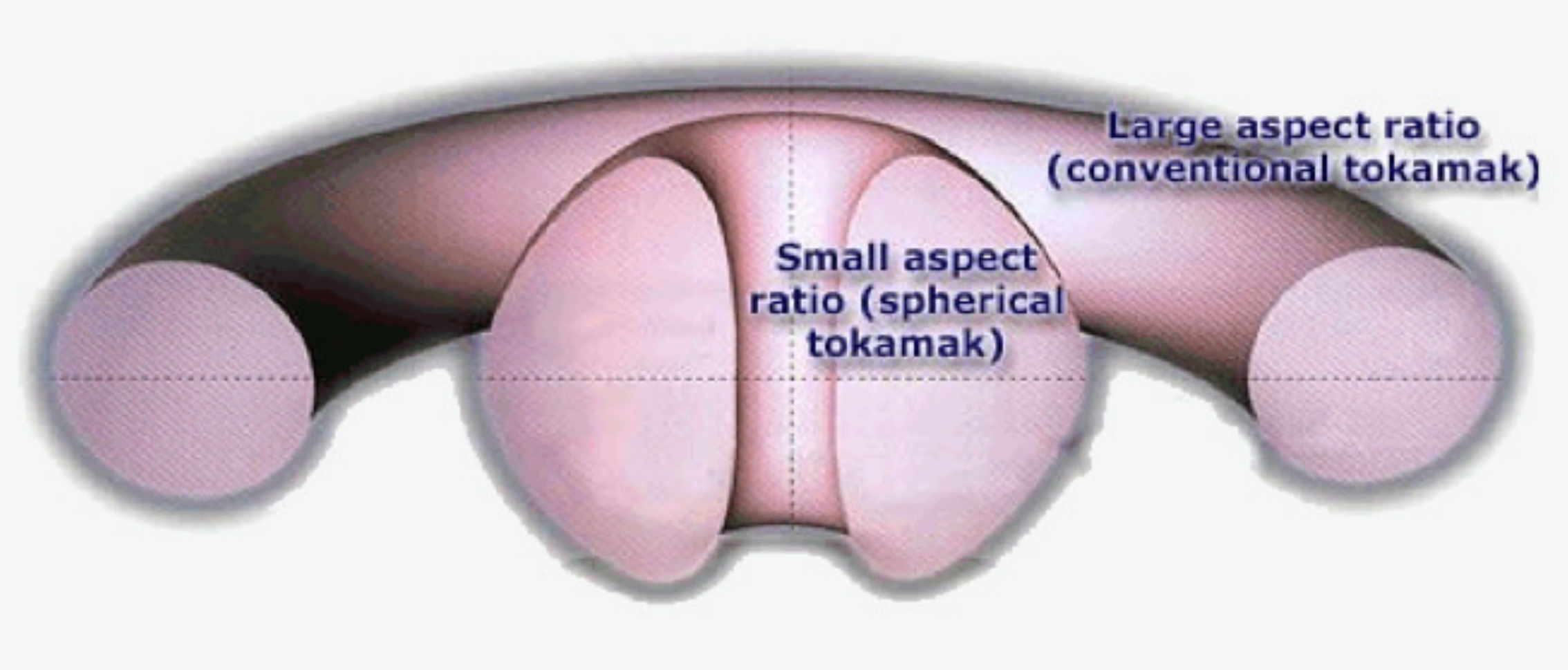}{Conventional vs. Spherical Tokamaks}{Comparisons between spherical and conventional tokamaks.  Geometrical configurations between the two tokamaks differ in their aspect ratio. Figure from CCFE.}
\newline\indent
These successful experiments on START led the team to upgrade START to the Mega Amp Spherical Tokamak (MAST) \cite{cox_fed_1999} with $R_0\sim0.8\:m$ and $a\sim0.6\:m$ ($\varepsilon_0^{-1}\sim1.3$) with plasma currents up to $\sim2.0\:MA$.  The first plasma in MAST was reported in 2001 \cite{sykes_nf_2001}.  Then, H-mode \cite{wagner_prl_1982} operation on MAST was reported \cite{akers_prl_2002}.\footnote{H-mode was also achieved on START \cite{sykes_prl_2000}.}  H-mode compared to L-mode (where H- and L- stand for high and low confinements) has a longer energy confinement time with steeper edge density and temperature gradients resulting in a larger stored energy in plasmas. In other words, H-mode plasmas establish an edge confinement ``barrier'' that reduces transport. H-mode operation has been achieved in many tokamaks (perhaps most of existing tokamaks), and they are accompanied by a reduction of turbulence (\cite{doyle_nf_2007}; and references therein).  This, again, provides the motivation of this work on plasma turbulence: \textit{what are the mechanisms that suppress the turbulence and how do we achieve it?}\footnote{H-mode operation is an empirical achievement without much physical understanding of it.}
\newline\indent
Before we sketch a simple picture of plasma turbulence, let us provide a picture of MAST (\figref{mast_plasma}) and a table summarizing some of its parameters (\tableref{table:mast_param}) from which the experimental data in this work were obtained.
\myfig[3.5in]{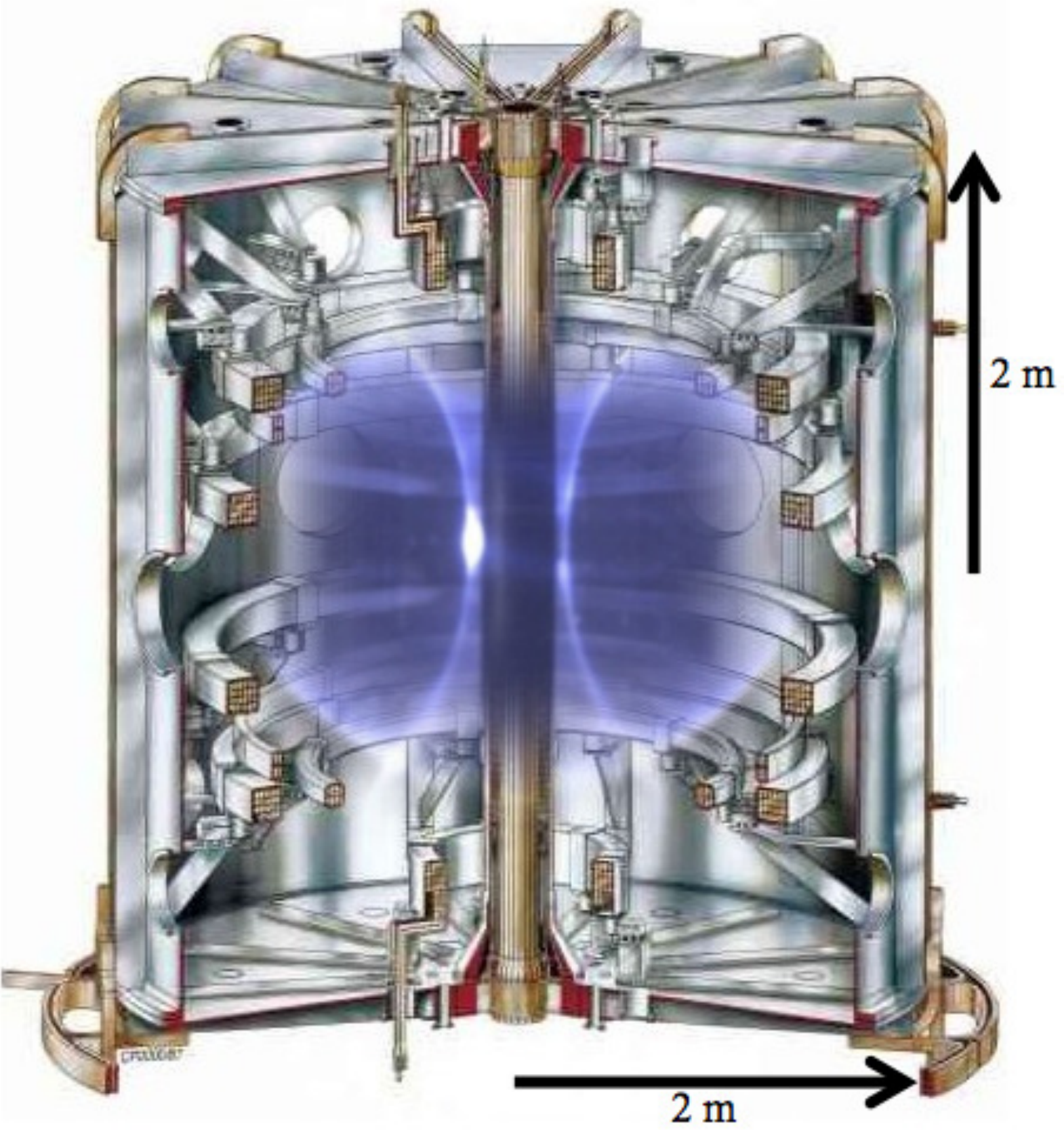}{A picture of MAST}{A picture of MAST.  The vacuum vessel and other structures are CAD drawings, and the overlaid plasma inside is a real picture from a MAST discharge.  Figure from CCFE.}
\begin{table}[t]\caption[Typical MAST parameters]{Typical MAST parameters \cite{freethy_phd_2012}} 
\centering
\label{table:mast_param}
\begin{tabular}{c | c }
Parameter & Value \\
\hline
Magnetic field on axis & $0.5\:T$ \\
Core temperature & $1.5\:keV$ \\
Core density & $5\times10^{19}\:m^{-3}$ \\
Plasma current & $1.3\:MA$ \\
Pulse length & $0.5\:s$ \\
Aspect ratio & $1.3$ \\
\hline
\end{tabular}
\end{table}

\section{Plasma turbulence in tokamaks}
Our theoretical level of understanding of turbulence, or turbulent flow, is so minimal (but interesting\footnote{Turbulence is intellectually so intriguing that proving existence and smoothness (no singularity) of the Navier-Stokes solution in three dimensional space is set to be one of the seven most important open mathematical problems by the Clay Mathematics Institute (www.claymath.org).}) that even the definition of turbulence is not well established \cite{davidson_oxford_2004}.  Consequently, experimental measurements of turbulence are crucial for us to understand it better, or at least its physical properties.  For this reason, a 2D ($8$ radial $\times$ $4$ vertical (poloidal) channels) beam emission spectroscopy (BES) diagnostic was installed on MAST to measure density fluctuations associated with turbulence.  We know that the nonlinearity of a system and the linear drive of unstable modes are vital for the system to be in a turbulent state, otherwise the system is just filled with well-behaved (stable) waves. In this work, we restrict the description of plasma turbulence to its observed properties such as linear drives, spatial and temporal characteristics and nonlinearly saturated levels of turbulence (density fluctuations), and how these quantities are correlated with local equilibrium quantities.

\subsection{Drift waves and shear flows}
Tokamak plasmas are far away from a 'global' thermal equilibrium state, but rather in a force-balanced state: $\vct{J}\times\vct{B}=\grad p$ with the plasma current density $\vct{J}$ and pressure $p$ and the background magnetic field $\vct{B}$.  Note that plasmas are in thermal equilibrium 'locally'.  Hence, there can exist gradients in the moments of phase-space distribution functions such as densities $n$, fluid velocities $\vct{U}$ and temperatures $T$.  Gradients in density and temperatures of the ions $T_i$ and electrons $T_e$ drive drift waves, which, in turn, can cause plasmas to be in a turbulent state via nonlinear self-interactions of drift waves \cite{carreras_ieee_1997, horton_rmp_1999, tynan_ppcf_2009, horton_ppcf_1980, waltz_pf_1988, fonck_prl_1989, wootton_pfb_1990, cowley_pfb_1991, kotschenreuther_pop_1995, dimits_pop_2000, dorland_prl_2000, jenko_pop_2000, dannert_pop_2005}.  More detailed descriptions of drift waves are provided in \appendixref{ch:drift_wave}.
\newline\indent
While the gradients of density and temperatures are regarded as the linear drives, sheared mean plasma flows can both drive and suppress the turbulence: shear in the parallel component of the plasma flow $\grad\Upar$ can drive turbulence \cite{catto_pf_1973, newton_ppcf_2010, schekochihin_ppcf_2012, highcock_prl_2012} (as described in \appendixref{sec:temp_vel_driven_dr_wave}), while the perpendicular component of shear $\grad\Uper$ is known to suppress the turbulence \cite{barnes_prl_2011, highcock_prl_2010, roach_ppcf_2009, camenen_pop_2009, kinsey_pop_2005, dimits_nf_2001, waltz_pop_1994, mantica_prl_2009, mantica_prl_2011, burrell_pop_1999, burrell_pop_1997}. Here, $\Upar$ and $\Uper$ are the parallel and perpendicular (with respect to the background magnetic field) components of the mean plasma flows, respectively. Thus, the ratio of the two shearing rates, $\grad\Upar/\grad\Uper$, plays an important role on the plasma turbulence in tokamaks \cite{highcock_prl_2012}.
\newline\indent
Let us digress briefly from the shear flow issue to introduce a new quantity called the 'safety factor' $q$.  This will turn out to be one of the major parameters controlling the ratio $\grad\Upar/\grad\Uper$. The safety factor $q$ is defined as \cite{wesson_clarendon_2004}
\begin{equation}\label{eq:q_definition}
q = \frac{1}{2\pi}\oint \mathrm{d}l\frac{1}{R}\frac{\Btor}{\Bpol},
\end{equation}
where $\Btor$ and $\Bpol$ are the toroidal and poloidal components of the magnetic field, respectively.  The closed line integral ($\oint \mathrm{d}l$) is carried over one poloidal rotation at a fixed flux surface. $q$ is interpreted as the number of required toroidal rotations for a magnetic field line to close itself, i.e., to complete one poloidal rotation.  For a large aspect ratio tokamak with a circular poloidal cross-section, $q$ can be approximated as: 
\begin{equation}\label{eq:q_approx}
q\approx\frac{r}{R}\frac{\Btor}{\Bpol}=\varepsilon\frac{\Btor}{\Bpol},
\end{equation}
where $\varepsilon=r/R$ (cf. $\varepsilon_0=a/R_0$).  \figref{b_field_structure}(a) shows an example of a flux surface with $q=4$. Note that $q$ is called the safety factor because of its close relation with the MHD (magnetohydrodynamic) stability: the higher the $q$, the more MHD stable it is \cite{wesson_clarendon_2004}.  How $q$ is related to the MHD stability is beyond the scope of this work.
\myfig[5.0in]{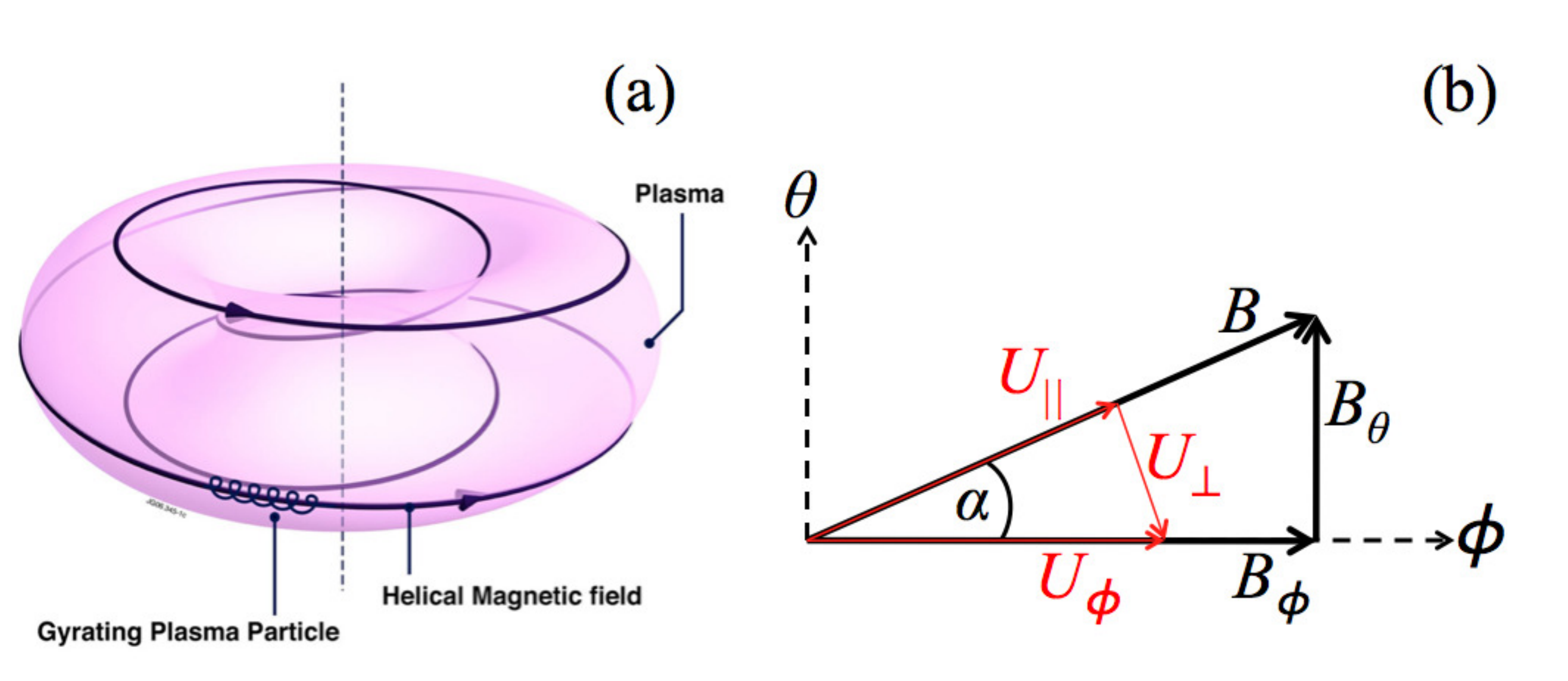}{Magnetic field line on a flux surface}{(a) A schematic of a helical magnetic field line with $q=4$ on a flux surface. Figure from www.efda.org.; (b) A schematic of local magnetic field vectors (black arrows) in toroidal $\Btor$ and poloidal $\Bpol$ directions with the pitch angle $\alpha$.  Red arrows illustrate local mean plasma flows in parallel ($\Upar$), perpendicular ($\Uper$) and toroidal ($\Utor$) directions.}
\newline\indent
Suppose that the mean plasma flow is purely in the toroidal direction\footnote{The mean flows in the poloidal direction are strongly damped by the (neoclassical) collisions \cite{connor_ppcf_1987, catto_pf_1987, hinton_pf_1985, cowley_clr_1986}, and \chref{ch:eddy_motion} in this work shows experimental signatures consistent with the neoclassical damping of the poloidal flow.} denoted as $\Utor$, then $\Upar/\Uper$ is equal to $\Btor/\Bpol$ (see \figref{b_field_structure}(b)) which is, in turn, just $q/\varepsilon$ from \eqref{eq:q_approx}.  Thus, the ratio $\grad\Upar/\grad\Uper$, i.e., the ratio of turbulence drive and suppression, can be approximated as $q/\varepsilon$, and we expect that the smaller $q/\varepsilon$, the less turbulence.  \textit{Experimental observations on the turbulence level as a function of $q/\varepsilon$, one of the major topics of this work, is discussed in \chref{ch:larger_RLTi}}. 

\subsection{Spatial and temporal characteristics of plasma turbulence}
One of the common features in wide ranges of turbulent flows observed in nature, such as smoke from a cigarette, water flows in tunnels, oceanic currents, hurricanes, clouds, and the solar wind, is that they all have broadband spectra both in the frequency and the wavenumber domain.\footnote{However, the scalings can be quite different for different physical phenomena.}  Plasma turbulence in a tokamak is no exception to this. Indeed, the plasma turbulence observed using the 2D BES diagnostic installed on MAST exhibits broadband frequency spectra (see \figref{bes_spec}). Note that the BES turbulence data are not analysed in terms of wavenumber spectra due to the limited number of spatially separated channels of the diagnostic.\footnote{We may be able to construct wavenumber spectra from the data using the maximum entropy method \cite{tanaka_rsi_2008, skilling_mnras_1984}, but such a technique has not been used. This is left as a future work.}  \textit{The principle of this diagnostic, without which this work would not exist, is explained in \chref{ch:bes_principle}.}
\newline\indent
The broadband spectra of turbulence means that there exist many different spatial and temporal scales from the energy-containing (or energy-injection) range down to the dissipation range through the inertial range if these ranges are well separated in their scales.  In this work, we concentrate on the spatial and temporal scales of the plasma turbulence in the energy-containing range, $\kper\rhoi < 1$, due to the limited spatial resolution of the 2D BES diagnostic, i.e., the sensitivity of the diagnostic decreases for the higher wavenumber ranges.  Here, $\kper$ is the perpendicular wavenumber of the turbulence and $\rhoi$ the ion Larmor radius.  Nevertheless, because the energy-containing range gives the largest contribution to the turbulent transport due to the large scale of the fluctuations, studying and understanding the turbulence characteristics in this range is the most critical to building a more efficient fusion power plant.
\newline\indent
Spatial structures of the plasma turbulence in a tokamak can be thought about in terms of time scales, a consequence of the $\vct{v}\cdot\grad$ operator which arises in the evolution equations where $\vct{v}$ can be related to various physical effects.   For instance, $v_\parallel\grad_\parallel$ acting on some turbulence distribution function $h$ in phase-space is associated with the parallel dynamics of the turbulence, hence the parallel structure of the turbulence.  Thus, it can be identified what physical effects influence (or possibly determine) the spatial structures of the turbulence by comparing various time scales in a tokamak.  In fact, we do find from our experimental data that \textit{the parallel streaming time, the magnetic drift time and the linear drive time associated with drift waves driven by density- or ion-temperature-gradient scale with the turbulence correlation time consistently.  These results together with the consequences of these ``balanced time scales'' on the turbulence spatial structures are discussed in \chref{ch:critical_balance}.}

\chapter{Structure of this work}
Our goal is to observe and understand the turbulence in MAST, hence we installed a 2D BES diagnostic on MAST, and this work is based on the experimental data from the diagnostic.  We structure our work in the following logical steps:
\newline
(1) We need to obtain useful information, such as the properties of turbulence, from raw BES data.  Thus, we develop statistical techniques to do so in \chref{ch:bes_principle}.
\newline
(2) The statistical techniques must be examined for their reliability, hence we develop a way to generate synthetic 2D BES data in \chref{ch:synthetic_bes} for which all the properties of turbulence are known in advance so that the statistical techniques can be examined.
\newline
(3) Then, we obtain turbulence information from the 2D BES data using the statistical techniques, and make conclusions about physics of turbulence in subsequent chapters: what causes fluctuating density patterns to move in the poloidal direction in the lab frame while the plasmas are rotating toroidally (\chref{ch:eddy_motion}), and what equilibrium quantities are correlated (or determine) spatial and temporal characteristics of turbulence (\chref{ch:critical_balance}).
\newline
(4) Finally, we describe how we can improve the performance of tokamaks with a careful correlation analysis among many local equilibrium parameters in \chref{ch:larger_RLTi} by producing a critical manifold separating turbulent and non-turbulent state.
\newline\indent
We first describe in more detail how the 2D BES diagnostic can be used to measure density fluctuations up to a few $100$ kHz range with the spatial resolution of $\sim2\:cm$, and how we obtain the characteristics of turbulence from the BES data in \chref{ch:bes_principle}.  Then, we explain in \chref{ch:synthetic_bes} how to generate synthetic BES data, a forward model of the 2D BES diagnostic, which can be used in many different aspects such as examining the reliability of statistical analyses as in \chref{ch:eddy_motion} and performing direct comparisons between outputs of numerical turbulence simulations and experimental turbulence data \cite{field_ttf_2012, shafer_pop_2012, holland_pop_2011}.
\newline\indent
We find that the fluctuating density patterns move in the poloidal (vertical) direction in the lab frame.  This seemingly contradicting result with the mean plasma flows being dominantly toroidal is discussed in \chref{ch:eddy_motion}.  We show, via a careful ordering of the density continuity equation, that such an apparent poloidal motion of the patterns arises due to the fact that the elongated patterns in the parallel direction are advected by the dominant mean toroidal plasma flows provided there exists a finite angle between the parallel and toroidal directions, i.e., the projection effect analogous to the apparent up-down motion of helical strips of a 'rotating barber-pole' \cite{munsat_rsi_2006}.
\newline\indent
Next, we compare the turbulence correlation time, the particle (ion) parallel streaming time, the drift time associated with ion temperature or density gradients and the magnetic drift time finding that they are all comparable in \chref{ch:critical_balance}.  This result suggests that the observed turbulence in MAST is ``critically balanced'' and its characteristics are determined by the local equilibrium, from which spatial correlation lengths of turbulence are derived and examined.  Furthermore, we infer the turbulence nonlinear time from the density fluctuations, and we find that the ratio of the inferred nonlinear time to the turbulence correlation time is a function of ion-ion collisionality.  We argue that this observation is consistent with the decorrelation of turbulence being dominantly controlled by zonal flows \cite{diamond_ppcfreview_2005, fujisawa_nf_2009}. 
\newline\indent
The final result we present in this work provides a way to achieve better tokamak performance.  We show in \chref{ch:larger_RLTi} statistically that the normalized ion-temperature-gradient scale length $R/\LTi$ is inversely correlated with local $q/\varepsilon$, the ratio between the shearing rates of parallel and perpendicular flows; while $R/\LTi$ and the local shearing rate of the mean toroidal plasma flows are positively correlated.  The dependence of $R/\LTi$ on $q/\varepsilon$ is strong which implies that $R/\LTi$ can be increased at a fixed shearing rate of the mean flows by lowering $q/\varepsilon$. Furthermore, we present a critical manifold in local equilibrium parameter space separating turbulent and non-turbulent state based on the fact that the observed turbulent heat flux is inversely correlated with $\RLTi$.
\newline\indent
Then, we close this work with the conclusions in \chref{ch:conclusions}.
\newline\newline\noindent
The author of this work generated point-spread-functions of the 2D BES system, developed software to generate synthetic 2D BES data and to perform statistical analyses on raw BES data. A. Field, D. Dunai and S. Zoletnik designed, installed the 2D BES system on MAST and took the data. C. Michael generated EFIT data which contain the equilibrium magnetic field information. A. Schekochihin, I. Abel, M. Barnes, G. Colyer, S. Cowley, F. Parry, E. Highcock provided theoretical insights.

\mypart{Turbulence (Density Fluctuation) Measurements}
\chapter{Beam Emission Spectroscopy (BES) diagnostic: Measuring density fluctuations}\label{ch:bes_principle}
\begin{center}
\textit{This chapter is largely based on Refs. \cite{ghim_rsi_2010, ghim_ppcf_2012}.}
\end{center}
\section{Principle of 2D BES turbulence diagnostic}\label{sec:how_bes_works}
The 2D BES system on MAST utilizes an avalanche photodiode (APD) 2D array camera \cite{dunai_rsi_2010} with eight columns and four rows of channels, which have an active area of $1.6\times 1.6\:mm^2$ each.  It measures the Doppler-shifted $D_\alpha$ emission from the collisionally excited neutral-beam atoms\footnote{Neutral beams are injected mainly to heat and to provide toroidal momentum to MAST plasmas.} (deuterium) with a temporal resolution of $0.5\:\mu s$.  The optical system is designed such that the observed locations can be radially scanned along the path of the neutral beam (South-Neutral Beam Injection (S-NBI)) whose $1/e$ half-width is $8\:cm$, while the optical focal point follows the axis of the beam (see \figref{mast_top_plan}). More detailed descriptions on the optical system of the 2D BES are available in Ref. \cite{field_rsi_2012}.  The nominal location of the BES system, i.e., where the optical line-of-sight (LoS) is best aligned with the local magnetic field,\footnote{The BES measurements are spatially localized to a good degree because the LoS at the intersection of the beam are approximately parallel to the local magnetic field lines and the intersection length is much shorter than the parallel correlation length of the turbulence.} is at major radius $R=1.2\:m$.  At this location, a magnification factor of $8.7$ at the axis of the beam results in each channel observing an area of $1.5\times 1.5\:cm^2$ with $2\:cm$ separation between the centres of adjacent channels (see \figref{mast_side_plan}). The poloidal (vertical) locations of the views are fixed at $Z=-0.03, -0.01, 0.01$ and $0.03\:m$. 
\mydoublesidefig{2.2in}{3.2in}{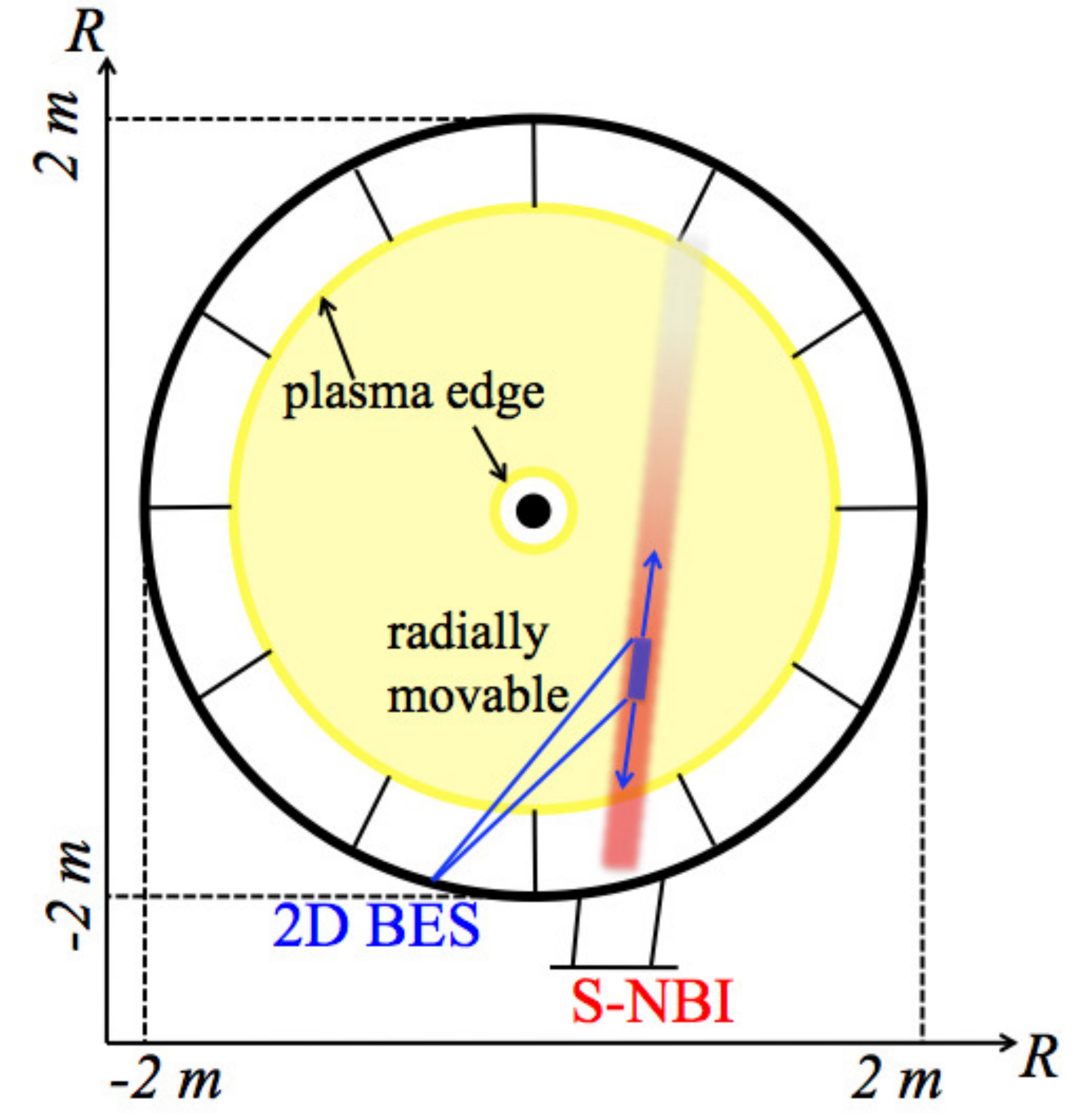}{MAST top plan view}{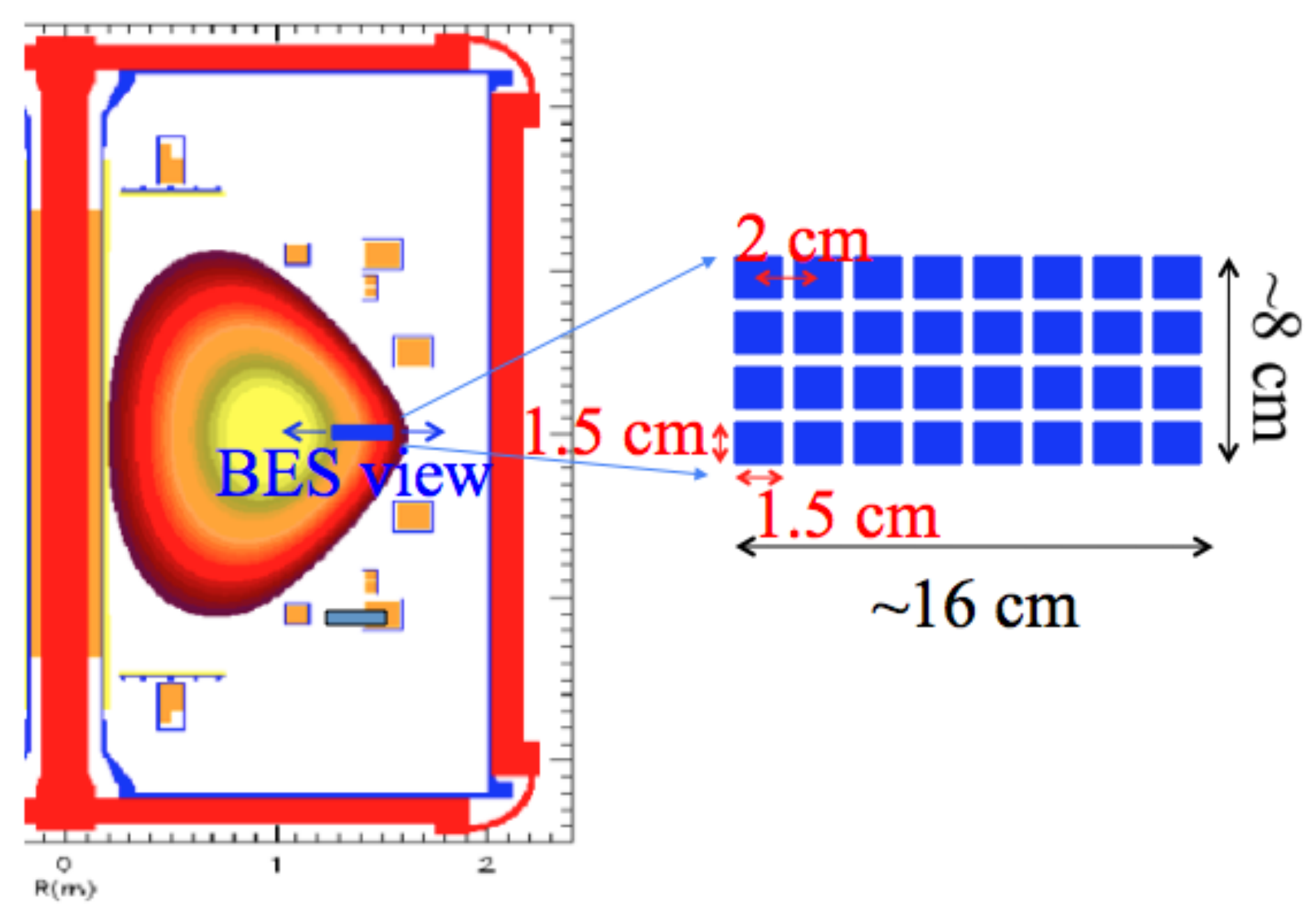}{MAST side plan view}{Top view of MAST. South-NBI is injected to heat and rotate (toroidally) the plasmas (red strip), and the optical system of the 2D BES is focused to the axis of the S-NBI (blue rectangle) to collect Doppler-shifted $D_\alpha$ emissions from the collisionally excited neutral-beam atoms.  The 2D BES can be scanned radially following the axis of the S-NBI.  A typical injection energy of the S-NBI is $60-70\:keV$ with the $2-3\:MW$ of input power.}{Side view of MAST.  Radial viewing position of the 2D BES can be scanned from the core to the edge of plasmas.  The 2D BES system consists of $8$ radial $\times$ $4$ poloidal channels with $2\:cm$ separation between the neighboring channels in both directions.  The area of each channel is about $1.5\times 1.5\:cm^2$ which comes from the magnification factor of $8.7$.  This varies slightly depending on the radial viewing position of the 2D BES system.}
\newline\indent
The angle between the LoS of the 2D BES system and the neutral beam with the injection energy of $60-70\:keV$ results in a Doppler shift of the $D_\alpha$ emission approximately $3\:nm$ to the red from the background $D_\alpha$ at $656.28\:nm$  (see \figref{bes_wave_spec}).\footnote{There exists strong background $D_\alpha$ emission from the plasma edge.}  The background $D_\alpha$ can be removed with a suitable optical filter \cite{field_rsi_2012}, and so that the $D_\alpha$ emission detected by the 2D BES system comes only from the neutral beam, hence the measurement is localized to the beam. A more detailed description of the 2D BES system on MAST can be found elsewhere \cite{field_rsi_2012}.
\newline\indent
The measured intensity of the $D_\alpha$ beam emission is directly related to the background plasma density because the latter is the cause of the excitation of the neutral-beam atoms. The beam atoms are excited by collisions with the electrons, ions and impurities, but at energies greater than $40\:keV$, the electron contribution can be ignored \cite{fonck_rsi_1990}. The fluctuating part of the plasma (ion) density $\dn$ can be determined according to 
\begin{equation}\label{eq:photon_dens_relation}
\frac{\dn}{n} = \frac{1}{\betabes}\frac{\dI}{I},
\end{equation}
where $n$ is the mean plasma density, and $\dI$ and $I$ denote the fluctuating and mean parts of the photon intensity, respectively.  $\betabes$ is a coefficient depending on the population of the excited state and is a weak function of the background plasma density with values in the range $0.3 < \betabes < 0.7$ with the negligible temperature dependence \cite{hutchinson_ppcf_2002}.  $\betabes$ is calculated according to a collisional-radiative model, and we use values from the work of Hutchinson in \cite{hutchinson_ppcf_2002}.  A careful modeling of the 2D BES system shows that the system is capable of measuring the density fluctuation levels down to a few $0.1\:\%$ \cite{field_rsi_2012}.\footnote{Typical fluctuation levels at the edge of plasmas are of the order of $1\:\%$ and $0.1\:\%$ in the core for L-mode discharges, while core regions in H-mode discharges usually have smaller fluctuation levels.}
\newline\indent
\myfig[5.5in]{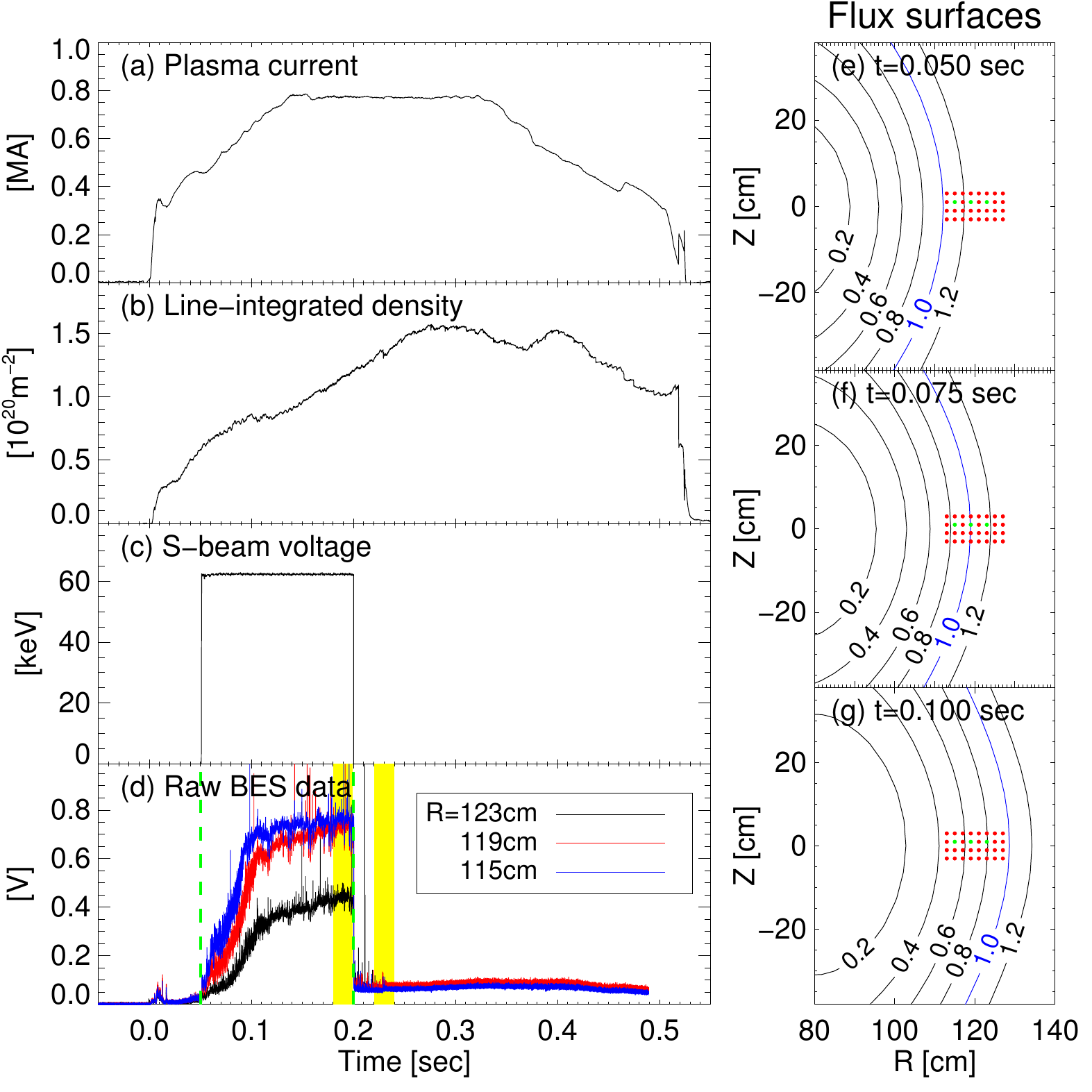}{Examples of raw BES data}{Evolution of (a) the plasma current (b) line-integrated electron density (c) S-beam voltage and (d) BES intensity data at $R=123\:cm$ (Ch.\#19: black), $119\:cm$ (Ch.\#21: red) and $115\:cm$ (Ch.\#23: blue).  The vertical green dashed lines show the times when the 2D BES system obtains localized density fluctuations from the S-NBI.  The two yellow strips ($0.18-0.20\:s$ and $0.22-0.24\:s$) indicate the time durations when the spectra of beam emission are measured shown in \figref{bes_wave_spec}.  (e), (f) and (g) show the contour of flux surfaces $\psi$ at $t=0.05\:s$, $0.075\:s$ and $0.1\:s$, respectively.  Circles show the viewing positions of the BES system (the green circles indicate the locations where BES signals are obtained for (d)).}
\myfig[5.0in]{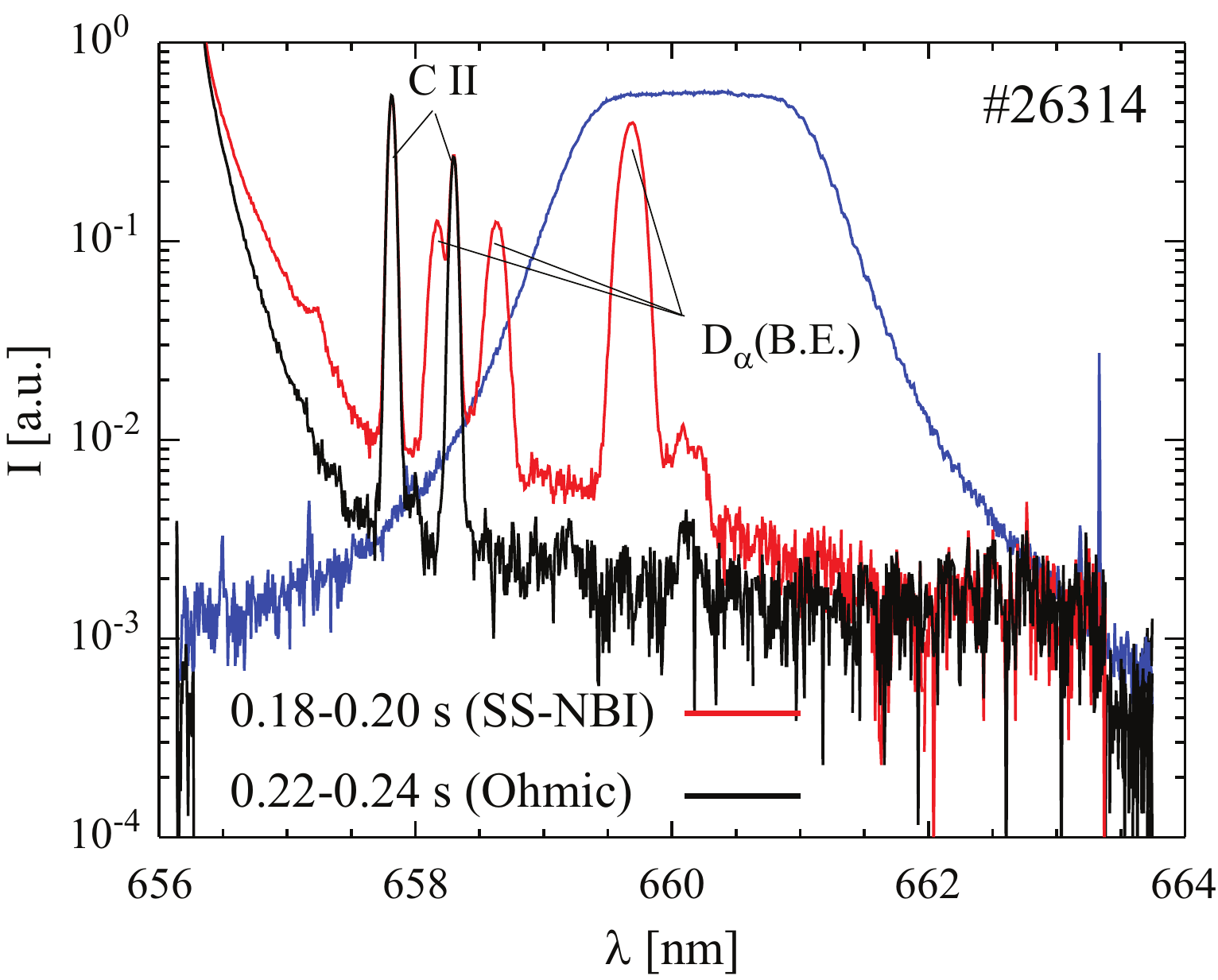}{Spectra of beam emission signals}{Spectra of beam emission measured both before (red) and after (black) the S-beam cut-off time using the Motional Stark Effect (MSE) spectrometer at a viewing radius of $1.2\:m$ together with the transmission of the BES filter (blue).  Figure taken from Ref. \cite{field_rsi_2012}.}
\myfig[4.5in]{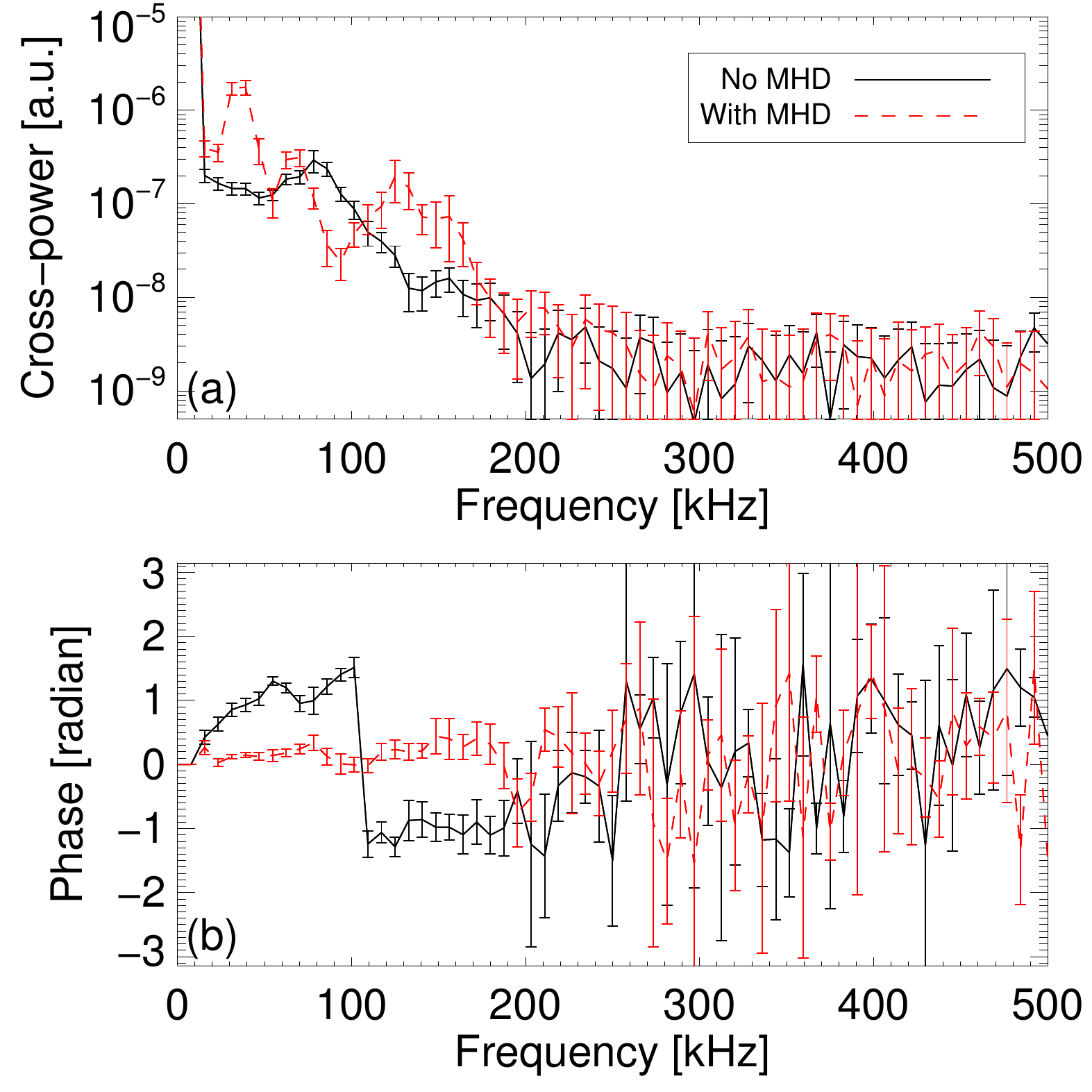}{Crosspower and crossphase spectra of BES data}{Spectra of crosspower (top panel) and crossphase (bottom) of a measured fluctuating density from 2D BES system on MAST when the signal contains MHD modes (red dashed) or no MHD modes (black solid).}
Thus, the 2D BES system on MAST directly measures fluctuations of plasma density in the radial ($8$ channels)-poloidal ($4$ channels) plane at a fixed toroidal location with the spatial resolution of $\sim 2\:cm$ in both directions when the S-NBI is injected.\footnote{BES signals are not spatially localized when S-NBI is not injected (see \figref{bes_basic_char}(d)).}  Note that the actual spatial resolution is somewhat broadened due to various physical effects which are explained in more detail in \secref{sec:bes_psf}.
\newline\indent
\figref{bes_basic_char} shows the evolution of (a) the plasma current, (b) line-integrated electron density, (c) S-beam voltage and (d) BES intensity data from three different radial locations, $R=123\:cm$, $119\:cm$ and $115\:cm$, at a fixed poloidal location $Z=1.0\:cm$.  It shows that the levels of BES signals rapidly increase after the S-NBI is injected at $t=0.05\:s$ as the size of plasmas increases until around $0.1\:s$.  This increase in the size of plasmas can be seen from \figref{bes_basic_char}(e)-(f) showing the contour of normalized flux surfaces $\psi$, i.e., $\psi=1.0$ at the LCFS, at $t=0.05\:s$, $0.075\:s$ and $0.1\:s$.  The circles show the viewing locations of the BES system (three green circles correspond to the positions where the BES signals are obtained for \figref{bes_basic_char}(d)).  Then, the BES signals slowly increase as the electron density increases until the S-NBI cuts off at $t=0.2\:s$. \figref{bes_wave_spec} shows measured spectra of beam emission from $Z=0\:cm$ at the same toroidal location as the BES system using the Motional Stark Effect (MSE) diagnostic viewing at a radial location of $R=1.2\:m$.  The observed Doppler shifts of the $D_\alpha$ lines are almost the same as those for the BES system.  Spectra are shown during a period with the S-beam voltage on (red) and during the ohmic phase just after the beam is switched off (black) (see the yellow strips in \figref{bes_basic_char}(d)).  There are three Doppler shifted $D_\alpha$ lines corresponding to the emissions from $D$, $D_2$ and $D_3$ corresponding the full, half and third energy components of the neutral beam.  There are non-zero levels of BES signals even after the S-NBI cut-off (\figref{bes_basic_char}(d)) which is due to the CII ($657.81, 658.29\:nm$) lines as the carbon impurity density is not negligible in MAST. \figref{bes_spec} shows a typical cross-power spectrum and cross-phase between the two poloidally separated channels with and without MHD modes. Effects of MHD modes in determining turbulence characteristics are discussed in \secref{sec:get_turb_info}.

\section{Point-spread-functions of the 2D BES system}\label{sec:bes_psf}
BES is a volume-sampling diagnostic, therefore in order to generate synthetic BES data (see \chref{ch:synthetic_bes}) one really requires LoS integration based on three-dimensional input data of the fluctuating plasma density which would require very large data files.  However, by availing the spatial structure of the turbulence which is elongated along the magnetic field lines synthetic BES data can be constructed with 2D input data on the poloidal cross-section using the 2D point-spread-functions (PSFs) of the detectors.  The relevant physical effects such as magnetic field-line curvature, LoS geometry, finite excited-state lifetime of $D_\alpha$, the beam attenuation and divergence must all be taken into account in the calculation of the PSFs.  Without considering these effects, the perpendicular spatial resolution will be limited to $\sim 2\:cm$.  
\newline\indent
In order to calculate the 2D PSFs, successive image planes are constructed along the LoS \cite{shafer_rsi_2006}. The size of the detector images at these planes is set by the optical magnification factors which vary along the LoS. Light-cones whose sizes are determined by the optical system are convolved with these detector images.  The resulting blurred images are then convolved with an exponentially decaying function whose $1/e$ distance is calculated by considering the beam velocity projection of the perpendicular to the LoS and the half-life of $D_\alpha$.  Hutchinson \cite{hutchinson_ppcf_2002} calculated the half-life of $D_\alpha$ denoted as $\tau_3$, and it is reported that at a plasma density of $\sim10^{19}\:m^{-3}$, $\tau_3$ is $\sim3-10\:ns$.  The resultant images along the LoS are then moved to the optical focal plane by following the magnetic field-lines.\newline\indent
\myfig[5.5in]{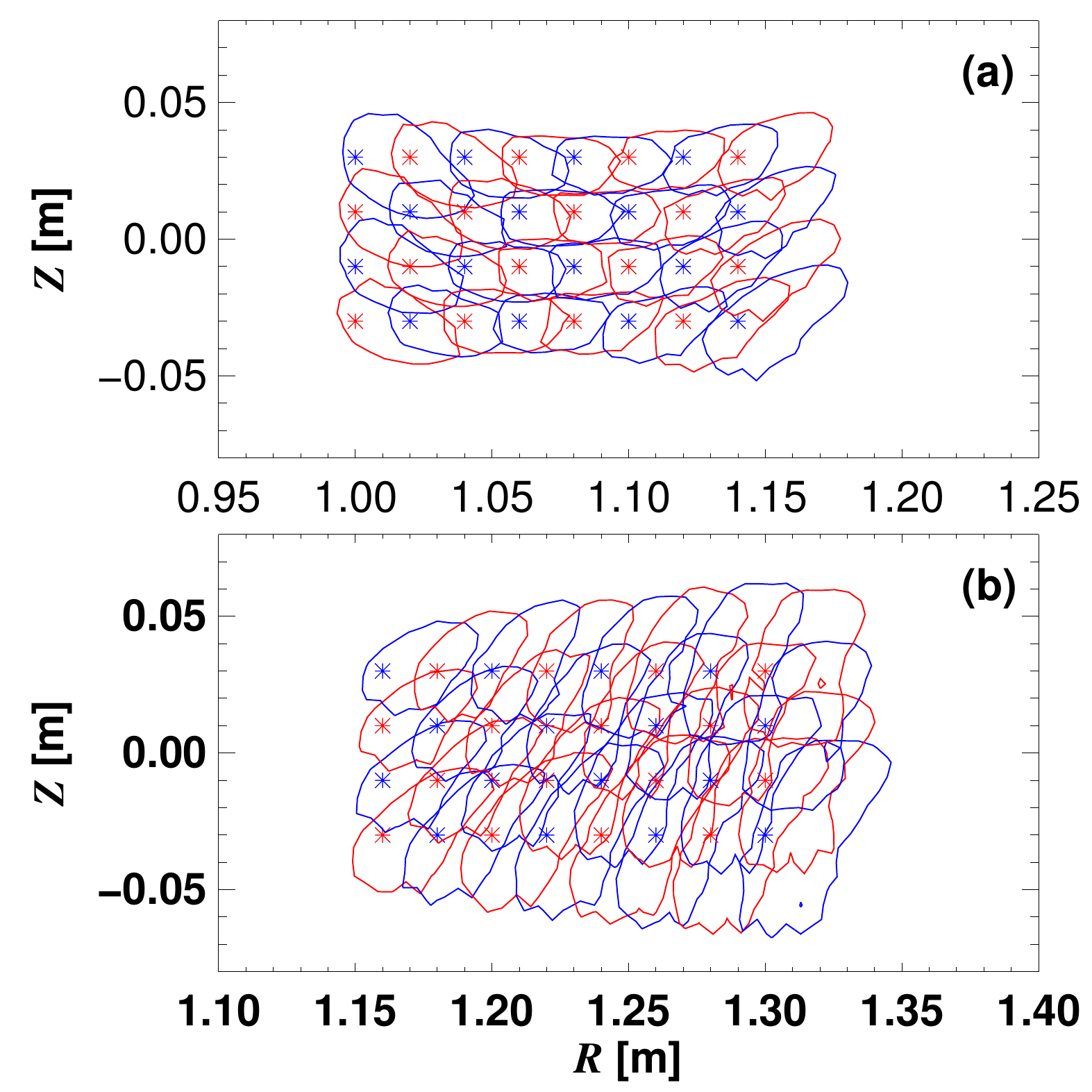}{Examples of PSFs of the 2D BES system}{Example of PSFs of the 2D BES system covering (a) $1.00\:m < R < 1.14\:m$ and (b) $1.16\:m < R < 1.30\:m$.  The lines (red and blue) are the $1/e$ contour lines of the PSFs, and the asterisks show the optical focal points.} 
\figref{bes_psf} shows examples of generated PSFs at two different radial locations covering (a) $1.00\:m < R < 1.14\:m$ and (b) $1.16\:m < R < 1.30\:m$.  The radial smearing effects are mainly due to the finite half-life of $D_\alpha$ as the excited neutral-beam atoms can travel (radially) finite distances before they emit $D_\alpha$ fluorescence.  On the other hand, the poloidal smearing is due to the pitch angle $\alpha$ of the magnetic field.  The pitch angle changes much more significantly from the core to the edge of plasmas in a spherical tokamak compared to a conventional tokamak, and this effect can be seen in the strong dependence of the poloidal width of the PSFs as a function of viewing radius.  Consequently, deconvolution of the measured signal using the calculated PSFs on MAST is not trivial as the functions vary in space which is not the case in a conventional tokamak such as in DIII-D tokamak \cite{shafer_rsi_2006}.  More sophisticated algorithms will be required to enable the deconvolution with spatially varying PSFs \cite{lauer_spie_2002}.  Utilizing such an algorithm is beyond the scope of this work.  In fact, we find that most of measured turbulence has poloidal correlation lengths longer than the poloidal width of PSFs, and we do not use the data points whose radial correlation lengths are less than $2.0\:cm$.

\section{Extracting turbulence characteristics from the BES data}\label{sec:get_turb_info}
Once the raw BES data are obtained, we need to extract useful information, i.e., spatial and temporal characteristics of turbulence, from the data.  This is done by calculating the spatio-temporal covariance function and its normalized version, the correlation function, throughout this work.  The covariance function is calculated as
\begin{equation}
\cov\lp\DR, \DZ, \Dt\rp =\lab \dI\lp R, Z, t\rp\dI(\lp R+\DR, Z+\DZ, t+\Dt\rp\rab,
\end{equation}
and the correlation function
\begin{equation}
\corr\lp\DR, \DZ, \Dt\rp = \frac{\cov\lp\DR, \DZ, \Dt\rp}
{\sqrt{\lab\dI^2\lp R, Z,t\rp\rab\lab\dI^2\lp R+\DR, Z+\DZ, t+\Dt\rp\rab}},
\end{equation}
where $R$ and $Z$ denote the radial and poloidal (vertical) coordinates with their channel separation distances $\DR$ and $\DZ$, respectively, and $t$ the time and $\Dt$ the time lag; and $\lab\cdot\rab$ denotes a time average. Note that we use time-averaging instead of ensemble-averaging throughout the work by assuming that our signal is ergodic.\footnote{Ideally, we want to repeat the same experiment many times and average the data from the repeated experiments. However, as repeating the experiment is expensive (not to mention whether repeating it is possible or not), such ensemble-averaging is not possible in practice.}
\newline\indent
\myfig[5.0in]{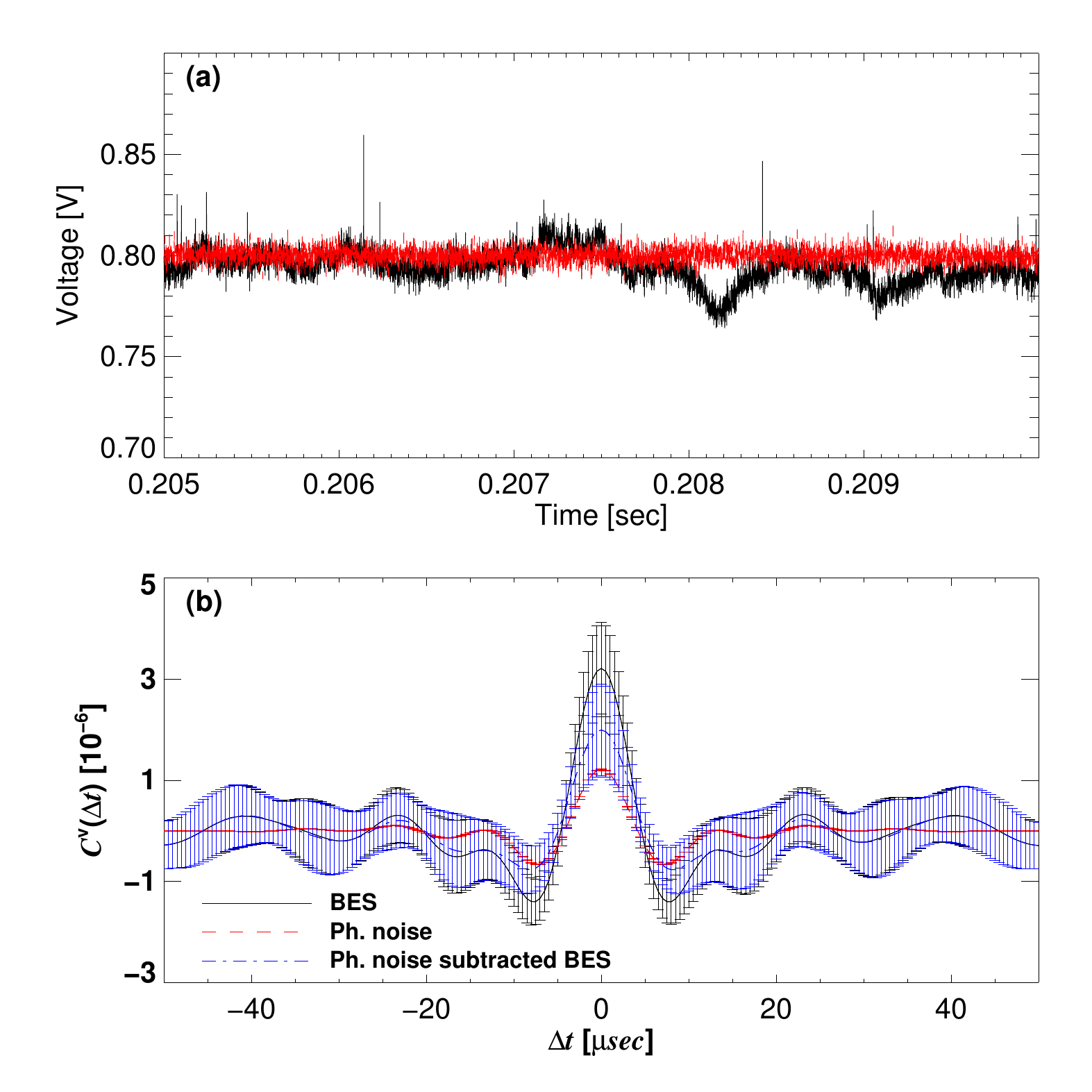}{Auto-covariance functions of BES signal and photon noise}{(a) Time trace of measured raw BES signal (black) and measured photon noise (red) whose mean voltage matches with that of the BES signal; (b) Calculated auto-covariance functions of the BES signal (black), the photon noise signal (red) and the photon noise subtracted BES signal (blue).  Note that both BES and photon noise signals are filtered from $20$ to $100$ kHZ in (b).}
One must be careful obtaining the fluctuation level of the BES signal as the information is acquired from an auto-covariance function at $\Dt=0$, i.e., $\lab\dI^2\lp t\rp\rab=\cov\lp\DR=0, \DZ=0, \Dt=0\rp$, which contains the noise of the BES signal as well.  To remove the noise from the signal, we independently measure photon noise levels by illuminating the BES sensors with an LED calibration source.\footnote{An LED calibration source is mounted inside the optics box of the BES system, so that we can apply LED light to the BES channels while all the other light sources are blocked by closing the shutter of the optics box.}  We obtain 150 different DC levels of noise signals from $0$ to $1.5\:V$ creating a database of noise signals.  Let us define a measured BES signal $I\lp t\rp$ from a MAST discharge consisting of a mean $\lab I\rab$, plasma turbulence signal $S\lp t\rp$ and noise $N\lp t\rp$: $I\lp t\rp = \lab I\rab + \dI\lp t\rp = \lab I\rab + S\lp t\rp + N\lp t\rp$.  The true (squared) fluctuation level of the turbulence is $\lab S^2\lp t\rp\rab = \cov\lp\DR=0, \DZ=0, \Dt=0\rp - \lab N^2\lp t\rp \rab$ assuming that $S\lp t\rp$ and $N\lp t\rp$ are uncorrelated.  As the noise source of the 2D BES system on MAST is dominated by the photon noise \cite{dunai_rsi_2010}, we find the corresponding $N\lp t\rp$ from the created noise database with the LED light whose DC level matches with the DC level of a measured BES signal, i.e., $\lab I\rab$.  This is shown in \figref{cov_bes_ph}(a) where the black line is a measured BES signal and red dashed line is the corresponding photon noise with the same DC levels.  If one wants to frequency-filter the measured signal, then it is necessary to filter the $N\lp t\rp$ with the same frequency band as well.  \figref{cov_bes_ph}(b) shows the $\cov\lp\DR=0, \DZ=0, \Dt\rp$ (black), $\lab N\lp t\rp N\lp t+\Dt\rp\rab$ (red) and $\cov\lp\DR=0, \DZ=0, \Dt\rp - \lab N\lp t\rp N\lp t+\Dt\rp\rab$ (blue) where both $\dI\lp t\rp$ and $N\lp t\rp$ are frequency-filtered from $20$ to $100$ kHz.  The fluctuation level of $S\lp t\rp$ is, then, the square-root of the $\cov\lp\DR=0, \DZ=0, \Dt=0\rp - \lab N^2\lp t\rp \rab$, i.e., the square-root of the blue curve at $\Dt=0$. Note that we remove beam noise by frequency filtering the signal below $20$ kHz.
\newline\indent
\myfig[5.5in]{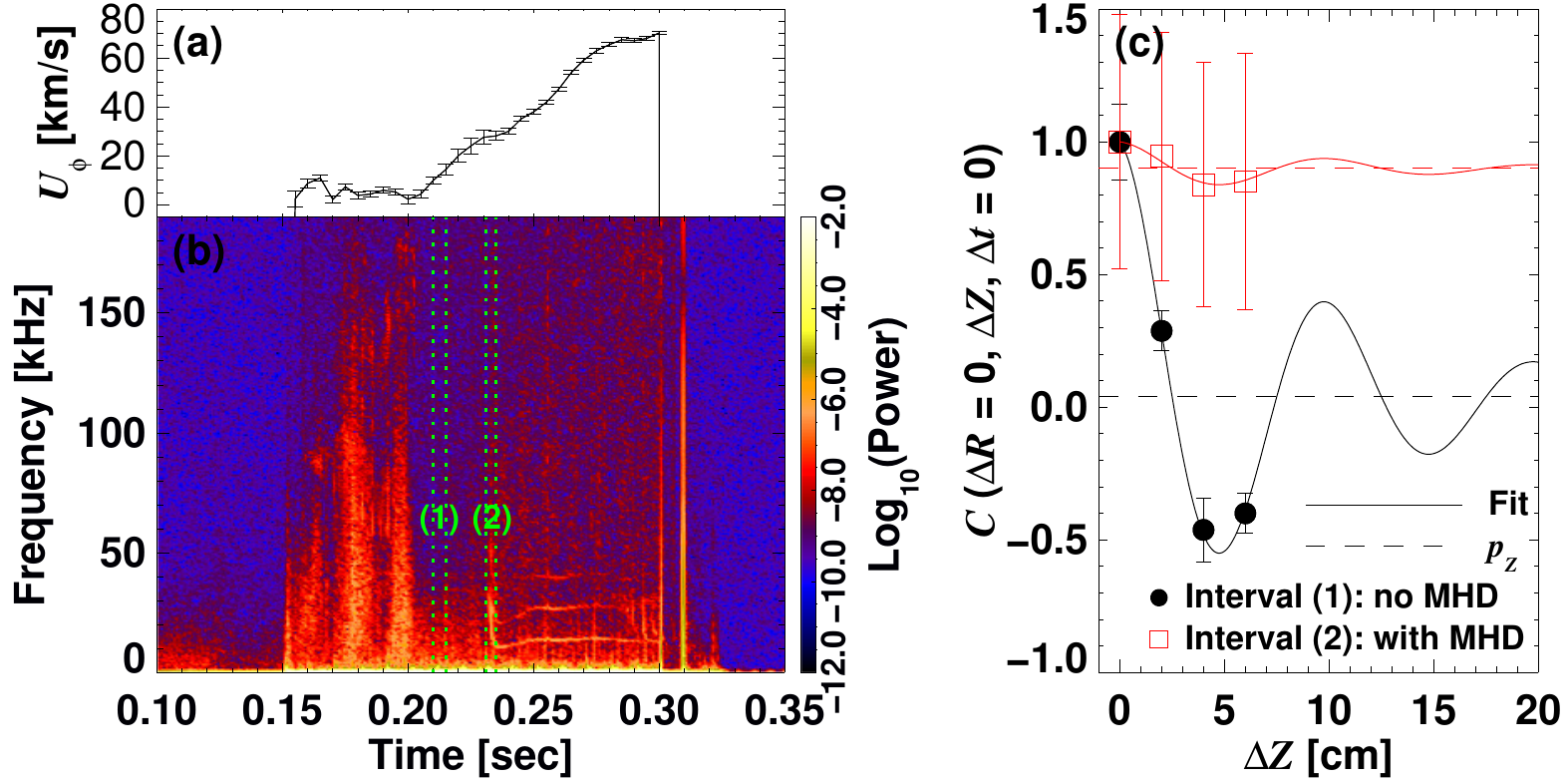}{Examples of $\corr\lp\DR=0, \DZ, \Dt=0\rp$ to obtain the poloidal correlation length}{Examples of $\corr\lp\DR=0, \DZ, \Dt=0\rp$ from MAST shot \#27267.  Evolution of (a) toroidal plasma flow $\Utor$ and (b) cross-power spectrogram of BES signal between two poloidally separated channels by $2\:cm$; (c) shows $\corr\lp\DR=0, \DZ, \Dt=0\rp$ from the time interval (1) in black and (2) in red.  Time intervals (1) and (2) are marked in (b). Note that the time interval (2) contains MHD activities (starting at $t\approx 0.23\:s$ with multiple harmonics) which forces $p_Z$ in \eqref{eq:pol_corr_fitting} to be large from the fitting procedure. In both cases, poloidal correlation lengths $\lZ$ are estimated to be $10.0\:cm$.}
Spatial correlation length is estimated using the correlation values at $\Dt=0$ with finite values of $\DZ$ or $\DR$.  We fit $\corr\lp\DR=0, \DZ, \Dt=0\rp$ to the function $f_Z\lp\DZ\rp$ defined as
\begin{equation}\label{eq:pol_corr_fitting}
f_Z\lp\DZ\rp = p_Z + (1-p_Z)\cos\lsb2\pi\frac{\DZ}{\lZ}\rsb\exp\lsb-\frac{\labs\DZ\rabs}{\lZ}\rsb,
\end{equation}
to estimate the poloidal correlation length $\lZ$.  Here, we assume wave-like fluctuations in the poloidal direction \cite{fonck_prl_1993} with the same wavelength and correlation length.  It is not possible to distinguish between the two with only four poloidal channels.  However, this assumption is verified using the 2D BES data from DIII-D tokamak which has $8$ poloidal channels.\footnote{The author of this work visited General Atomics where DIII-D tokamak is located in May 2012 and verified the assumption with Dr. George McKee who is in charge of its 2D BES system.}  The constant $p_Z$ is a fitting parameter to account for global structures such as MHD modes.  This is necessary as 2D BES data from MAST discharges usually include MHD signals (see \figref{sample_pol_corr}(b) and \secref{sec:exp_barberpole}).  Consider a fluctuating part of BES data $\dI_i\lp t\rp = S_i\lp t\rp + G_i\lp t\rp + N_i\lp t\rp$ where $S$, $G$ and $N$ denote for turbulence, global and noise signals, respectively. The subscript $i$ stands for a BES channel number.  Assuming that these three signals are not correlated to each other and noise signals from different channels are uncorrelated, we have $\lab \dI_i\lp t\rp\dI_j\lp t\rp\rab = \lab S_i\lp t\rp S_j\lp t\rp \rab + \lab G_i\lp t\rp G_j\lp t\rp \rab$.  With a fixed subscript $i$ and varying $j$ in the poloidal direction, $\lab G_i\lp t\rp G_j\lp t\rp \rab$ is a constant, i.e., $p_Z$ in \eqref{eq:pol_corr_fitting}, if there exists a global mode whose spatial structure is larger than the poloidal extent of the BES system; while $\lab S_i\lp t\rp S_j\lp t\rp\rab$ is assumed to have a wave-like spatial structure.  \figref{sample_pol_corr} shows examples of the fitting results: one from an MHD quiet period marked as (1) in (b) and the other from an MHD active period marked as (2) in (b).  The fitting gives $p_Z=0.04$ and $p_Z = 0.90$ for these two cases, respectively, as shown in \figref{sample_pol_corr}(c).  The estimated $\lZ$ is $10.0\:cm$ in both cases.  As a large $p_Z$ is likely to affect the estimation of $\lZ$, and uncertainty levels are larger when MHD modes are active, we set data points with large $p_Z$ unreliable and do not use them for the purpose of turbulence study.\footnote{However, these points are not only valid but also valuable for the study of MHD activities.}  One of the many criteria we use to reject data points with MHD modes is $p_Z > 0.5$.\footnote{Other criteria are listed in \chref{ch:critical_balance}.} Finally, we note that the BES cross-power spectrogram shows a drop of turbulence level when plasmas start to rotate significantly in the toroidal direction shown in \figref{sample_pol_corr}(a) and (b).  The influence of such rotation on the turbulence level is a subject of \chref{ch:larger_RLTi}.
\newline\indent
The radial correlation length $\lR$ is estimated by fitting $\corr\lp\DR, \DZ=0, \Dt=0\rp$ to the function  $f_R\lp\DR\rp$ defined as
\begin{equation}\label{eq:rad_corr_fitting}
f_R\lp\DR\rp = p_R + (1-p_R)\exp\lsb-\frac{\labs\DR\rabs}{\lR}\rsb,
\end{equation}
where $p_R$ plays the same role as $p_Z$ did for $f_Z$.  Note that we do not use a wave-like structure in the radial direction as observed in other tokamaks \cite{shafer_pop_2012, fonck_prl_1993}.  This is also vindicated for measured turbulence in MAST as $\corr\lp\DR, \DZ=0, \Dt=0\rp$ usually monotonically decreases from $\DR=0$, and their values are non-negative as in \figref{sample_rad_corr}.  Statistically, $p_R$ and $p_Z$ should be more or less the same, but they can differ as estimating a constant offset in $f_R\lp\DR\rp$ is non-trivial, i.e., there is no obvious baseline.  Thus, we estimate $\lR$ only for those points with $p_Z<0.5$.  If the fitting results have $p_R>0.5$ even for the points with $p_Z<0.5$, then we cut-out these points as well.  In other words, if a data point satisfies $p_Z>0.5$ \textit{or} $p_R>0.5$, then we set the point unreliable.  \figref{sample_rad_corr} shows the results for the two cases mentioned earlier: without MHD modes in black and with MHD modes in red.  For both cases, $p_R$ is estimated to be $0.0$ with $\lR=6.9\:cm$ without MHD and $2.5\:cm$ with MHD.  Again, it is clear that uncertainty levels are worse when there exist MHD modes.
\myfig[5.5in]{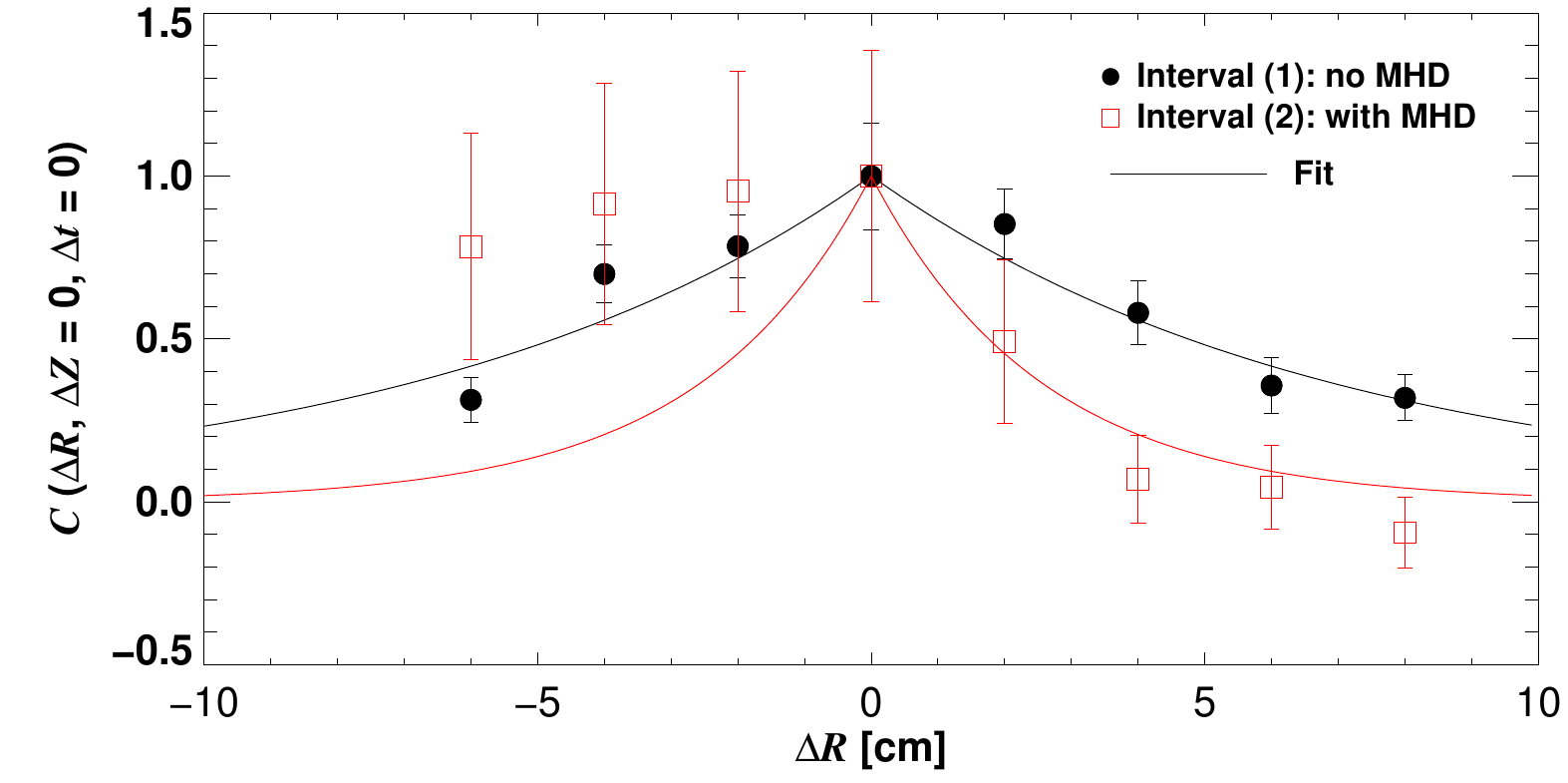}{Examples of $\corr\lp\DR, \DZ=0, \Dt=0\rp$ to obtain the radial correlation length}{Examples of $\corr\lp\DR, \DZ=0, \Dt=0\rp$ from the same time intervals as in \figref{sample_pol_corr}.  The fitting procedure gives the radial correlation lengths $\lR$ of $6.9\:cm$ and $2.5\:cm$ for interval (1) and (2), respectively.  When MHD activities are present, the fitting is much worse than that without any MHD activities.}
\newline\indent
In addition, a spurious long range radial correlation can be introduced to the signal due to edge-induced fluctuations known as beam common-modes \cite{durst_rsi_1992}. The beam common-mode can be identified by looking at the phase difference between the signals obtained from the core and edge of plasmas, and we find little signatures of such modes in our BES data.
\newline\indent
\myfig[5.5in]{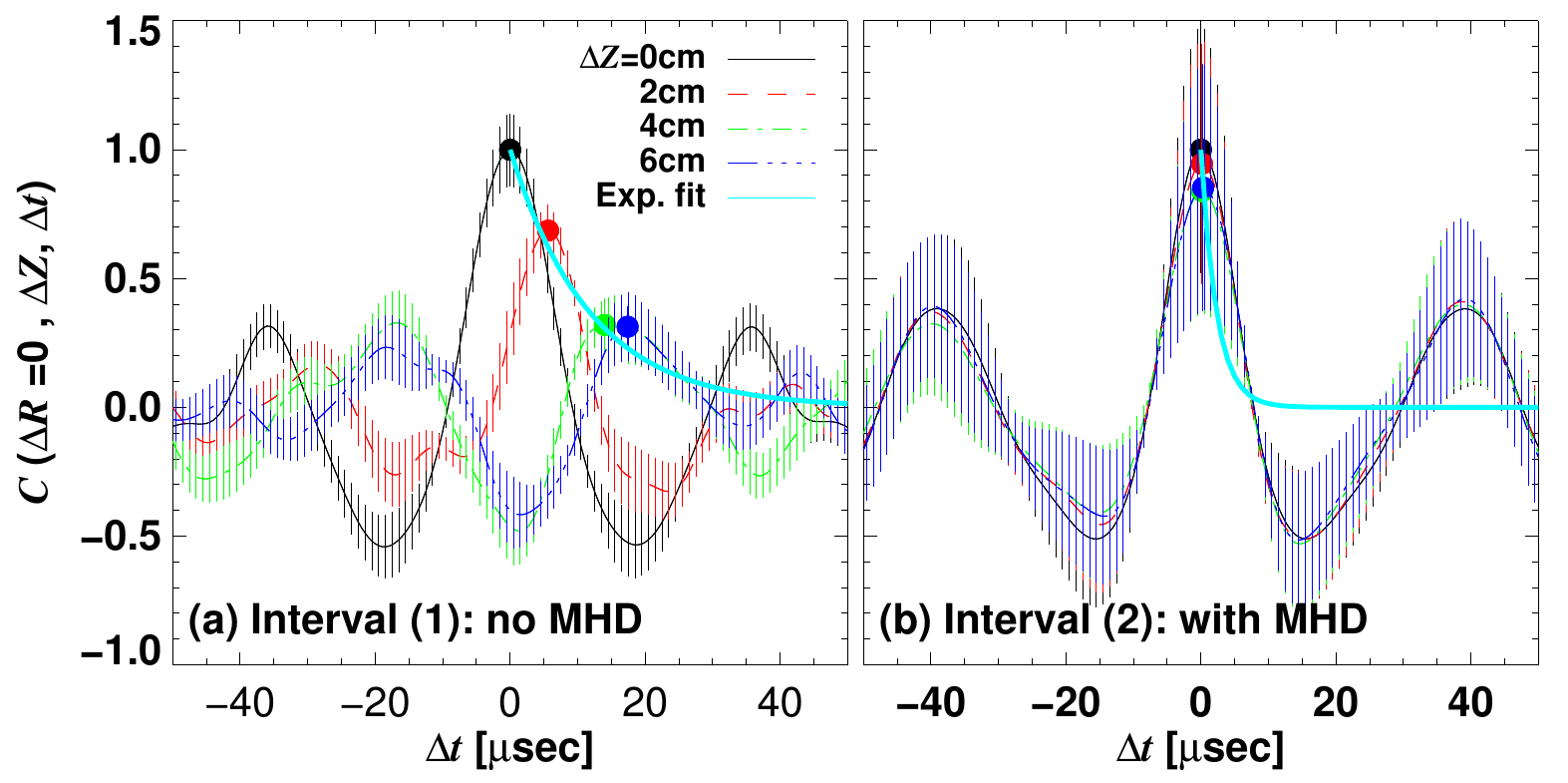}{Examples of $\corr\lp\DR=0, \DZ, \Dt\rp$ to obtain the correlation time}{Examples of $\corr\lp\DR=0, \DZ, \Dt\rp$ from the same time interval as in \figref{sample_pol_corr}.  Circles indicate the positions of $\taupeakcc\lp\DZ\rp$ and cyan lines are the $f_\tau\lp\DZ\rp$ fit with (a) $\tc=11.8\:\mu s$ and (b) $\tc=2.4\:\mu s$.  Note that $\taupeakcc\lp\DZ\rp$ is heavily influenced by the MHD activities in (b) which causes (possibly) an ambiguous estimation of $\tc$.  The effect of MHD modes on $\taupeakcc\lp\DZ\rp$ is studied in more detail in \secref{sec:the_cctd_method}.}
Estimating the correlation time $\tc$ of measured turbulence has to take account of the fact that turbulence is advected in the lab frame due to toroidal plasma flows.  Thus, we use the Lagrangian approach \cite{durst_rsi_1992}.  We fit $\corr\lp\DR=0, \DZ, \Dt=\taupeakcc\lp\DZ\rp\rp$, where $\taupeakcc\lp\DZ\rp$ is the time delay when $\corr\lp\DR=0, \DZ, \Dt\rp$ is maximum at a given $\DZ$, to the function $f_\tau\lp\DZ\rp$ defined as
\begin{equation}\label{eq:time_corr_fitting}
f_\tau\lp\DZ\rp = \exp\lsb-\frac{\labs \taupeakcc\lp\DZ\rp \rabs}{\tc}\rsb.
\end{equation}
\figref{sample_time_corr} shows examples of the fitting results: (a) without MHD and (b) with MHD modes.  The cyan lines show the $f_\tau\lp\DZ\rp$ fit with (a) $\tc=11.8\:\mu s$ and (b) $\tc=2.4\:\mu s$.  Again, when MHD modes are present, uncertainties are larger, and it can potentially affect the estimation of $\tc$.  The described Lagrangian approach relies on the fact that how precisely $\taupeakcc\lp\DZ\rp$ can be located; however, the existence of MHD modes can modify the true $\taupeakcc\lp\DZ\rp$ significantly, an effect which is described in \secref{sec:the_cctd_method} in more detail.  Finally, we note that the reliability of this method depends on the temporal decorrelation dominating over the parallel spatial decorrelation because we would see decreasing $\corr\lp\DR=0, \DZ, \Dt=\taupeakcc\lp\DZ\rp\rp$ as a function of $\DZ$ even if $\tc$ approaches infinite if the parallel correlation length is not long enough.  This fact is quantitatively discussed in \chref{ch:critical_balance}.
\newline\indent
All the fits described in this section are obtained via the \texttt{mpfit} procedure \cite{markwardt_mpfit}.

\chapter{Generating synthetic BES data}\label{ch:synthetic_bes}
\begin{center}
\textit{This chapter is largely taken from Ref. \cite{ghim_ppcf_2012}.}
\end{center}
As discussed in \chref{ch:bes_principle}, deconvolution of the measured signal using the calculated PSFs are not trivial.  However, we can generate synthetic BES data using the PSFs, which then can be used to find any effects of PSFs on statistical analyses as in \chref{ch:eddy_motion} and to make direct comparisons\footnote{These results are not included in this work partly because the work is mainly carried by Dr. Anthony Field using the calculated PSFs described in \chref{ch:bes_principle}.} of numerically simulated data with the experimental data as in Ref. \cite{field_ttf_2012}.  
\newline\indent
Note that we use a local coordinate system with $x$ denoting radial direction and $y$ denoting poloidal direction in this chapter.

\section{Gaussian eddies in space and time}\label{sec:gaussian_eddies}
To be able to numerically examine the reliability of any statistical analyses on turbulence data, it is necessary to know the exact characteristics of the turbulence.  For this purpose, we numerically generate artificial fluctuating density patterns, random both in space and time, then produce synthetic BES data using the PSFs of the 2D BES system on MAST (as described in \secref{sec:2d_synthetic_bes_data}) and compare the inferred characteristics from statistical analyses with the true characteristics.
\newline\indent
We follow a similar approach to the one suggested by Zoletnik \textit{et al.} \cite{zoletnik_rsi_2005}.  Let the density patterns be described by Gaussian structures both in space and time, namely,
\begin{eqnarray} \label{eq:ghim_eddy}
\dn\lp x, y, t \rp
	& = & \sum_{i=1}^{N} \dn_{0i} 
		\exp \Bigg[
		-\frac{\lp x-x_{0i} \rp^2}{2\lambda_x^2} 
-\frac{\lsb y+v_y\lp t \rp\lp t-t_{0i} \rp -y_{0i}
\rsb^2}{2\lambda_y^2} -\frac{\lp t-t_{0i} \rp^2}{2\tau_{life}^2} \Bigg] \times
		\nonumber \\
& & \hspace{1.45cm} \cos \Bigg[2 \pi \frac{\lsb y+v_y\lp t \rp\lp t-t_{0i}
\rp-y_{0i}
		              \rsb}{\lambda_y} \Bigg],
\end{eqnarray}
where $x$, $y$ and $t$ denote radial, poloidal and time coordinates, respectively.  These numerically generated density patterns are referred to as ``eddies'' in this work.  Here $N$ is the total number of eddies and the subscript $i$ denotes the $i^{th}$ eddy in the simulation; $\dn_{0i}$, $x_{0i}$, $y_{0i}$ and $t_{0i}$ are the maximum amplitude and central locations in the $x$, $y$ and $t$ coordinates of the $i^{th}$ eddy, respectively; $\lambda_x$, $\lambda_y$ and $\tau_{life}$ are the widths of our Gaussian eddies in the $x$, $y$ and $t$ directions; $\tau_{life}$ is the lifetime (or the correlation time) of the eddies in the moving frame; $v_y(t)$ is the apparent advection velocity of the eddies in the poloidal direction.  Although it is possible to introduce a finite radial velocity shear by making $v_y$ a function of $x$, the effect of such shearing rates is not investigated in this work, so we will only consider $v_y$ that are independent of $x$.  The $\cos$ term in the $y$ (poloidal) direction is introduced to model wave-like-structured eddies in the poloidal direction as observed in tokamaks \cite{fonck_prl_1993}. In choosing the wavelength in the poloidal direction, we assume that it is the same as the exponential decay length.\footnote{The author of the work visited General Atomics and confirmed that this assumption is reasonable with the 8 channel poloidal measurements on DIII-D.} This is because the two lengths cannot be separately and reliably measured using only four poloidal channels.  Note that the envelope (i.e., the $\exp$ term) and the wave structure (i.e., the $\cos$ term) of $\dn(x, y, t)$ have the same advection velocity $v_y(t)$.  The central locations of eddies, $x_0$, $y_0$ and $t_0$, are selected from uniformly distributed random numbers, whereas their amplitudes $\dn_0$ are selected from normally distributed random numbers whose standard deviation is one.\footnote{It is worth mentioning that there is another scheme of generating such eddies numerically, proposed by Jakubowski \textit{et al.} \cite{jakubowski_rsi_2001}.  They generated the time series of fluctuating density ($\dn_1$) using the inverse Fourier transform of a broadband Gaussian amplitude distribution in frequency space. Then, a second signal ($\dn_2$) was generated by imposing the desired time-delay fluctuation on the $\dn_1$ such that $\dn_2$ was a time-delayed version of $\dn_1$.  This method does not include spatial information for the signals.}
\newline\indent
The spatial domain of the simulation is $25\:cm$ and $20\:cm$ with the mesh size of $0.5\:cm$ in radial ($x$) and poloidal ($y$) directions, respectively. The time duration of the simulation is $20\:ms$ with a $0.5\:\mu s$ time step so as to have the same Nyquist frequency as the real 2D BES data from MAST.  The widths $\lambda_x$ and $\lambda_y$ are set so that the full width at half maximum (FWHM) in the radial direction and the wavelength in the poloidal direction are $\sim8\:cm$ (i.e., $\lambda_x=3.53\:cm$) and $\sim20\:cm$ (i.e., $\lambda_y=20.0\:cm$), respectively, which are similar to the measured correlation lengths with the 2D BES system on MAST.\footnote{Note that Smith \textit{et al.} \cite{smith_aps_2011} also reported that poloidal correlation lengths of the density patterns are $\sim20\:cm$ using their 2D BES system on NSTX.}  The eddy lifetime in the moving frame ($\tau_{life}$) is set to $15\:\mu s$.  However, some of the data sets in this work have different values of $\tau_{life}$, so the effect of $\tau_{life}$ on a statistical analysis, the cross-correlation time delay (CCTD) method to measure the velocity of fluctuating patterns, can be investigated in \secref{sec:the_cctd_method}.
\newline\indent
The total number of eddies is $N=20000$.  If the eddies are too sparse in the simulation domain, then we may not achieve steady statistical results, while overly dense eddies may cause an effective widening of the specified spatial ($\lambda_x$ and $\lambda_y$) and temporal ($\tau_{life}$) correlations as many eddies can merge into one larger eddy.  Thus, we introduce another control parameter, the spatio-temporal filling factor ($F$), defined as 
\begin{equation}\label{eq:filling_factor_def}
F=N\cdot\lp \frac{\lambda_x \lambda_y}{{\mathrm{total~simulation~area}}}\rp\cdot\lp \frac{\tau_{ac}}{{\mathrm{total~simulation~time}}} \rp,
\end{equation} 
where $\tau_{ac}$ is the autocorrelation time calculated as \cite{bencze_pop_2005}
\begin{equation}
\tau_{ac}=\frac{\tau_{life}\lp\lambda_y/v_y\rp}{\sqrt{\tau_{life}^2+\lp\lambda_y/v_y\rp^2}}
\end{equation}
for the generated eddies defined by \eqref{eq:ghim_eddy}.  All of our synthetic data was generated so as to $F\sim\mathcal{O}(1)$.
\newline\indent
The testing of the CCTD method in \secref{sec:the_cctd_method} will involve exploiting what happens if $v_y(t)$ has a mean and a temporally varying components.  Thus, we generate a temporal structure of $v_y$: at each $x$,
\begin{eqnarray}\label{eq:gam_generator}
v_y\lp t \rp & = & \lang v_y \rang + \delta v_y \lp t \rp \nonumber \\
                           & = & \lang v_y \rang + \tilde{v}_y\lp t\rp \ast \exp \lsb-\frac{t^2}{\tau_{fluc}^2} \rsb\sin \lp 2\pi f_{fluc}t \rp
\end{eqnarray}
where $\lang v_y \rang$ and $\delta v_y$ are the mean and temporally varying velocities, respectively, $\tau_{fluc}$ and $f_{fluc}$ are the lifetime and frequency of $\delta v_y(t)$, respectively, and $\tilde{v}_y(t)$ is generated from normally distributed random numbers.  The RMS (root-mean-square) value of $\delta v_y(t)$ denoted as $\delta v_y^{RMS}$ will be varied as well as $\lang v_y \rang$ to investigate the effects of these quantities on the CCTD method.  $\tau_{fluc}$ and $f_{fluc}$ allow one to introduce structured temporally varying velocities, while the randomness is kept by $\tilde{v}_y$.  As one of the causes for the temporal variation of the poloidal velocity is believed to be the existence of geodesic acoustic modes (GAMs)\footnote{We do not investigate whether the CCTD method is able to detect such a temporally structured $\delta v_y\lp t\rp$ (or GAMs) in this work, rather we investigate how the existence of these structures affects the CCTD-determined mean velocity.} \cite{winsor_pf_1968}, we choose $\tau_{fluc}=500\:\mu s$ and $f_{fluc}=10$ kHz to mimic the GAM features detected by Langmuir probes on MAST \cite{robinson_ppcf_2012}.
\newline\indent
The simulations have been run on a NVIDIA\textregistered\space GeForce GTS 250 GPU card using CUDA programming, which increases the computational speed owing to the highly parallelizable structure of \eqref{eq:ghim_eddy}.

\section{Synthetic 2D BES data}\label{sec:2d_synthetic_bes_data}
We generate the $i^{th}$ (1 to 8) radial and $j^{th}$ (1 to 4) poloidal channel of the synthetic BES data $I^{ij}\lp t \rp$ by using the calculated point-spread-functions (PSFs) of the actual 2D BES system on MAST \cite{ghim_rsi_2010} described in \chref{ch:bes_principle} and $\dn\lp x, y, t \rp$ from \eqref{eq:ghim_eddy} with an additional random noise.  In general, $\dn\lp x, y, t \rp$ can be taken from any turbulence numerical simulations as done in Ref. \cite{field_ttf_2012}. Furthermore, a large-scale (in space) coherent (in time) oscillation is included to imitate a global MHD mode.  Namely, $I^{ij}\lp t \rp$ is defined as
\begin{equation}\label{eq:syn_bes_data_total_def}
I^{ij}\lp t \rp=I_{DC}^{ij} + \dI^{ij}\lp t \rp + I_{MHD}^{ij}\lp t\rp + I_N^{ij}\lp t \rp,
\end{equation} 
where $I_{DC}^{ij}$ is the DC value -- a typical value of $0.8\:V$ is used for all channels \cite{field_rsi_2012}.  The rest of the terms are as follows.
\newline\indent
$\dI^{ij}\lp t\rp$ is the fluctuating part of the signal generated from the fluctuating density $\dn\lp x,y,t\rp$, i.e., the Gaussian eddies given by \eqref{eq:ghim_eddy} in this work, convolved with the PSFs of the 2D BES system:
\begin{equation}\label{eq:syn_bes_data_fluc_def}
\dI^{ij}\lp t \rp =\dI^{RMS} \int\int \dn\lp x,y,t \rp\mathcal{P}^{ij}\lp x, y \rp\,\mathrm{d}x\mathrm{d}y,
\end{equation}
where $\mathcal{P}^{ij}\lp x,y\rp$ is the PSF of the $i^{th}$ and $j^{th}$ channel of the 2D BES system, normalized so that RMS value of $\dI^{ij}\lp t\rp$ is $\dI^{RMS}$.  This value is set so that the ratio of $\dI^{RMS}$ to $I_{DC}^{ij}$ is 0.05.
\newline\indent
$I^{ij}_{MHD}\lp t\rp$ models an MHD (global) mode.  We assume that the spatial scale of the MHD modes is larger than the BES domain in the poloidal direction, so $I^{ij}_{MHD}\lp t\rp$ does not vary in the poloidal direction.  The model MHD signal is generated in a way similar to temporal behaviour of $v_y\lp t\rp$ using \eqref{eq:gam_generator}, except that the mean value of $I^{ij}_{MHD}\lp t\rp$ is zero and $\tau_{fluc}=250\:\mu s$.  The frequency of the mode $f_{MHD}$ and its RMS value, denoted $I_{MHD}^{RMS}$, will be varied in various tests.  The value of $\tau_{fluc}$ here is representative of MHD burst-like fishbone instabilities \cite{mcguire_prl_1983} or chirping modes \cite{gryaznevich_ppcf_2004} in tokamaks, for which the spectrum has a finite bandwidth. 
\newline\indent
$I_N^{ij}\lp t\rp$ represents the noise in the signal.  As the noise of the 2D BES system on MAST is dominated by the photon noise \cite{dunai_rsi_2010}, $I_N^{ij}\lp t\rp$ is generated using normally distributed random numbers.  Its RMS level is set such that the signal-to-noise ratio (SNR) is $300$, which is typical of the 2D BES system on MAST \cite{field_rsi_2012}.
\newline\indent
\myfig[4.5in]{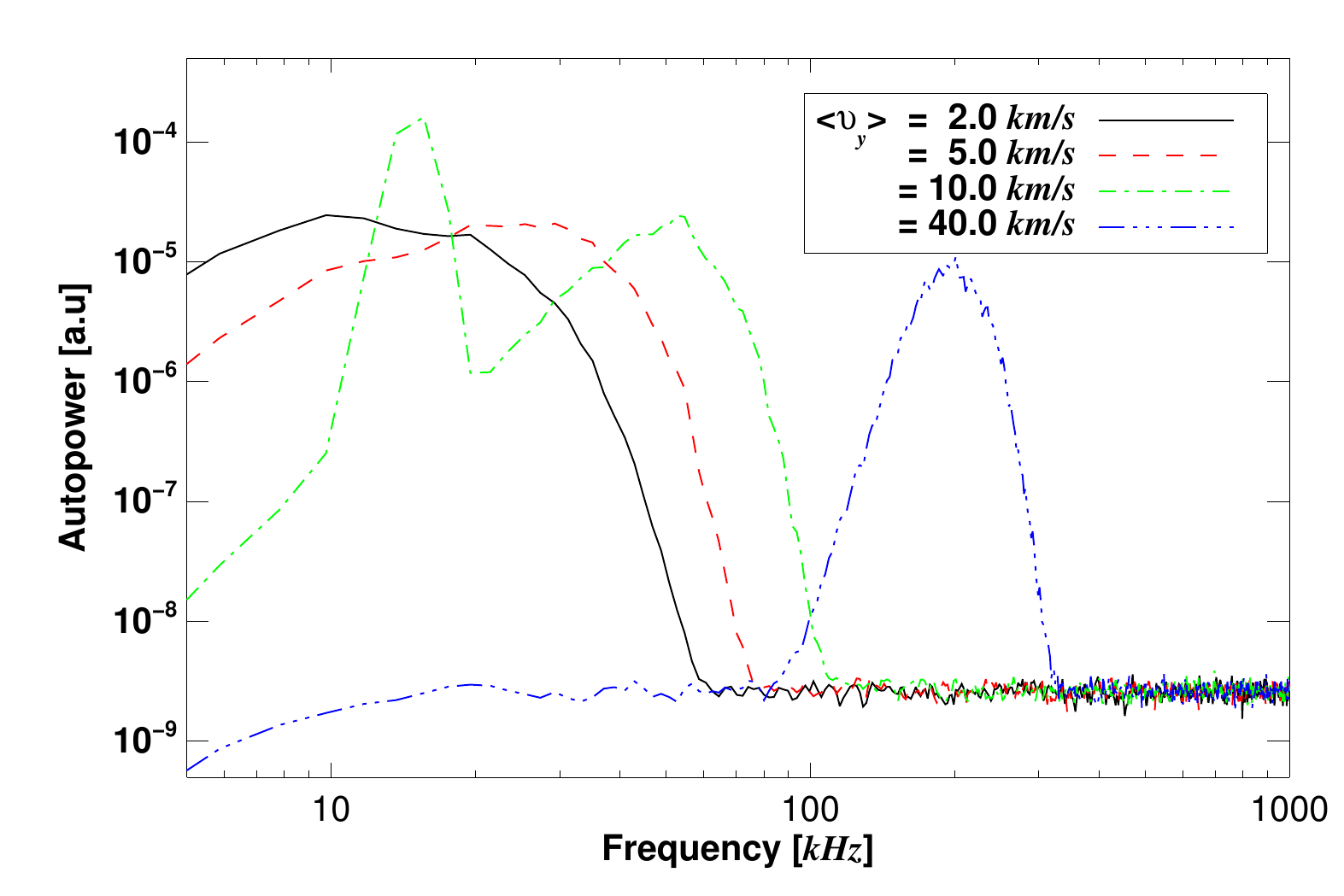}{Autopower spectra of synthetic 2D BES data}{Autopower spectra of synthetic 2D BES data for various $\lang v_y \rang$. Note that the spectrum for $\lang v_y \rang = 10.0\: km/s$ (green dash dot line) has finite $I_{MHD}^{ij}$ (i.e., temporal oscillations due to global modes) in \eqref{eq:syn_bes_data_total_def} at $15$ kHz, with fluctuation level of $5\:\%$ of the DC level.  For other cases, $I_{MHD}^{RMS}=0$.  All the spectra are generated using a high-pass filter with the frequency cutoff at $5$ kHz.}
\figref{syn_data_power} shows examples of autopower spectra of the synthetic 2D BES data for $\lang v_y \rang=2.0, 5.0, 10.0$ and $40.0\:km/s$.  The autopower spectrum is calculated as $\labs FT\lcb I^{ij}\lp t \rp\rcb\rabs^2$ where $FT\lcb\cdot\rcb$ is the Fourier transform in the time domain.  Increasing the value of $\lang v_y \rang$ has two effects: Doppler shift and broadening of the spectra, as expected.  Note that in \figref{syn_data_power}, the data for $\lang v_y \rang=10.0\:km/s$ contains the finite $I_{MHD}^{RMS}$ with $f_{MHD}=15$ kHz and $I_{MHD}^{RMS}/I_{DC}^{ij}=0.05$, while $I_{MHD}^{RMS}=0$ for other cases.
\newline\indent
\myfig[5.5in]{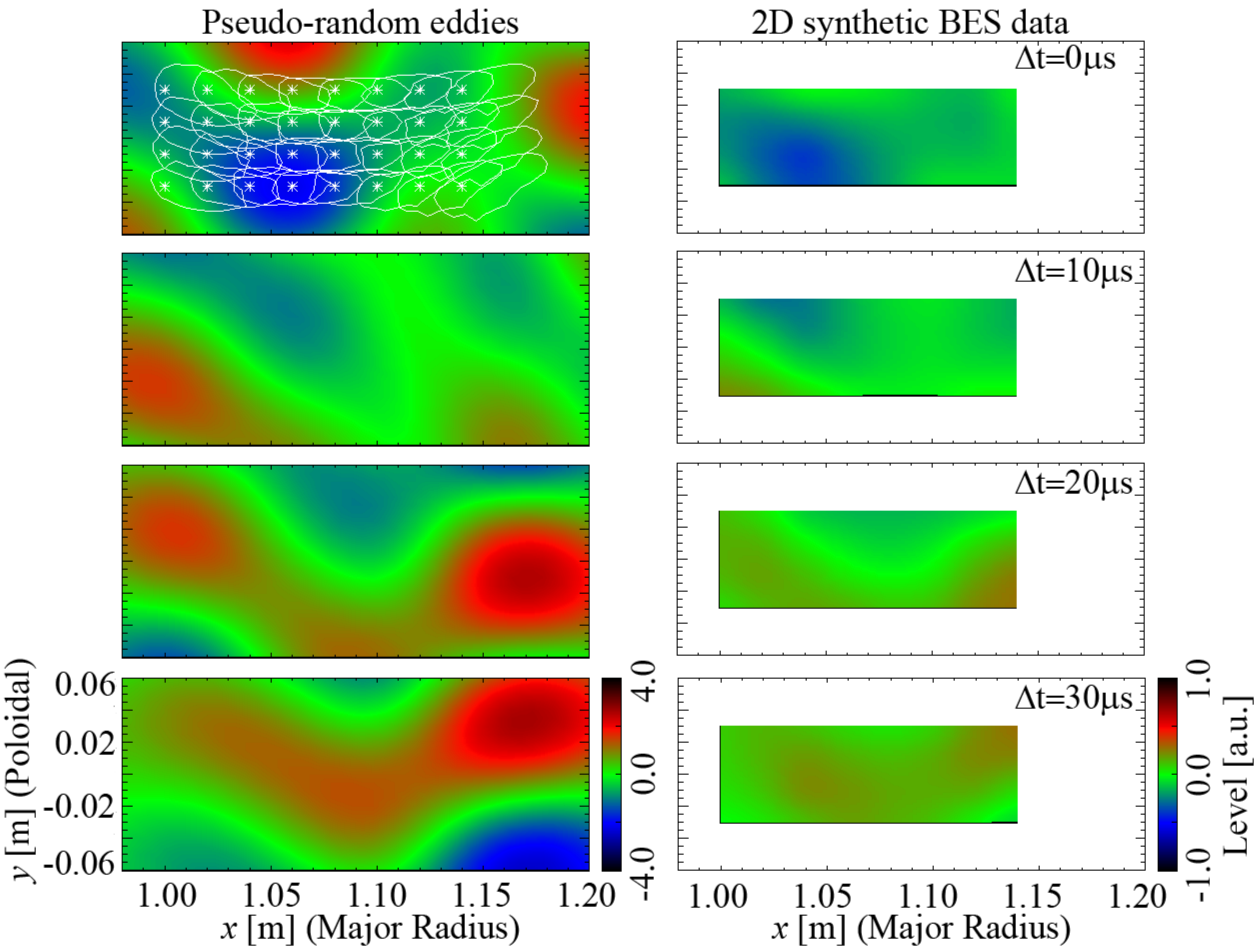}{Time snapshots of artificial Gaussian eddies}{Left column: four time snapshots of Gaussian eddies, $\dn\lp x, y, t \rp$ given by \eqref{eq:ghim_eddy}.  Right column: the corresponding normalized synthetic 2D BES data given by \eqref{eq:syn_bes_data_total_def} without the DC component.  White lines in the top left panel show the $1/e$ contour lines of the PSFs (see \figref{bes_psf}(a)), and the white asterisks show the optical focal points of the 32 channels of the 2D BES system.}
\figref{syn_data_movie} shows several time snapshots of artificial Gaussian eddies (\eqref{eq:ghim_eddy}) in the left column and the corresponding synthetic 2D BES data in the right column (with DC component removed from \eqref{eq:syn_bes_data_total_def}). The eddies are moving upward with $\lang v_y\rang = 5.0\:km/s$.  The top left panel in this figure also shows the $1/e$ contour lines of the PSFs for the 32 channels (see \figref{bes_psf}(a)). Snapshots for the synthetic 2D BES data are generated with the bandpass frequency filtering from $10$ to $70$ kHz to suppress the noise. As the synthetic 2D BES data have only 32 spatial points, spatial interpolation is performed using parametric cubic convolution technique \cite{park_image_processing_1982}.

\mypart{One Further Step in Understanding Plasma Turbulence}

\chapter{Measurements and physical interpretation of mean motion of turbulent density patterns}\label{ch:eddy_motion}
\begin{center}
\textit{This chapter is largely taken from Ref. \cite{ghim_ppcf_2012}.}
\end{center}
The mean motion of turbulent patterns detected by a two-dimensional (2D) beam emission spectroscopy (BES) diagnostic on the Mega Amp Spherical Tokamak (MAST) is determined using a cross-correlation time delay (CCTD) method.  Statistical reliability of the method is studied by means of synthetic data analysis.  The experimental measurements on MAST indicate that the apparent mean poloidal motion of the turbulent density patterns in the lab frame arises because the longest correlation direction of the patterns (parallel to the local background magnetic fields) is not parallel to the direction of the fastest mean plasma flows (usually toroidal when strong neutral beam injection is present).  This effect is particularly pronounced in a spherical tokamak because of the relatively large mean rotation and large magnetic pitch angle.  The experimental measurements are consistent with the mean motion of plasma being toroidal.  The sum of all other contributions (mean poloidal plasma flow, phase velocity of the density patterns in the plasma frame, non-linear effects, etc.) to the apparent mean poloidal velocity of the density patterns is found to be negligible.  These results hold in all investigated L-mode, H-mode and internal transport barrier (ITB) discharges.  The one exception is a high-poloidal-beta (the ratio of the plasma pressure to the poloidal magnetic field energy density) discharge, where a large magnetic island exists.  In this case BES detects very little motion.  This effect is currently theoretically unexplained.

\section{Introduction}
It is now widely accepted that turbulent transport in magnetically confined fusion plasmas can exceed the irreducible level of neoclassical transport by an order of magnitude or more \cite{carreras_ieee_1997}. However, both theoretical and experimental works of the past two decades \cite{barnes_prl_2011, highcock_prl_2010, roach_ppcf_2009, camenen_pop_2009, kinsey_pop_2005, dimits_nf_2001, waltz_pop_1994,  mantica_prl_2009, mantica_prl_2011, burrell_pop_1997, highcock_pop_2011, waltz_pop_1997,  connor_nf_2004, devries_nf_2009} suggest that sheared $\vct{E}\times\vct{B}$ flows can moderate such anomalous transport and hence improve the performance of magnetically confined fusion plasmas. 
\newline\indent 
With the aim of characterizing the microscale plasma turbulence and searching for correlations between it and the background plasma characteristics, a two-dimensional (8 radial $\times$ 4 poloidal channels) beam emission spectroscopy (2D BES) system \cite{field_rsi_2012} has been installed on the Mega Amp Spherical Tokamak (MAST).  It is able to measure density fluctuations at scales above the ion Larmor radius $\rho_i$, viz., $k_\perp\rho_i<1$, where $k_\perp$ is the wavenumber perpendicular to the magnetic field.  The 2D BES view plane lies on a radial-poloidal plane at a fixed toroidal location.  Following the detected turbulent density patterns on this view plane allows one to determine their mean velocity in the radial and poloidal directions.  Typically, there are no significant mean plasma flows in the radial direction in a tokamak, whereas considerable apparent poloidal motion is detected by the 2D BES system.
\newline\indent
In this Chapter, we show experimentally that this apparent poloidal motion is primarily due to the strong mean toroidal rotation of the plasma.  The fluctuating density patterns are highly elongated in the parallel direction, and their toroidal advection produces apparent poloidal motion due to the projection effect.  This effect is particularly pronounced in MAST because of a relatively large pitch angle of the magnetic field compared to conventional tokamaks. The BES measurements are shown to be consistent with a dominantly toroidal mean flow; the poloidal flows are of the order of the diamagnetic velocities.  These results are obtained using the cross-correlation time delay (CCTD) method, which is a frequently used statistical technique to determine the apparent velocity of density patterns \cite{durst_rsi_1992, cosby_master_1992}.  We also investigate the method itself thoroughly to determine the statistical uncertainties of the technique. This is done by generating synthetic 2D BES data with random Gaussian density patterns calculated on a graphical processing unit (GPU) card using CUDA (Compute Unified Device Architecture) programming.
\newline\indent
This Chapter is organized as follows.  In \secref{sec:vel_detect}, we explain how the apparent velocity of turbulent density patterns can be inferred from the 2D BES data.  We also show what physical effects contribute to the apparent velocity calculated by the CCTD method. The CCTD method to determine the velocity of the density patterns and its statistical reliability are studied in \secref{sec:the_cctd_method} using synthetically generated 2D BES data described in \chref{ch:synthetic_bes}.  In \secref{sec:eddy_motion_exp_results}, we present the experimental results with the aim of identifying the main cause of apparent motion of density patterns measured by the 2D BES system.  Our conclusions are presented in section \ref{sec:eddy_motion_conc}.

\section{Velocity of density patterns}\label{sec:vel_detect}
From the time-dependent 2D measurement of density fluctuations, one can infer the apparent velocity of the density patterns.  This has been the subject of much attention \cite{shafer_pop_2012, durst_rsi_1992, zweben_ppcf_2012, tal_pop_2011, xu_ppcf_2011, zweben_pop_2006, jakubowski_prl_2002} in the hope that this velocity can be related in a more or less straightforward way to the actual plasma flows.  We will first explore how the mean pattern velocity can be determined and then discuss the interpretation of this quantity.
\subsection{The cross-correlation time delay (CCTD) method}
The CCTD (cross-correlation time delay) method has been widely used to determine the apparent velocities of turbulent density patterns detected by BES systems, and it is well described in \cite{durst_rsi_1992} and \cite{cosby_master_1992}. Here, a brief summary of the method is provided. The normalized fluctuating intensity of the photons, $\hat I\equiv\dI/I$, measured by a 2D BES system (see \chref{ch:bes_principle}) is a function of the radial $x$, vertical (poloidal) $y$ and time $t$ coordinates\footnote{The coordinate system used in this Chapter is same as the one used in \chref{ch:synthetic_bes}.}: $\hat{I}=\hat{I}\lp x, y, t\rp$.  The cross-correlation function of this fluctuating signal is defined as
\begin{equation}\label{eq:cc_definition}
\mathcal{C}\lp \Dx, \Dy, \Dt \rp = 
\frac{\lang \hat{I} \lp x, y, t \rp 
\hat{I} \lp x+\Dx, y+\Dy, t+\Dt \rp\rang}{\sqrt{\lang \hat{I}^2 \lp x, y, t \rp \rang 
\lang \hat{I}^2 \lp x+\Dx, y+\Dy, t+\Dt \rp \rang}},
\end{equation}
where $\Dx$ and $\Dy$ are the radial and vertical (poloidal) channel separation distances, respectively, $\Dt$ is the time lag, and $\lang \cdot \rang$ denotes time average defined in \secref{sec:the_cctd_method}.  The apparent poloidal velocity $\vbes$ of the density patterns detected by the 2D BES system can be determined from the time lag $\Dt=\taupeakcc$ at which the cross-correlation function reaches its maximum for a given $\Dy$ and $\Dx=0$.\footnote{We concentrate on the apparent mean `poloidal' motion of the density patterns.  Thus, the information about the radial correlations of the 2D BES data is not used in this Chapter.}  If a straight line is fitted to the experimentally measured $\taupeakcc\lp\Dy\rp$, the inverse of its slope is the velocity $\vbes$.  Although any two poloidally separated channels are sufficient to determine $\vbes$, using just two channels is insufficient to estimate the uncertainties in the linear fit.  Thus, in this Chapter, all four available poloidal channels are used to determine these quantities. This assumes that the mean velocity does not change over the time the density patterns take to move past the four poloidal channels and that the lifetime of these patterns is sufficiently long, so the same patterns are observed by all four channels.
\newline\indent
\myfig[5.0in]{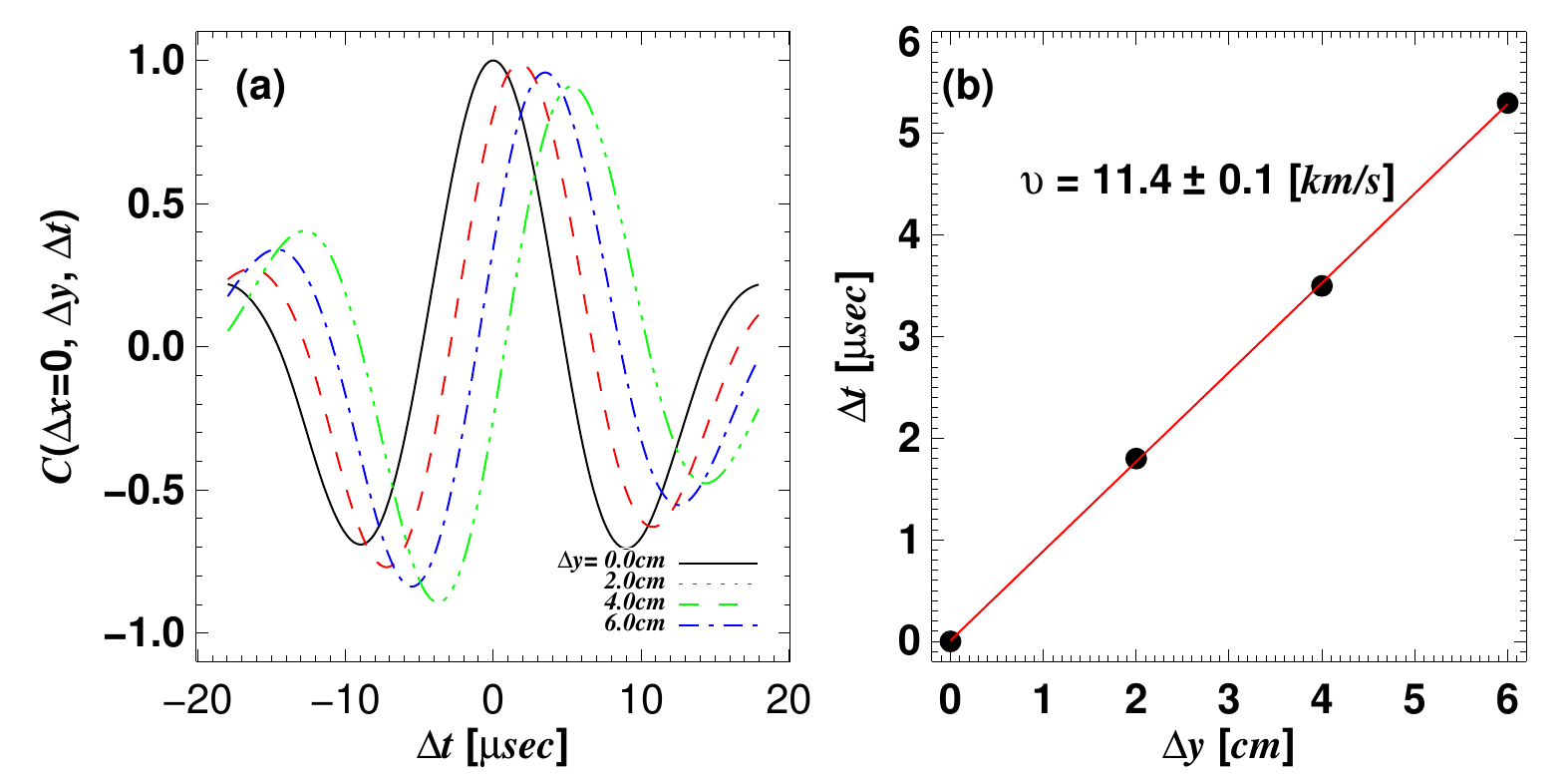}{An example of the CCTD method}{(a) Cross-correlation functions calculated using \eqref{eq:cc_definition}) for $\Dy = 0.0\:cm$ (black solid line), $2.0\:cm$ (red dash line), $4.0\:cm$ (blue dash dot line) and $6.0\:cm$ (green dash dot dot line). $\taupeakcc\lp\Dy\rp$ is the position of maximum of the cross-correlation function. (b) Position of maximum $\taupeakcc\lp\Dy\rp$ and a linear fit.  The measured velocity is $11.4\pm0.1\:km/s$.}
\figref{cctd_method} shows an example of this procedure. This example is based on a synthetic data set consisting of Gaussian-shaped random ``eddies'' moving with the poloidal velocity of $10.0\:km/s$, which are then used to produce artificial 2D BES data (see \chref{ch:synthetic_bes} for the description of the synthetic data).  With the four available poloidal channels, cross-correlation functions are calculated using \eqref{eq:cc_definition} and shown in \figref{cctd_method}(a); $\taupeakcc$ is plotted as a function of $\Dy$ in \figref{cctd_method}(b). The inverse of the slope of a fitted straight line is the velocity $\vbes$.   Note the slight discrepancy between the actual and CCTD-determined velocities.  The origin and size of this discrepancy are discussed in \secref{sec:the_cctd_method}.

\subsection{Physical meaning of the CCTD-determined velocity}\label{sec:physical_meaning_cctd}
Using the described CCTD method, the 2D BES system on MAST is expected to be able to determine $\vbes$ as has been done previously on TFTR \cite{durst_rsi_1992} and DIII-D \cite{jakubowski_prl_2002} using their BES systems \cite{paul_rsi_1990, mckee_rsi_1999}. However, as McKee \textit{et al.} \cite{mckee_pop_2003, mckee_rsi_2003} pointed out, one must distinguish between the poloidal velocity measured by 2D BES system ($\vbes$) and the actual velocity of the poloidal plasma flow ($U_y$).
\newline\indent
The mean plasma flow can be decomposed into toroidal ($U_z$) and poloidal ($U_y$) components.  For typical tokamak plasmas where strong neutral beams are injected, $\labs U_z\rabs \gg \labs U_y\rabs$ is satisfied as any mean poloidal flows are strongly damped \cite{connor_ppcf_1987, catto_pf_1987, hinton_pf_1985, cowley_clr_1986}, leaving $U_y$ of the order of the diamagnetic velocity $\sim\rhostar\vth$, where $\rhostar=\rhoi/a$, $a$ is the plasma minor radius, and $\vth$ is the ion thermal velocity.  Note that $U_z$ can be on the order of $\vth$ for the neutral-beam-heated plasmas.  Thus, $U_y$ can be ignored compared to $U_z$, except possibly in regions with strong pressure gradients.
\newline\indent
\myfig[4.5in]{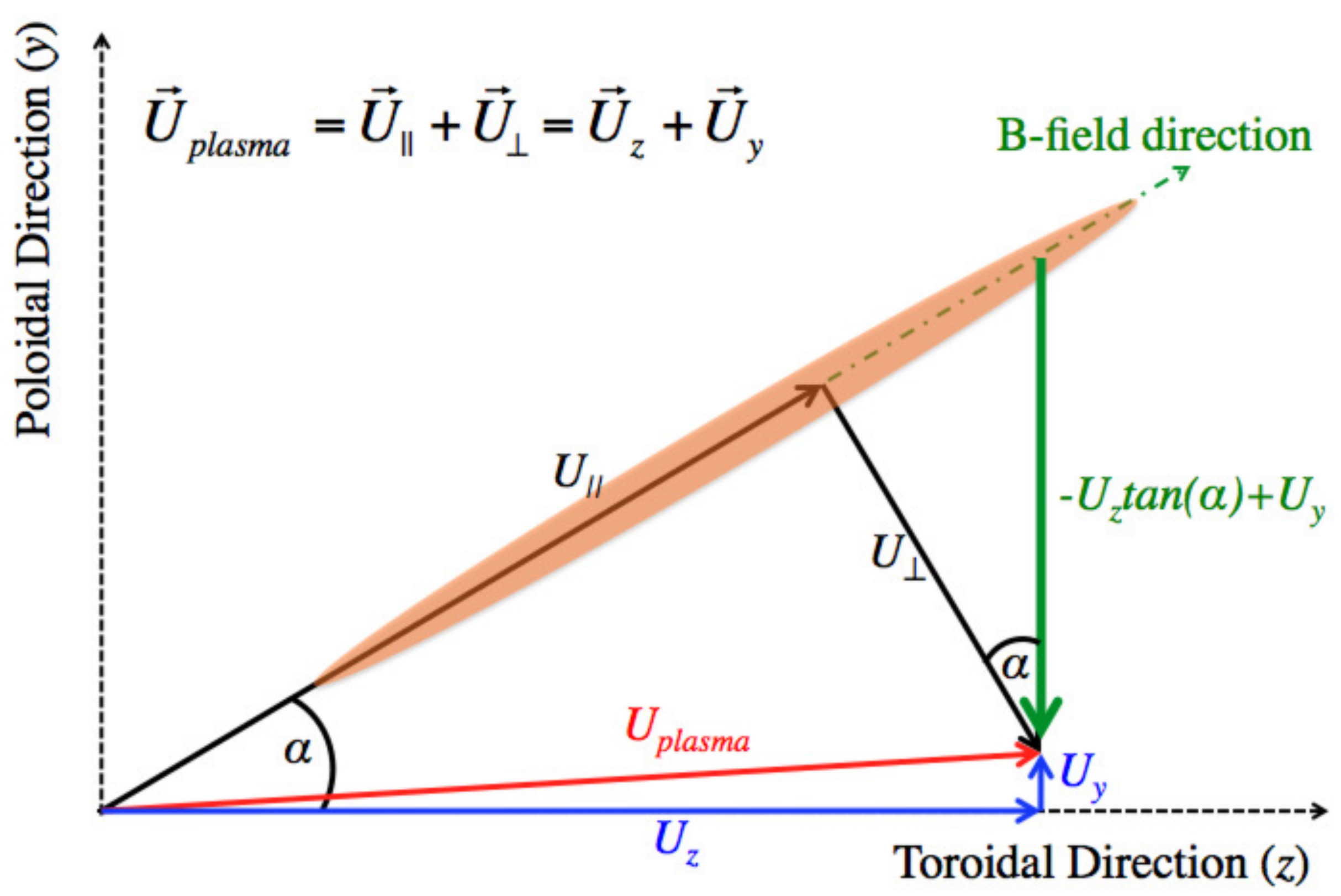}{Cartoon illustrating 'rotating barber-pole' effect}{Cartoon illustrating how the mean toroidal plasma flow ($U_z$) induces an apparent mean poloidal motion.  An elongated density pattern (shaded oval) along the magnetic field line (green dash dot) is advected by the toroidal flow (blue arrow).  Because the longest correlation direction of the density pattern is not in the toroidal direction, the apparent mean poloidal flow (green arrow) arises.  The apparent velocity is $-U_z\tan\alpha+U_y\approx -U_z\tan\alpha$, where $\alpha$ is the local magnetic pitch angle.}
As the 2D BES system on MAST observes the density patterns advected by $U_z$, there will be an apparent motion of the patterns in the poloidal direction, as shown in \figref{barber_pole}. This effect is analogous to the apparent up-down motion of helical strips of a `rotating barber-pole' (cf. \cite{munsat_rsi_2006}).  The magnitude of this apparent velocity can be readily calculated via elementary geometry: namely, we expect the BES system to ``see'', to lowest order in $\rhostar$, 
\begin{equation}\label{eq:simple_v_bes_relation}
\vbes\approx-U_z\tan\alpha,
\end{equation}
 where $\alpha$ is the pitch angle of the local magnetic field line.
\newline\indent
\eqref{eq:simple_v_bes_relation} is experimentally verifiable because all three physical quantities are readily obtained by separate diagnostics: $\vbes$ from the 2D BES system, $U_z$ from the Charge eXchange Recombination Spectroscopy (CXRS) system \cite{conway_rsi_2006}, and $\alpha$ either from \texttt{EFIT} equilibrium reconstruction \cite{lao_nf_1985} or directly from the Motional Stark Effect (MSE) system \cite{kuldkepp_rsi_2006, debock_rsi_2008} on MAST.  Although the CXRS system measures the toroidal flow of the $C^{6+}$ ions, the difference between the velocity of the $C^{6+}$ ions and the bulk plasma ions, $D^+$, is predicted to be on the order of $\rhostar$ in a strongly beam-heated plasma \cite{kim_pfb_1991}.  In \secref{sec:eddy_motion_exp_results}, \eqref{eq:simple_v_bes_relation} will be experimentally verified for various types of discharges.  Agreement will indicate consistency of the experiment with the assumptions behind \eqref{eq:simple_v_bes_relation}.  Such agreement will indeed be obtained, except in one intriguing case.
\newline\indent
Let us now consider what are the assumptions necessary for \eqref{eq:simple_v_bes_relation} to hold by analysing how the estimated $\vbes$ depends on actual physical quantities associated with plasma flows and fluctuations in a tokamak.  The cross-correlation function \eqref{eq:cc_definition} of the normalized fluctuating photon intensity $\hat I$ can, in view of \eqref{eq:photon_dens_relation}, be considered proportional to the cross-correlation function of the relative ion density fluctuation $\dn/n$ (by definition, $\lang\dn\rang=0$).  Therefore, the CCTD-determined velocity of the density patterns can be related to the actual physical quantities in a tokamak by invoking the ion continuity equation.  Splitting also the ion velocity into mean and fluctuating parts, $\vct{u}=\vct{U}+\delta\vct{u}$, $\lang\delta\vct{u}\rang=0$, we have
\begin{equation}\label{eq:continuity_equation_full}
\frac{\partial n}{\partial t}+\frac{\partial\dn}{\partial t}+\grad\cdot\lp n\vct{U}+n\delta\vct{u}+\dn\vct{U}+\dn\delta\vct{u}\rp=0.
\end{equation}
Averaging this equation and subtracting the averaged equation from \eqref{eq:continuity_equation_full}, we obtain
\begin{equation}\label{eq:continuity_equation}
\frac{\partial\dn}{\partial t}=-\grad\cdot\lp n\delta\vct{u}+\dn\vct{U}+\dn\delta\vct{u}-\lang\dn\delta\vct{u}\rang\rp.
\end{equation}
We will now order various terms in this equation in terms of the small parameter $\rhostar=\rhoi/a$ which is approximately $1/50-1/100$ in MAST.
\newline\indent
Assuming that the spatial scale of all mean quantities is $\sim\mathcal{O}(a)$ while the spatial scale of all fluctuating quantities is $\sim\mathcal{O}(\rhostar a)$, and also $\dn/n \sim \delta u/\vth \sim \rhostar$, we get
\begin{eqnarray}\label{eq:continuity_equation_ordered}
\frac{\partial}{\partial t}\frac{\dn}{n} & = & -\vct{U}\cdot\grad\frac{\dn}{n}-\grad\cdot\delta\vct{u} \nonumber \\
&  & -\delta\vct{u}\cdot\grad\ln n - \grad\cdot\lp\frac{\dn}{n}\delta\vct{u}\rp-\frac{\dn}{n}\lp\grad\cdot\vct{U}+\vct{U}\cdot\grad\ln n\rp \nonumber \\ 
& & + \mathcal{O}\lp\rhostar^2\rp,
\end{eqnarray}
where we have dropped all terms $\sim\mathcal{O}(\rhostar^2)$ and smaller.  The first two terms on the right-hand-side are $\sim\mathcal{O}(\vth/a)$ and the following three terms are $\sim\mathcal{O}(\rhostar\vth/a)$.  Note that we have not yet made any assumptions about the nature of the mean flow $\vct{U}$ (beyond it being large-scale) or about time scale of the fluctuations.
\newline\indent
In fact, the $\rhostar$ ordering, which is the standard gyrokinetic ordering \cite{frieman_pf_1982}, can take us further. First of all, the mean flow turns out to be purely toroidal to lowest order \cite{abel_rpp_2012}.  Operationally, this occurs because of the strong collisional (neoclassical) damping of mean poloidal flows \cite{connor_ppcf_1987, catto_pf_1987, hinton_pf_1985, cowley_clr_1986}.  Thus, $\vct{U}=U_z\hat z+\vct{U}_1$, where $z$ is the toroidal direction (locally) and $\vct{U}_1\sim\mathcal{O}(\rhostar)$ including all poloidal flows\footnote{Note that the poloidal velocity $U_y$ of the bulk plasma ions has been measured with the CXRS system to be only a few $km/s$ on MAST \cite{field_ppcf_2009}, which is consistent with $U_y\sim\mathcal{O}(\rhostar)$.  Such measurements are, however, not routinely available for MAST, and one of the goals for this study is to confirm that $U_y$ is indeed small.} and first-order corrections to $U_z$ (radial flows, associated with particle fluxes, are, in fact, even smaller).  Coupled with the fact that mean quantities have no toroidal variation in a tokamak, this means that the fifth term on the right-hand-side of \eqref{eq:continuity_equation_ordered} is also $\sim\mathcal{O}(\rhostar^2)$, while the first term can be expressed as
\begin{eqnarray}\label{eq:delta_n_along_U}
\vct{U}\cdot\grad\frac{\dn}{n}&=&U_z\frac{\partial}{\partial z}\frac{\dn}{n} + \vct{U}_1\cdot\grad\frac{\dn}{n} \nonumber \\
&=&-U_z\frac{b_y}{b_z}\frac{\partial}{\partial y}\frac{\dn}{n}+\frac{U_z}{b_z}\hat b\cdot\grad\frac{\dn}{n}+\vct{U}_1\cdot\grad\frac{\dn}{n},
\end{eqnarray}  
where $\hat b=(0, b_y, b_z)$ is the unit vector in the direction of the magnetic field in a local orthogonal Cartesian system ($x$: radial, $y$: poloidal and $z$: toroidal), and we have used the identity $\mbox{ $\hat{b}\cdot\grad=b_y\partial/\partial y + b_z\partial/\partial z$}$.  Making a further assumption, again standard in gyrokinetics, that the parallel spatial scale of the fluctuating quantities is $\sim\mathcal{O}(a)$, we conclude that the second term in the second line of \eqref{eq:delta_n_along_U} is $\mathcal{O}(\rhostar)$.  Finally, the gyrokinetic ordering also implies that compressibility effects (the second term on the right-hand-side of \eqref{eq:continuity_equation_ordered}) are also order $\rhostar$.  This is because the lowest-order fluctuating velocity perpendicular to the magnetic field is the incompressible $\vct{E}\times\vct{B}$ drift and the parallel scale of the fluctuations is long, so $\grad\cdot\delta\vct{u}\sim\grad_\parallel\delta u_\parallel\sim\mathcal{O}\lp\rhostar\rp$.
\newline\indent
Combining \eqref{eq:continuity_equation_ordered} and \eqref{eq:delta_n_along_U} together with the estimates described above, we find 
\begin{equation}\label{eq:continuity_equation_BES}
\frac{\partial}{\partial t}\frac{\dn}{n} + U_{eff}\frac{\partial}{\partial y}\frac{\dn}{n}=\gamma\frac{\dn}{n},
\end{equation}
where $U_{eff}=-U_zb_y/b_z=-U_z\tan\alpha$ is the dominant apparent velocity of the density patterns ($\alpha$ is the local pitch angle of the magnetic field line).  The term containing $U_{eff}$ is the only $\mathcal{O}(\rhostar^0)$ term in \eqref{eq:continuity_equation_BES}.  The $\mathcal{O}(\rhostar)$ and higher terms have been assembled in the right-hand-side: by definition, $\gamma$ is such that 
\begin{eqnarray}\label{eq:definition_gamma}
\gamma\frac{\dn}{n}&=&-\frac{U_z}{b_z}\hat b\cdot\grad\frac{\dn}{n}-\vct{U_1}\cdot\grad\frac{\dn}{n} \nonumber \\
& &- \grad\cdot\delta\vct{u}-\delta\vct{u}\cdot\grad\ln n - \grad\cdot\lp\frac{\dn}{n}\delta\vct{u}\rp + \mathcal{O}(\rhostar^2).
\end{eqnarray}
This contains, in order of terms, the effects associated with
\newline\noindent
(1) parallel variations of the fluctuations,
\newline\noindent
(2) mean poloidal flows of bulk plasma ions,
\newline\noindent
(3) compressibility of the fluctuations,
\newline\noindent
(4) linear response to mean density gradient (drift waves),
\newline\noindent
(5) nonlinear effects (turbulence),
\newline\noindent
and a slew of higher-order effects of varying degree of obscurity.
\newline\indent
Thus, the right-hand-side of \eqref{eq:continuity_equation_BES} contains all the nontrivial physics of waves and turbulence in the plasma.  The apparent velocity of the density patterns detected by the 2D BES system will not be influenced by these effects to dominant order --- if the orderings assumed above are correct.  What it does contain is the poloidal signature $U_{eff}$ of the dominant toroidal rotation of the plasma --- the `rotating barber-pole' effect discussed at the beginning of this section.  Indeed, if \eqref{eq:continuity_equation_BES} holds and its right-hand-side is small, then, to lowest order, the density patterns just drift in the $y$-direction (poloidal) with the velocity $U_{eff}$, so the maximum of the cross-correlation function \eqref{eq:cc_definition} will be achieved at $\tau=\Delta y/U_{eff}$.  Hence \eqref{eq:simple_v_bes_relation} for the BES-measured velocity.
\newline\indent
If we are able to confirm \eqref{eq:simple_v_bes_relation} experimentally, this means that the theoretical considerations employed above are consistent with the experiment.  This is important because most of the theories of tokamak turbulence rely on such considerations.  Note that there are no separate diagnostics capable of measuring individually all the $\mathcal{O}(\rhostar)$ terms in \eqref{eq:definition_gamma}.  Therefore, the only conclusion one can formally draw from \eqref{eq:simple_v_bes_relation} holding is that the sum of these terms is small.

\section{Assessment of the cross-correlation time delay (CCTD) method}\label{sec:the_cctd_method}
Before we present our experimental results in \secref{sec:eddy_motion_exp_results}, let us describe the CCTD method and its statistical reliability in more detail.  In this section, errors involved in determining the mean velocity of the density patterns by the CCTD method are examined using the synthetic 2D BES data generated according to the procedure explained in \chref{ch:synthetic_bes}. The velocity measured via the correlation function (\eqref{eq:cc_definition}) is denoted $\vbes$ and compared with the prescribed value $\lang v_y \rang$ that appears in \eqref{eq:gam_generator}, i.e., the mean poloidal velocity of the synthetic data.  \ref{sec:desc_cctd_method} provides detailed description of the CCTD method used in this work, then four types of error are identified for the quantitative comparisons.  These errors are evaluated in \ref{sec:mean_vel_detect} and \ref{sec:influence_tau_life} for different values of $\lang v_y \rang$ and the eddy correlation time $\tau_{life}$.  Subsequent sections are devoted to investigating how the existence of global (MHD) modes and temporally varying poloidal velocity affect the errors.

\subsection{Description of the CCTD method}\label{sec:desc_cctd_method}
As defined by \eqref{eq:cc_definition}, cross-correlation functions are calculated as time averages of the data.  For a $20\:ms$-long synthetic data set containing $\Ntotal=40,000$ time data points with the sampling time $\Dtsam=0.5\:\mu s$, we want to determine $\vbes$ with a time resolution $\tres=1\:ms$.  First, a cross-correlation function \eqref{eq:cc_definition} is calculated on a sub-time window of the synthetic 2D BES data containing $\Ncorr$ points, where $\Ncorr < \tres/\Dtsam$.  Then, such cross-correlation functions are averaged over $\Navg$ consecutive sub-time windows where $\Navg=(\tres/\Dtsam)/\Ncorr$ so that an averaged cross-correlation function is obtained at every $\tres$.  In this Section, we use $\Ncorr=80$, so $\Navg=25$.
\newline\indent
Denoting $f(t)$ and $g(t)$ the time series over a sub-time window from two poloidally separated synthetic 2D BES channels, the cross-correlation function \eqref{eq:cc_definition} for this sub-time window is:
 \begin{equation}\label{eq:modified_cc_def}
\corrsub\lp r\Dtsam\rp=\frac{\frac{1}{\Ncorr}\displaystyle\sum_{k=0}^{\Ncorr-1}f\lp k\Dtsam\rp\:g\lp\lp k+r\rp\Dtsam\rp}{\frac{1}{\Ncorr-1}\sqrt{\displaystyle\sum_{k=0}^{\Ncorr-1}f^2\lp k\Dtsam\rp\displaystyle\sum_{k=0}^{\Ncorr-1}g^2\lp\lp k+r\rp\Dtsam\rp  }},
\end{equation}
for any integer $r$ with $|r| < \Ncorr-1$.  Finally, by averaging $\corrsub$ for $\Navg$ consecutive sub-time windows we obtain the smoothed averaged cross-correlation function $\mathcal{C}(r\Dtsam)$ from $1\:ms$-long data points.
\newline\indent
The CCTD method has a serious limitation due to the fact that the sampling time $\Dtsam$ is finite.  In order to calculate $\vbes$ using only two poloidally separated channels, a line is fitted through two points on a $\lp\Dy,\taupeakcc\rp$ plane as shown in \figref{cctd_method}(b).  The first point is located at $\lp\Dy, \taupeakcc\rp=\lp 0,0\rp$ by definition, and the second point at $\lp\Dy, r\:\Dtsam\rp$.  Then, possible values of $\vbes$ are restricted to $\Dy/\lp r\Dtsam\rp$ where $r$ is an integer.  For the 2D BES system on MAST, using two adjacent poloidal channels ($\Dy=2.0\:cm$) with a sampling time $\Dtsam=0.5\:\mu s$, the possible values of $\vbes$ are limited to $40.0,\:20.0,\:13.3,\ldots\:km/s$ for $r=1,\:2,\:3,\ldots$.  Such a limitation may be mitigated by using four poloidally separated channels.  However, using four channels is not always possible if the channels that are farthest apart are not correlated.  To resolve this issue, we use a second-order polynomial fit on the cross-correlation function $\mathcal{C}(r\Dtsam)$ to locate its global maximum: if $r_\mathrm{peak}$ is the point where the discrete cross-correlation function $\mathcal{C}(r\Dtsam)$ is maximum, we use the three values of $\mathcal{C}(r\Dtsam)$ at $r=r_\mathrm{peak}$, $r_\mathrm{peak}-1$ and $r_\mathrm{peak}+1$ to fit a second-order polynomial.  The ``true'' maximum is found from this fit.  We denote the time delay at which this maximum is reached by $\taupeakcc$.

\subsection{Definition of errors}\label{sec:def_uncertainties}
For a given set of $20\:ms$-long synthetic 2D BES data, we calculate $\vbes$ with the time resolution of $1\:ms$ (\secref{sec:desc_cctd_method}).  Furthermore, we do this at three different radial locations\footnote{As described in \chref{ch:synthetic_bes}, $v_y\lp t\rp$ are identical at all radial locations.  One column in the middle and two columns from the edges of the 2D BES channels are used.} so that the average of $\vbes$, denoted $\lang \vbes\rang$, can be calculated using $60$ values of $\vbes$.  To make quantitative comparisons between $\lang \vbes\rang$ and $\lang v_y \rang$ defined in \eqref{eq:gam_generator}, we define four types of error.
\newline\indent
The normalized bias error
\begin{equation}\label{eq:norm_bias_err}
\hsbias=\frac{\lang \vbes \rang - \lang v_y\rang}{\lang v_y \rang}
\end{equation}
is a quantitative measurement of the systematic discrepancy between the measured and the true value.  The normalized random error
\begin{equation}\label{eq:norm_rand_err}
\hsrand=\frac{\sqrt{\lang\lp\vbes-\lang\vbes\rang \rp^2 \rang}}{\labs\lang \vbes\rang\rabs}
\end{equation}
quantifies the degree of fluctuation in the measured $\vbes$ with respect to $\lang\vbes\rang$. This value may depend on the MHD contribution in \eqref{eq:syn_bes_data_total_def} and the temporally varying poloidal velocity $\delta v_y(t)$ in \eqref{eq:gam_generator}.
\newline\indent
Furthermore, as linear fitting is done to determine $\vbes$ (see \figref{cctd_method}), two other types of error are present.  The slope of a linear fit can be denoted as $\vbes \pm \delta v_\mathrm{fit}$ where $\delta v_\mathrm{fit}$ is a degree of the uncertainty of the least-square fit.\footnote{In \figref{cctd_method}, we plotted $\taupeakcc$ as a function of $\Dy$ and determined $\vbes$ as the inverse of the slope of a fitted line.  Operationally, we actually plot $\Dy$ as a function of $\taupeakcc$ so the slope of a fitted line is the $\vbes$.}  Then, the normalized mean of $\delta v_\mathrm{fit}$ is
\begin{equation}\label{eq:norm_mean_fit_err}
\hsmeanfit=\frac{\lang \delta v_\mathrm{fit} \rang}{\labs\lang \vbes \rang\rabs},
\end{equation}
and the normalized random error in $\delta v_\mathrm{fit}$ is
\begin{equation}\label{eq:norm_rand_fit_err}
\hsrandfit=\frac{\sqrt{\lang\lp\delta v_\mathrm{fit}-\lang\delta v_\mathrm{fit}\rang\rp^2\rang}}{\labs\lang \vbes \rang\rabs}.
\end{equation}
These two uncertainties together provide an estimation of how well a linear line is fitted to given data points.  For example, if the assumption that $\tau_{life}$ is long enough so that all four poloidally separated channels observe the same eddies is not satisfied, then $\hsmeanfit$ becomes large.  On the other hand, if this assumption is occasionally satisfied, then $\hsrandfit$ exhibits such events.  Note that error bars of the CCTD-determined apparent velocities in Figures \ref{fig:27278_time_evolution}-\ref{fig:27385_time_evolution} in \secref{sec:eddy_motion_exp_results} show $\lang \delta v_\mathrm{fit} \rang$.
\newline\indent
In the following sections, these four types of error will be evaluated for various values of $\lang v_y \rang$ and $\tau_{life}$, and various ranges of $I_{MHD}^{RMS}$, $f_{MHD}$ and $\delta v_y^{RMS}$.

\subsection{Measuring mean velocity}\label{sec:mean_vel_detect}
To investigate the reliability of the CCTD method described in \secref{sec:desc_cctd_method} for estimating $\vbes$, we generate a number of synthetic 2D BES data sets with various values of $\lang v_y \rang$ while keeping all the other parameters in Eqs. (\ref{eq:ghim_eddy}), (\ref{eq:gam_generator}), (\ref{eq:syn_bes_data_total_def}) and (\ref{eq:syn_bes_data_fluc_def}) constant.  In real experiments, there is almost always some temporal variation of $v_y$, thus the RMS value of $\delta v_y$ in \eqref{eq:gam_generator} is set to $5\:\%$ of $\lang v_y \rang$ in this subsection.  The synthetic 2D BES data are frequency-filtered to suppress the noise before the cross-correlation functions are calculated. \figref{vz_filtered_data} shows examples of (a) $v_y\lp t \rp$ generated according to \eqref{eq:gam_generator} with $\lang v_y \rang = 10.0\:km/s$ and (b) the original (black) and frequency-filtered (red) autopower spectra of a generated synthetic signal.  Here, the noise cut-off level is set to be the $5$ times the averaged autopower level above $900$ kHz (green dashed line).
\myfig[5.5in]{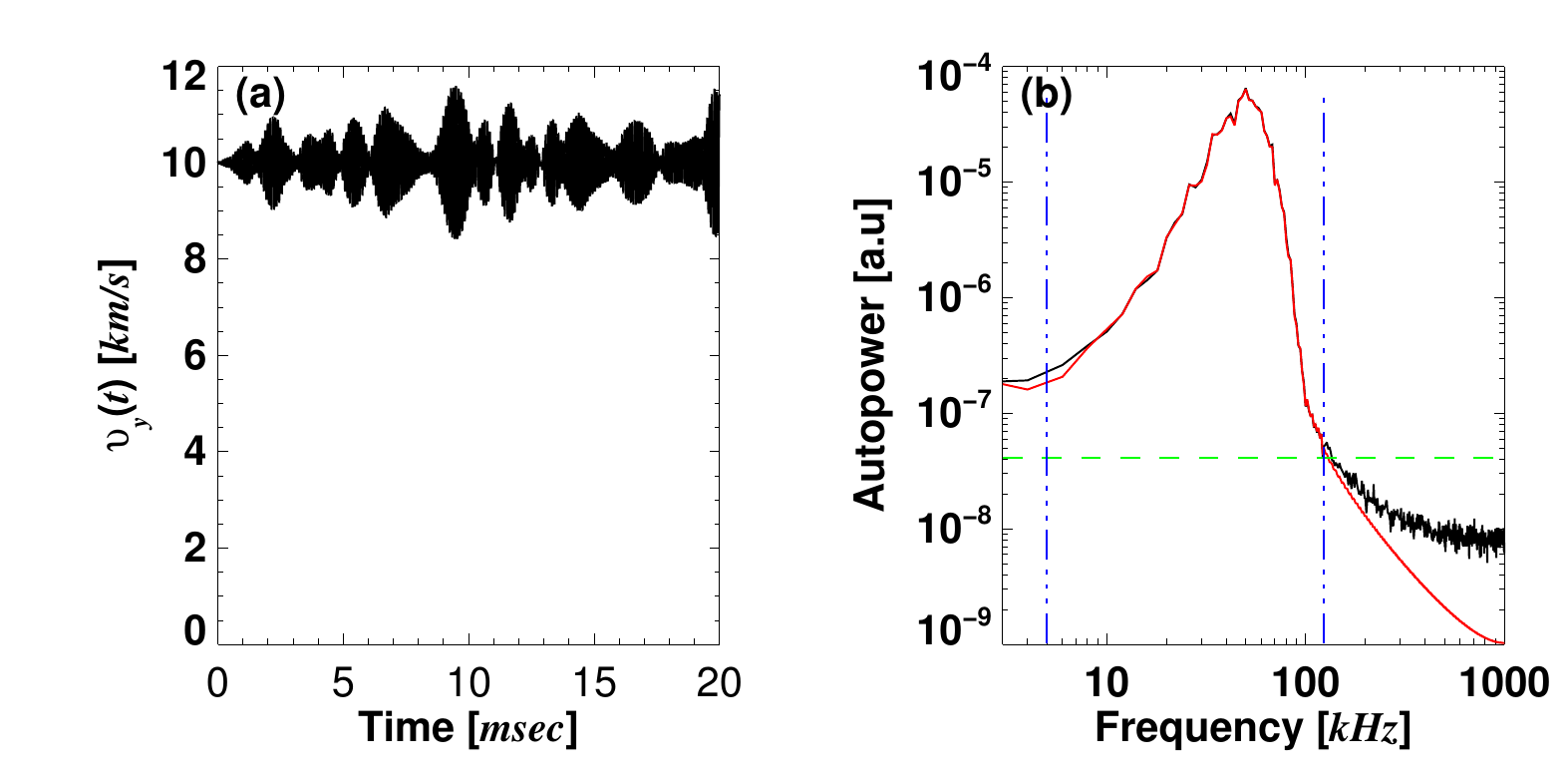}{Examples of $v_y\lp t\rp$ and fluctuating density signal in synthetic 2D BES data}{(a) Poloidal velocity $v_y\lp t \rp$ generated using \eqref{eq:gam_generator} with $\lang v_y \rang = 10.0\:km/s$.  (b) Autopower spectra of the original (black) and frequency-filtered (red) synthetic BES signals.  The green horizontal dashed line shows the noise cut-off level, defined to be $5$ times the averaged autopower level above $900$ kHz, and vertical blue dash-dotted lines indicate the low- and high-frequency cutoffs.}
\newline\indent
\myfig[5.5in]{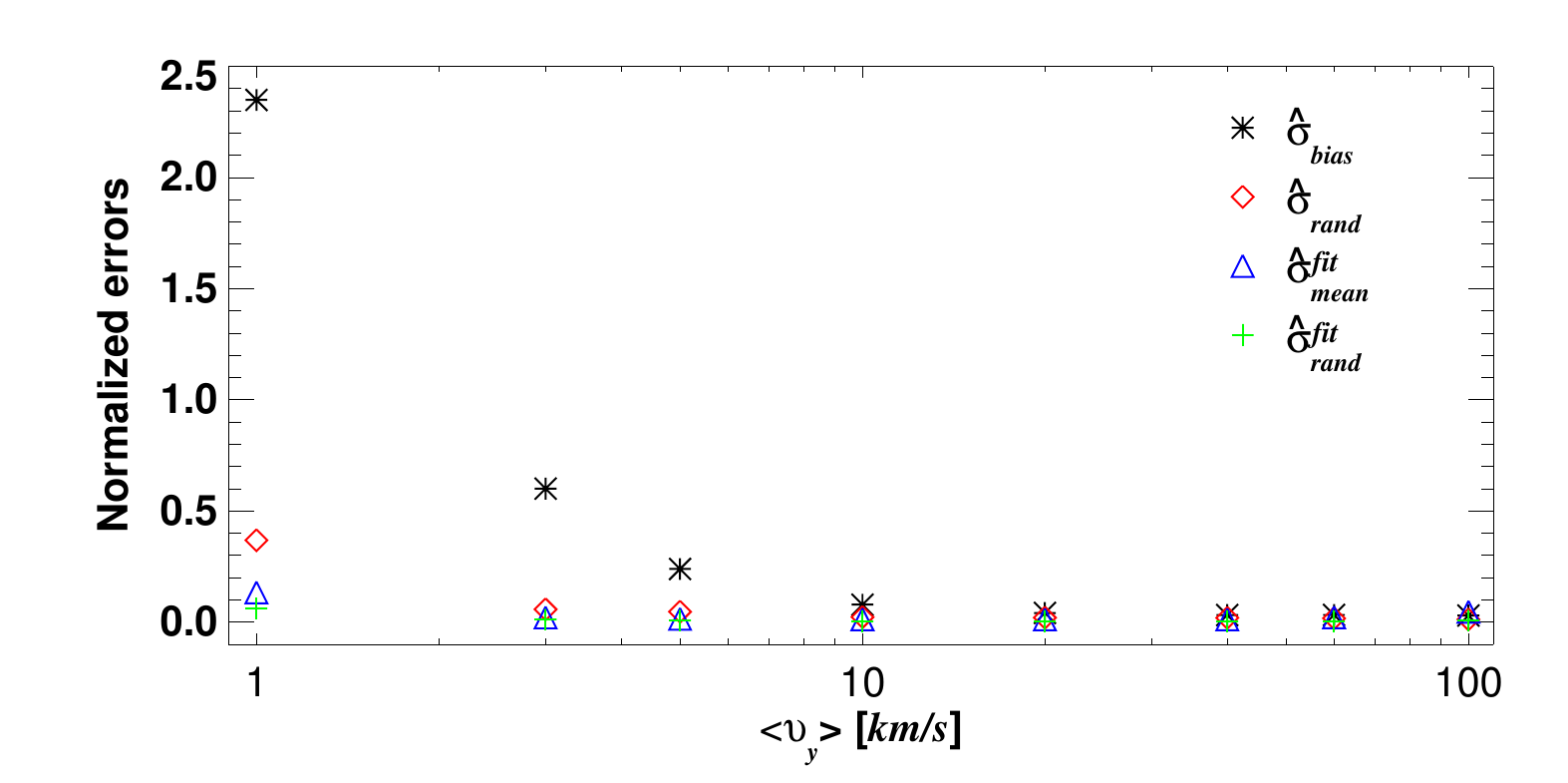}{Dependence of statistical reliability of the CCTD method for the mean velocity}{The four types of error defined in Eqs. (\ref{eq:norm_bias_err})-(\ref{eq:norm_rand_fit_err}) calculated for values of $\lang v_y \rang$ ranging from $1$ to $100\:km/s$.}
\figref{mean_vel_check} shows $\hsbias$, $\hsrand$, $\hsmeanfit$ and $\hsrandfit$ defined in \secref{sec:def_uncertainties} and calculated for values of $\lang v_y \rang$ ranging from $1$ to $100\:km/s$.  The basic conclusions that can be made based on these results are as follows:
\newline\noindent
(1) For $\lang v_y \rang \lesssim 5.0\:km/s$, the CCTD method is not reliable.  This is due to the fact that eddies do not live long enough to be detected by all the poloidally separated channels.  Indeed, it was a priori clear that $\lang v_y \rang<\Dy / \tau_{life}$ could not be measured. This translates to $\lang v_y \rang<4.0\:km/s$ for $\Dy=6.0\:cm$ and $\tau_{life}=15.0\:\mu s$, so our results are consistent with this simple criterion.
\newline\noindent
(2) The CCTD method usually overestimates $\lang v_y \rang$ (i.e., $\hsbias>0$). This can be explained by the effective channel separation distance ($\Dy$) being in fact slightly less than $2.0\:cm$ because of the overlapping of the PSFs, as shown in \figref{bes_psf}. 
\newline\noindent
(3) The limitation of the CCTD method due to the finite $\Dtsam$ is successfully overcome by fitting a second order polynomial to the cross-correlation function, as explained in \secref{sec:desc_cctd_method}.

\subsection{Effect of the eddy lifetime}\label{sec:influence_tau_life}
As explained in \secref{sec:physical_meaning_cctd}, the CCTD method for determining $\lang v_y\rang$ is based on the idea that the peak of the cross-correlation function occurs at $\taupeakcc=\tauprop$, where $\tauprop=\Dy/\lang v_y\rang$ is the propagation time of the fluctuating density patterns between detectors poloidally separated by the distance $\Dy$. However, $\taupeakcc$ will not coincide with $\tauprop$ if the lifetime $\tau_{life}$ of the fluctuations is not long compared to $\tauprop$.  The failure of the method for $\lang v_y \rang < 5.0\:km/s$ illustrated in \figref{mean_vel_check} is an example of what happens when $\tauprop$ is too large.  Here, we investigate the effect of $\tau_{life}$ on $\taupeakcc$ quantitatively, via a systematic $\tau_{life}$ scan of the synthetic BES data.
\newline\indent
\myfig[5.5in]{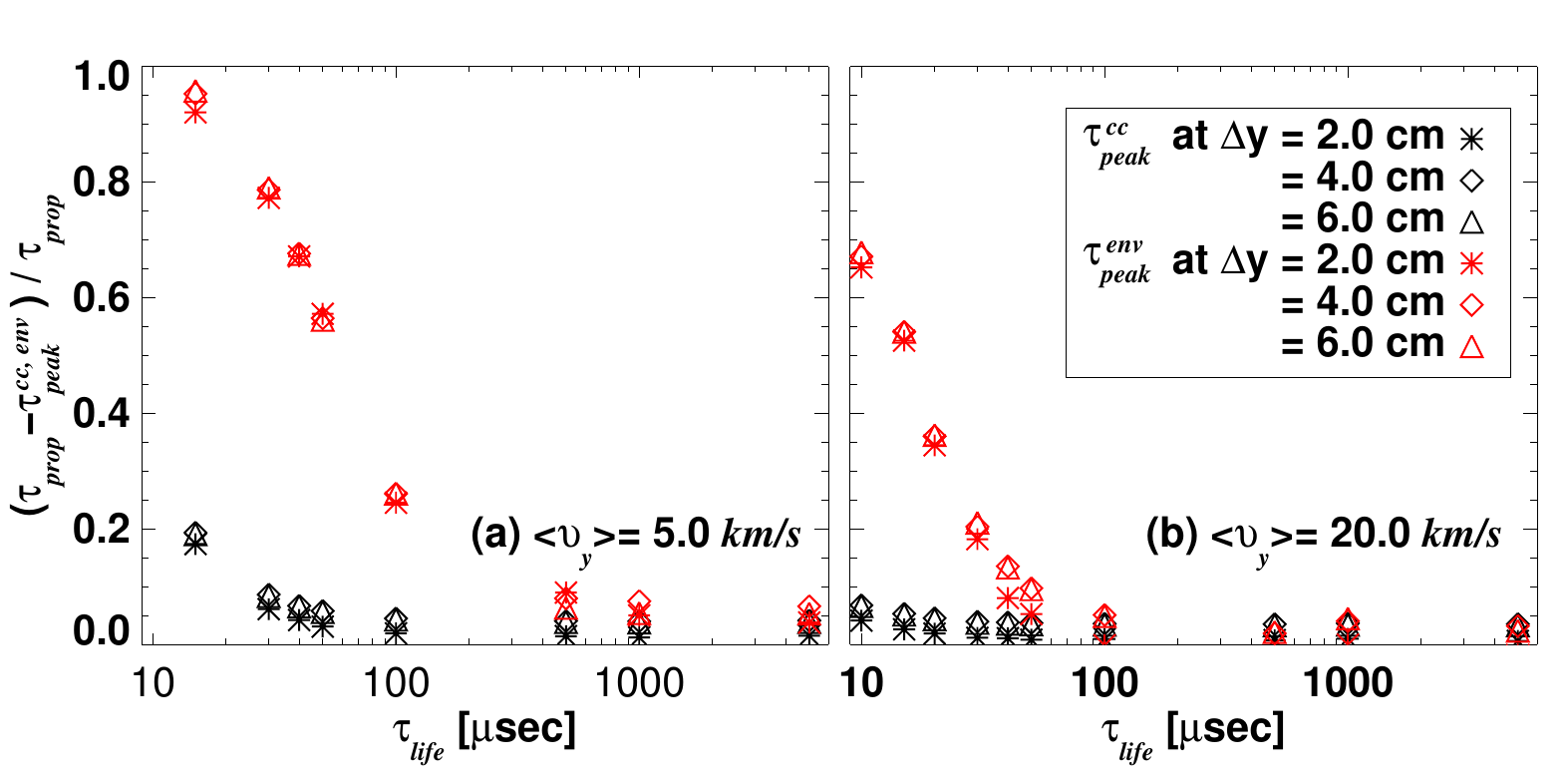}{Dependence of statistical reliability of the CCTD method on the lifetime of eddies}{Relative discrepancy between the propagation time $\tauprop=\Dy/\lang v_y\rang$ and the times $\taupeakcc$ (black) or $\taupeakenv$ (red) at which the cross-correlation function or its envelope reaches their peaks for (a) $\lang v_y \rang=5.0\:km/s$ and (b) $20.0\:km/s$.}
Two values $\lang v_y \rang=5.0$ and $20.0\:km/s$ are chosen for this study.  For $\lang v_y \rang=5.0\:km/s$, $\tauprop=4.0$, $8.0$ and $12.0\:\mu s$ with $\Dy=2.0$, $4.0$ and $6.0\:cm$, respectively; for $\lang v_y \rang=20.0\:km/s$, they are $1.0$, $2.0$ and $3.0\:\mu s$.  The peak time $\taupeakcc$ is found using the polynomial fitting method described in \secref{sec:desc_cctd_method}, and $\lp\tauprop-\taupeakcc\rp/\tauprop$ as a function of $\tau_{life}$ is plotted for three different values of $\Dy$ in \figref{lifetime_effects}.  It shows that $\taupeakcc$ underestimates the true $\tauprop$ for small values of $\tau_{life}$, leading to an overestimation of the $\lang v_y \rang$, consistent with the results shown in \figref{mean_vel_check}.  It is encouraging, however, that even relatively low velocities of just a few $km/s$ can be determined by the CCTD method with reasonable accuracy ($\sim\:20\%$).
\newline\indent
It is also possible to consider the global maximum of the envelope of the cross-correlation function.  We use a Hilbert transform to determine the time delay $\taupeakenv$ at which the envelope of the cross-correlation function is maximum \cite{durst_rsi_1992} as shown in \figref{fig_corr_func}.  The comparison between $\taupeakenv$ and $\tauprop$ is shown in \figref{lifetime_effects}.  It is clear that $\taupeakenv$ has a much stronger dependence on $\tau_{life}$ than $\taupeakcc$, so this measure will not be used to estimate $\lang v_y \rang$ in this work. We note, however, that the strong dependence of $\taupeakenv$ on the eddies' lifetime $\tau_{life}$ and of $\taupeakcc$ on their propagation time $\tauprop$ may provide a way to measure correlation times in the plasma frame.  Such an investigation is currently being pursued and will be reported elsewhere.
\myfig[4.0in]{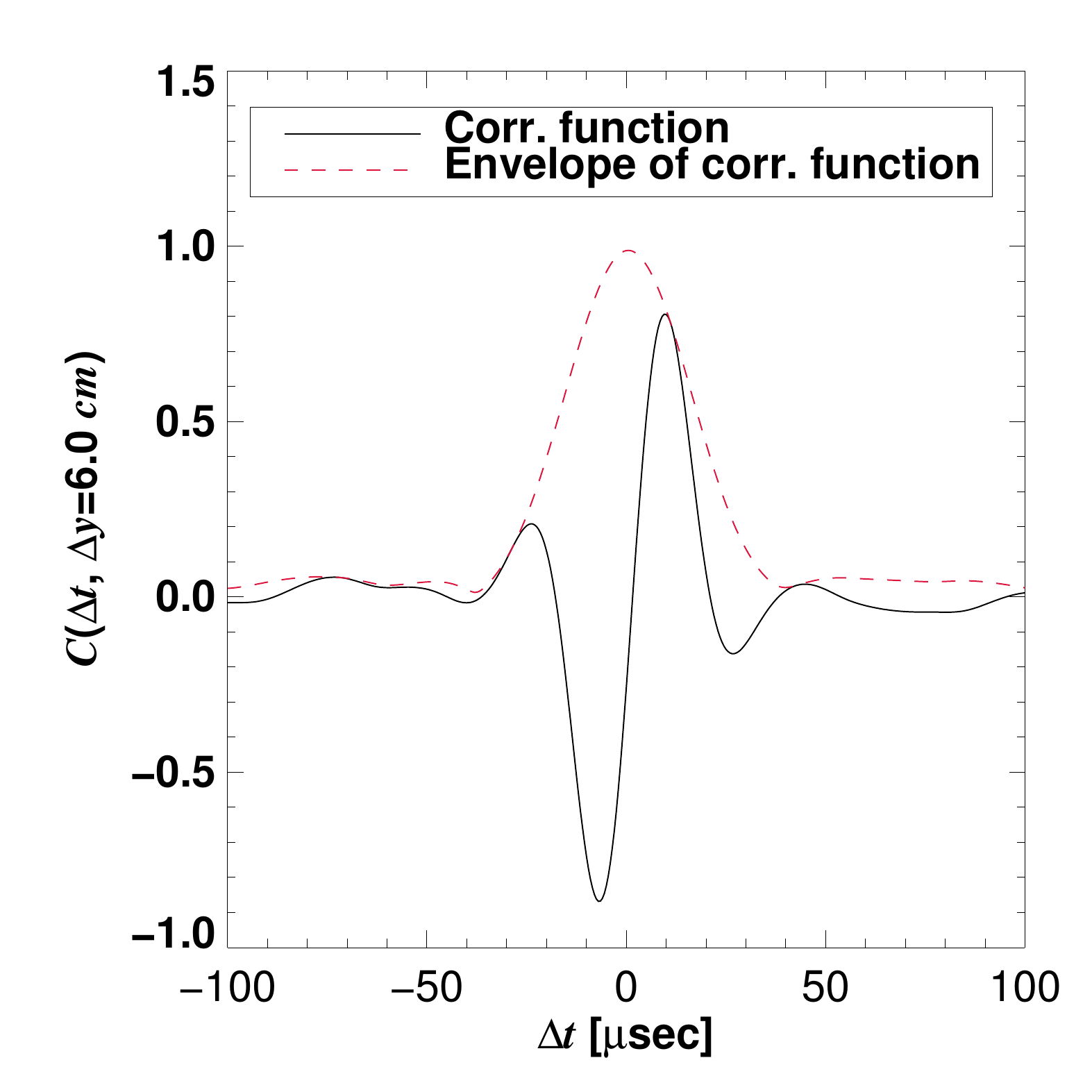}{Comparison of correlation function and its envelope}{Comparison of correlation function and its envelop at $\Dy=6.0\:cm$ with $\lang v_y \rang=5.0\:km/s$ and $\tau_{life}=15.0\:\mu s$. It is clear that the peak position of the envelope function $\taupeakenv$ is smaller than that of the correlation function $\taupeakcc$.}

\subsection{Effect of coherent MHD modes}\label{sec:influence_global_mode}
Many experimental 2D BES data sets on MAST exhibit strong MHD (global mode) activity in addition to the small-scale turbulence.  Removing such global modes in the frequency domain is not straightforward as they can have multiple harmonics extending into higher frequencies.  While they could be filtered out relatively easily in the wavenumber domain, constructing wavenumber spectra with a very limited number of spatial data points is difficult.  Thus, it is useful to investigate how the presence of such modes affects the quality of our measurement of $\lang v_y \rang$.  In this section, this is done by using synthetic BES data sets with different RMS levels $I_{MHD}^{RMS}$ and frequencies $f_{MHD}$ of the global oscillations (the $I_{MHD}^{ij}$ term in \eqref{eq:syn_bes_data_total_def}).
\newline\indent
\myfig[5.5in]{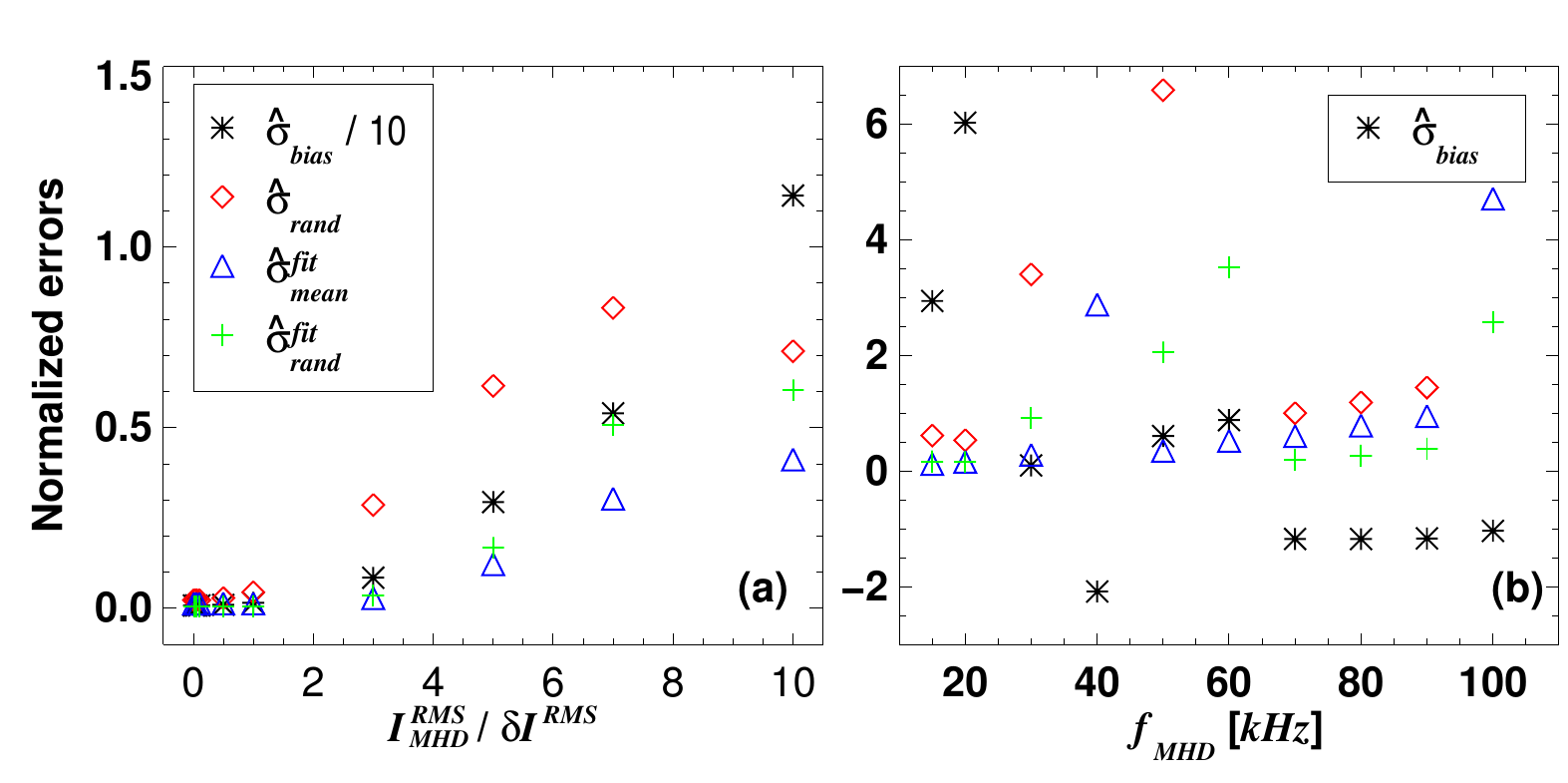}{Dependence of statistical reliability of the CCTD method on the amplitude and frequency of MHD}{Four types of error (a) as functions of the RMS levels of a global mode $I_{MHD}^{RMS}$ relative to that of turbulence signal $\delta I^{RMS}$; the frequency is fixed at $f_{MHD}=15.0$ kHz; (b) as functions of the global mode frequency $f_{MHD}$ at fixed $I_{MHD}^{RMS}/\delta I^{RMS}=5.0$.  Note that $\hsbias$ in (a) is scaled down by a factor of 10, and some points are missing in (b) because they are out of the plot range.}
The four errors ($\hsbias$, $\hsrand$, $\hsmeanfit$ and $\hsrandfit$) are calculated for various ratio of $I_{MHD}^{RMS}$ to the RMS value of $\delta I^{ij}\lp t \rp$ (i.e., $\delta I^{RMS}$ in \eqref{eq:syn_bes_data_fluc_def}).  These errors are plotted in \figref{check_mean_vel_with_MHD}(a) for the $I_{MHD}^{RMS}$ scan.  Here, the frequency of the global mode $f_{MHD}=15.0$ kHz and $\lang v_y \rang = 10.0\:km/s$.  It is clear that if the power level of the mode is larger than that of the turbulence signal, then the CCTD method produces large bias errors $\hsbias$.  To examine how the frequency of a global mode affects the errors, $f_{MHD}$ is varied with a fixed value of $I_{MHD}^{RMS}/\delta I^{RMS}=5.0$.  The results of this scan are shown in \figref{check_mean_vel_with_MHD}(b).  It shows that $\hsbias$ can be either positive or negative with different values of $f_{MHD}$ meaning that global modes in real experimental data can cause both over- and under-estimation of the true $\lang v_y \rang$.
\newline\indent
\myfig[5.5in]{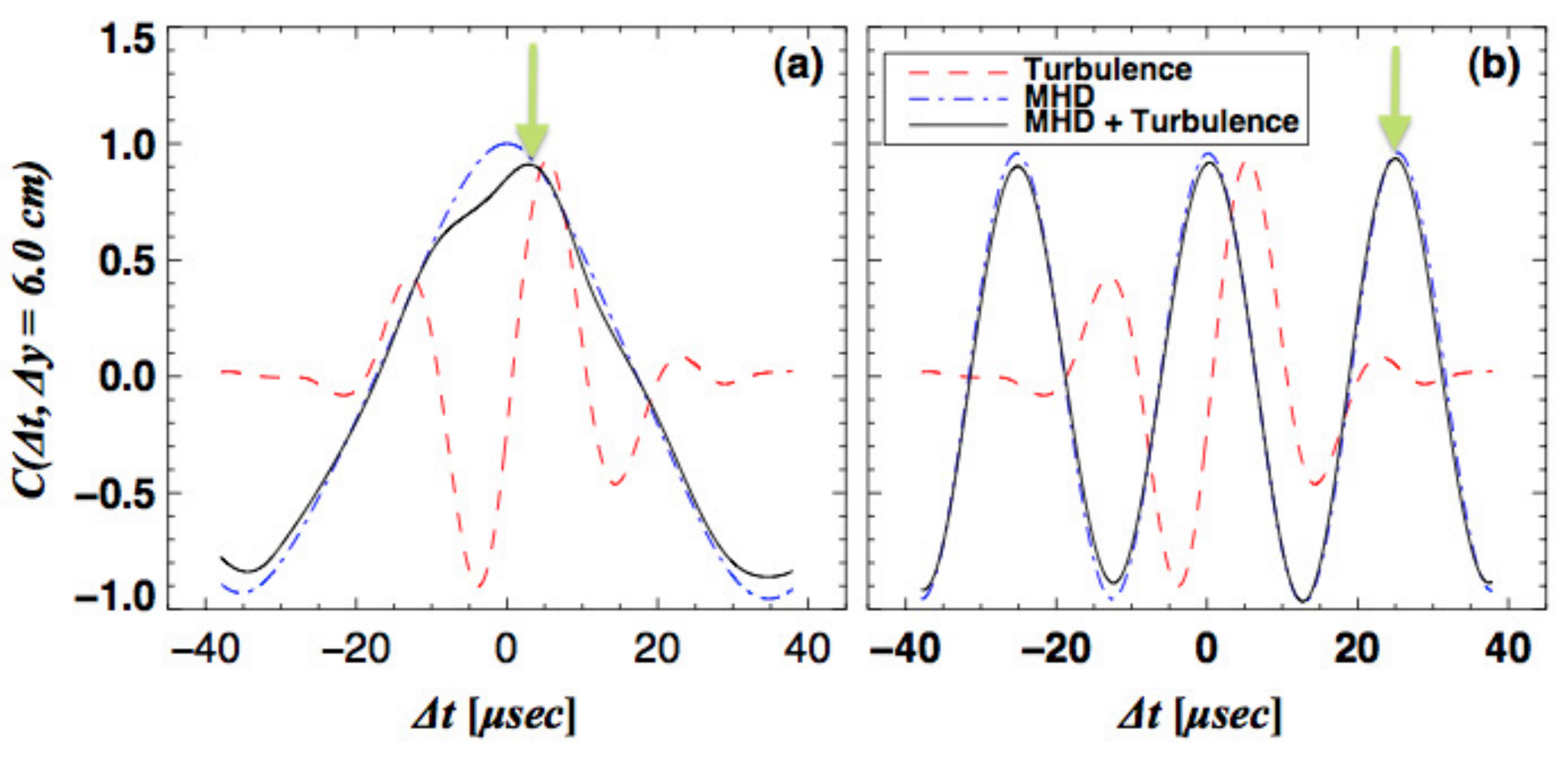}{Effects of MHD on cross-correlation functions}{Cross-correlation functions of the random eddies only (red dash), the global mode only (blue dash dot) and the eddies with the global mode (black solid) with (a) $f_{MHD}=15.0$ and (b) $f_{MHD}=40.0$ kHz.  Green arrows indicate the position of $\taupeakcc$, which does not coincide with the maximum of the cross-correlation function of the eddies only (red dash).}
\figref{cc_with_MHD} shows how different frequencies $f_{MHD}$ can cause such an over- or under-estimation of the $\lang v_y \rang$. Two identical sets of synthetic BES data with $\lang v_y\rang=10.0\:km/s$ are generated, one with and another without a global mode at (a) $f_{MHD}=15.0$ kHz and (b) $f_{MHD}=40.0$ kHz, with $I_{MHD}^{RMS}/\delta I^{RMS}=5.0$.  Without the global modes, the cross-correlation functions with $\Dy=6.0\:cm$ (red dashes in \figref{cc_with_MHD}) have the expected value $\taupeakcc\approx 6.0\:\mu s$ for both cases.  In contrast, the presence of the global mode in the synthetic BES data shifts $\taupeakcc$ towards (a) smaller time-lag (over-estimation) or (b) larger time-lag (under-estimation).  
\newline\indent
We conclude that a global (MHD) mode with $I_{MHD}^{RMS}>\delta I^{RMS}$ affects the structure of the cross-correlation functions (both the shape and the position of $\taupeakcc$) rendering the CCTD method unreliable.

\subsection{Effect of temporally varying poloidal velocity}\label{sec:influence_fluc_vel}
\myfig[5.5in]{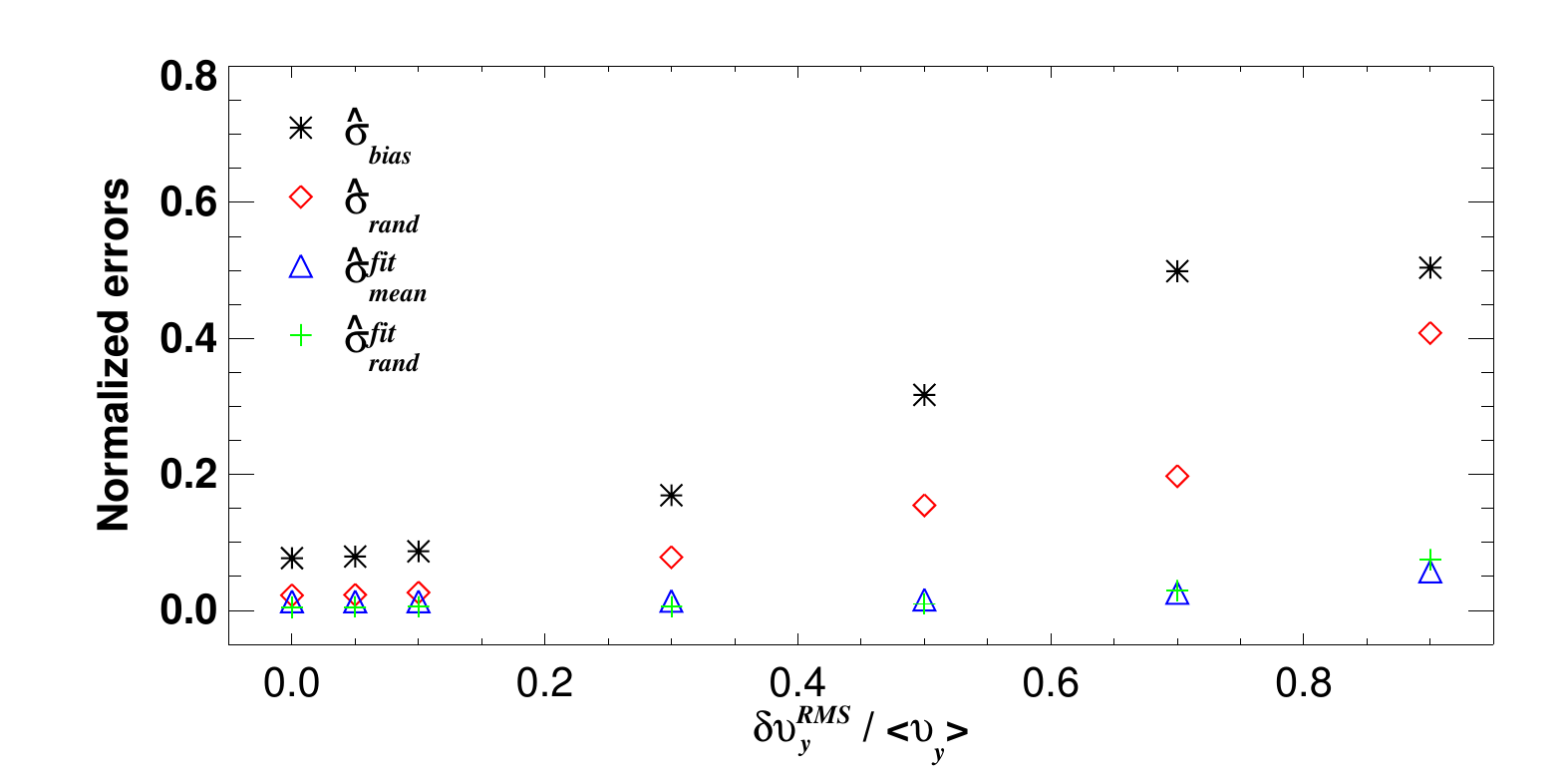}{Dependence of statistical reliability of the CCTD method on the amplitude of fluctuating velocity}{Four types of error for various RMS levels $\lang v_y\rang$ of temporally varying poloidal velocities.}
No physical quantities are absolutely quiet in real experiments, thus it is necessary to investigate how the RMS level $\delta v_y^{RMS}$ of the temporal variation of the poloidal velocity (see \eqref{eq:gam_generator}) influences the measurement of $\lang v_y \rang$. 
\newline\indent
\figref{check_mean_vel_with_vfluc} shows how finite $\delta v_y^{RMS}/\lang v_y \rang$ (with $\lang v_y\rang=10.0\:km/s$) affect the four errors defined in \secref{sec:def_uncertainties}.  It appears that $\hsbias$ saturates at around $50\:\%$ for the scenarios we have investigated, while other three errors increase without showing any sign of saturation.  Thus, the CCTD method to measure $\lang v_y\rang$ is subject to a non-negligible bias error (up to $\sim\:50\%$) if the RMS level of temporal variation of the poloidal velocity is greater than a half of the mean poloidal velocity.

\section{Experimental results}\label{sec:eddy_motion_exp_results}
In this section, we apply the CCTD method to 2D BES data from MAST discharges to determine the apparent mean poloidal motion ($\vbes$) of the ion density patterns.  Then, $\vbes$ is compared with the `rotating barber-pole' velocity ($U_z\tan\alpha$) where the toroidal plasma velocity $U_z$ is obtained from the CXRS system \cite{conway_rsi_2006} and the local magnetic pitch angle $\alpha$ either from \texttt{EFIT} equilibrium reconstruction \cite{lao_nf_1985} or the MSE system \cite{kuldkepp_rsi_2006, debock_rsi_2008}.
\newline\indent
The 2D BES data are first bandpass-filtered from $20.0$ to $100.0$ kHz to reduce the noise level.  The low-pass filter removes the high-frequency noise component from the photon shot noise and electronic noise as the signal hits the noise level above $100.0$ kHz in general, while the high-pass filter reduces the contribution to the signal from low-frequency, coherent MHD modes as well as the beam noise.  The apparent mean poloidal velocity of the density patterns $\vbes$ is determined from average correlation functions calculated over $25$ time intervals of $40\:\mu s$ duration, resulting in total $1\:ms$ averaging.  Second-order polynomial fitting is applied around the maximum of the correlation function so that $\taupeakcc$ can be obtained from a continuous time domain rather than a discretized one due to the $0.5\:\mu s$ sampling time as described in \secref{sec:desc_cctd_method}.  Finally, five consecutive values of $\vbes$ obtained in this manner are averaged, so the total averaging time is $5\:ms$ which is the effective time resolution of $\vbes$.  Using these five values of $\vbes$, the time average of various errors defined in Eqs. (\ref{eq:norm_rand_err})-(\ref{eq:norm_rand_fit_err}) in \secref{sec:def_uncertainties} are also computed.
\newline\indent
We present measurements of $\vbes$ from four different discharges: shot \#27278 (L-mode), shot \#27276 (H-mode), shot \#27269 (ITB) and shot \#27385 (high-poloidal-beta).  All four discharges had double-null diverted (DND) magnetic configurations and co-current NBI (neutral  beam injection).  In all of these discharges, the 2D BES system viewed at nominal major radial position of $R=1.2\:m$ corresponding to normalized minor radii $r/a=0.2-0.3$ for L- and H-modes, and $r/a=0.3$-$0.4$ for ITB and high-poloidal-beta discharges.  The evolution of key parameters for these discharges is shown in \figref{basic_info}.  The evolution of  plasma current, line-integrated electron density and poloidal beta characterize the overall behaviour of plasmas, while the non-zero S-beam voltage corresponds to times when the 2D BES system obtains localized density fluctuation.  The $D_\alpha$ intensity trace is used to identify when the H-mode discharge (shot \#27276) goes into its H-mode: namely, at $t=0.21 - 0.28\:s$.  Note that the ITB discharge (shot \#27269) starts developing a strong temperature gradient at $\sim\:0.2\:s$ and the peak ion ($C^{6+}$ from the CXRS) temperature keeps increasing until the NBI cuts off at $0.3\:s$.  The viewing position of the 2D BES system is in the middle of the strong temperature gradient region for this discharge.
\myfig[4.5in]{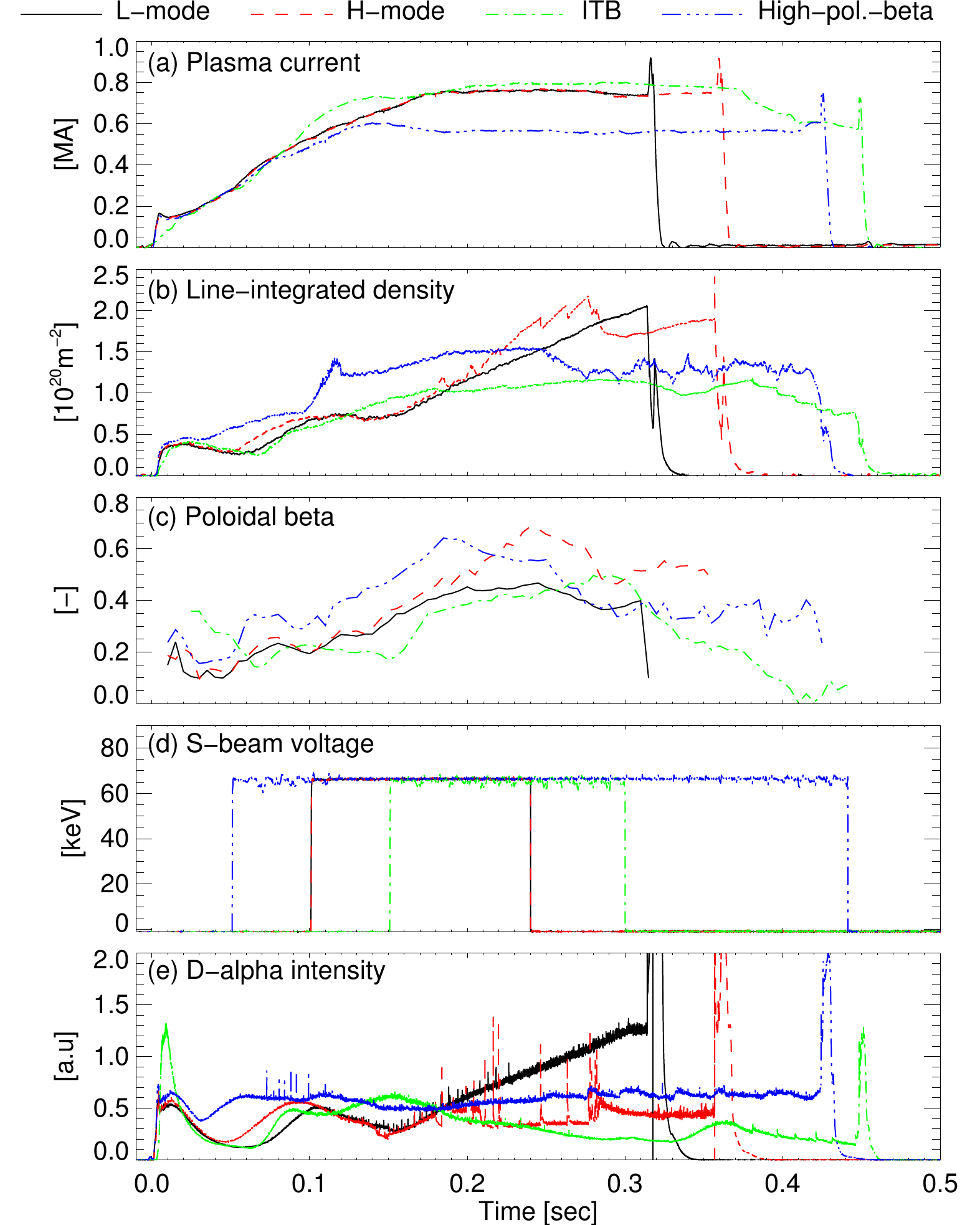}{Evolution of basic plasma parameters}{Evolution of (a) plasma current (b) line-integrated electron density (c) poloidal beta (d) NBI (S-beam) injection energy and (e) edge $D_\alpha$ intensity of L-mode (shot \#27278, black solid),  H-mode (shot \#27276, red dash), ITB (shot \#27269, green dash dot) and high-poloidal-beta (shot \#27385, blue dash dot dot) discharges.}

\subsection{L-mode (shot \#27278), H-mode (shot \#27276) and ITB (shot \#27269) discharges: 
$\vbes\approx -U_z\tan\alpha$}\label{sec:exp_barberpole}
\myfig[3.2in]{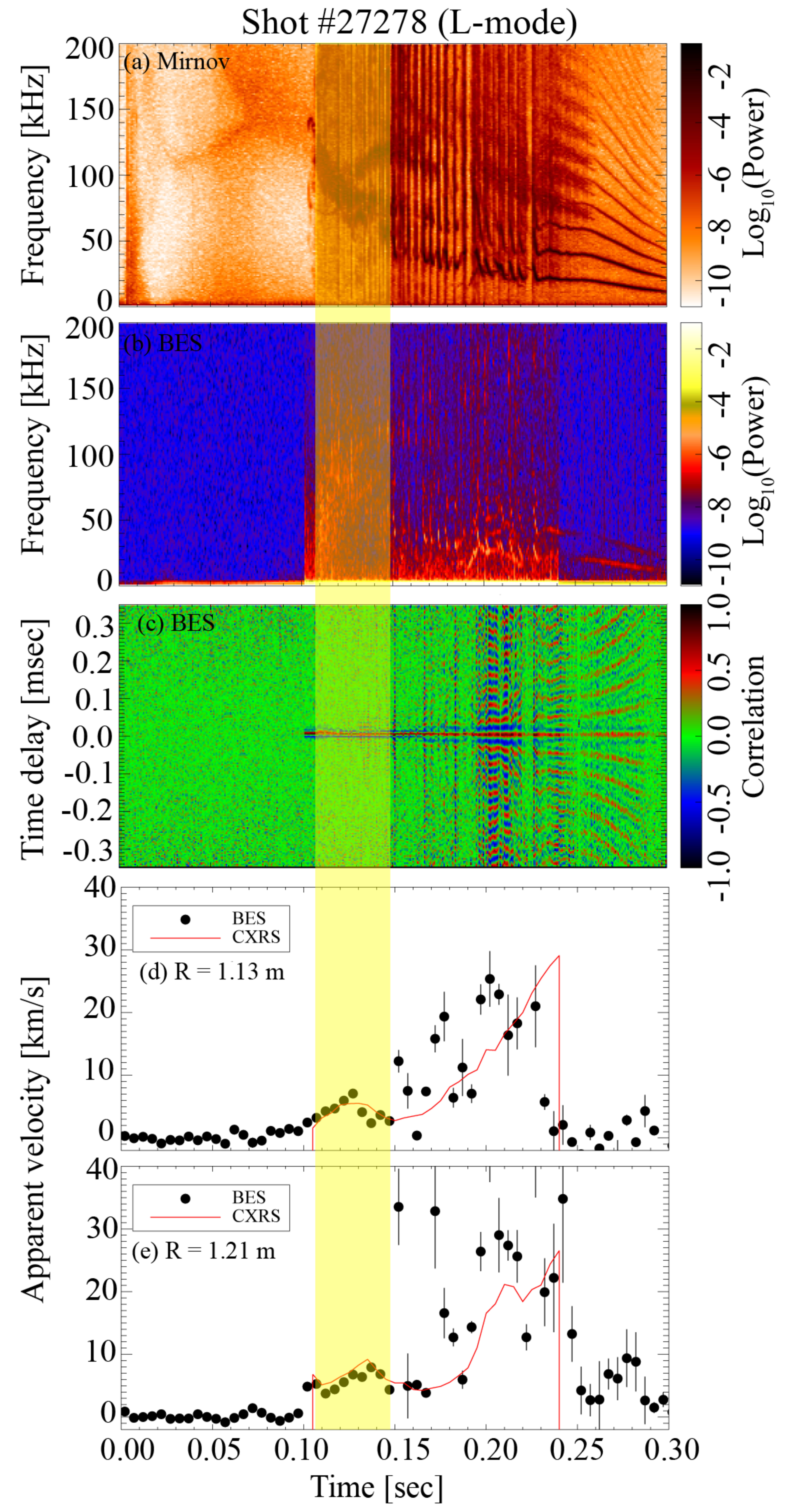}{Time evolution of shot \#27278 (L-mode)}{The evolution of shot \#27278 (L-mode) showing (a) cross-power spectrogram of the fluctuating B-field signal, (b) cross-power spectrogram and (c) cross-correlation of the density fluctuations from BES at $R = 1.21\:m$.  The time evolution of $-\vbes$ (circles) and the `rotating barber-pole' velocity ($U_z\tan\alpha$, red solid line) at (d) $R = 1.13\:m$ and (e) $R = 1.21\:m$.  Shaded region shows where $-v_y^{BES}\approx U_z\tan\alpha$. How we generate cross-correlation functions and power spectra are discussed in \appendixref{ch:ex_corr_power}.}
Time evolution of (a) cross-power of the fluctuating magnetic field signal from two toroidally separated outboard Mirnov coils, (b) cross-power and (c) temporal cross-correlation of density fluctuations from two poloidally separated BES channels (two mid-channels separated by $2\:cm$) located at $R = 1.21\:m$ are shown in Figures \ref{fig:27278_time_evolution} (L-mode), \ref{fig:27276_time_evolution} (H-mode) and \ref{fig:27269_time_evolution} (ITB discharge).  Here, a cross-power is defined as the Fourier transform (in the time domain) of the cross-correlation function (\eqref{eq:cc_definition}) with finite channel separation. The (minus) apparent mean poloidal velocity ($-\vbes$, circles) determined by the CCTD method and the `rotating barber-pole' velocity ($U_z\tan\alpha$, red solid lines) are also shown in panels (d) at $R = 1.13\:m$ and (e) at $R = 1.21\:m$ for these three discharges.  The error bars represent the mean error $\lang\delta v_{fit}\rang$ of the least-squares fit, as discusses in \secref{sec:def_uncertainties}.
\mydoublesidefig{3.0in}{3.0in}{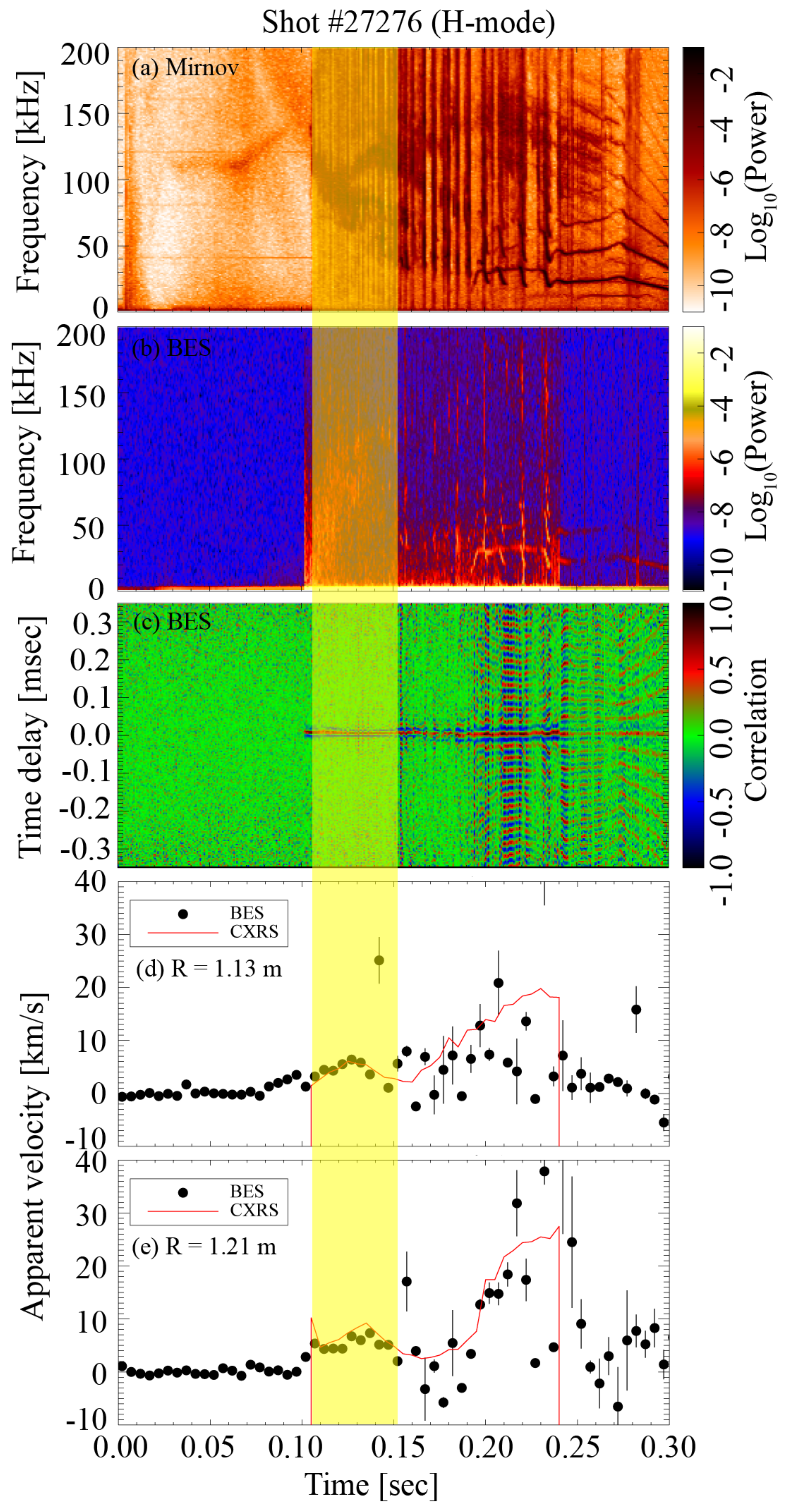}{Time evolution of shot \#27276 (H-mode)}{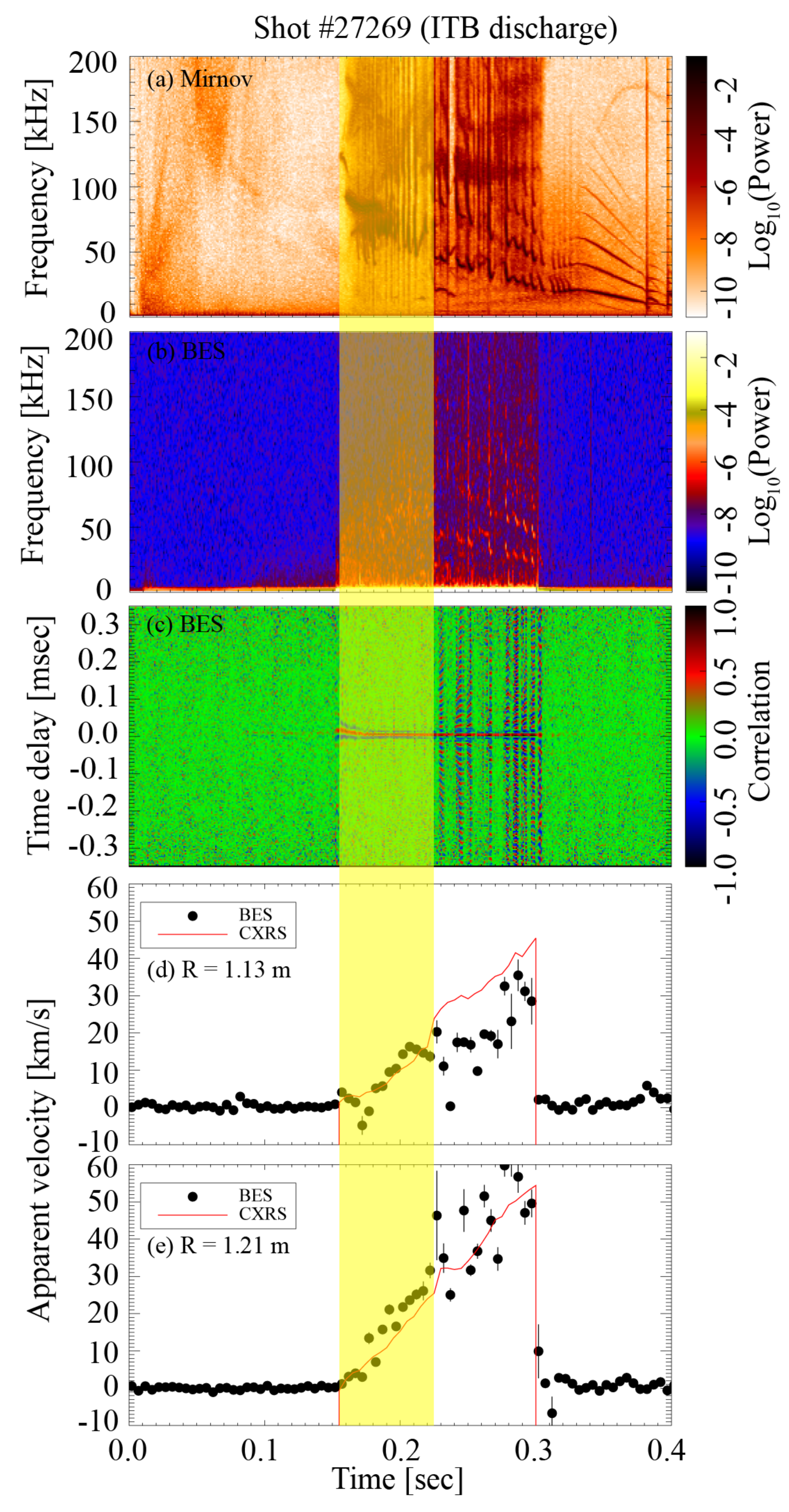}{Time evolution of shot \#27269 (ITB discharge)}{Same as \figref{27278_time_evolution} for shot \#27276 (H-mode).}{Same as \figref{27278_time_evolution} for shot \#27269 (ITB discharge).}
\newline\indent
Despite the fact that these three discharges belong to three very different classes, there are common features in the apparent mean poloidal velocity:
\newline\noindent
(1) $\vbes$ is not reliable (i.e., has large error bars) when strong MHD activity is present.  The cross-power spectrograms from BES show clear signatures of MHD modes with many harmonics, which hamper filtering the BES signal over the frequency domain.  The temporal cross-correlations, i.e., Figures \ref{fig:27278_time_evolution}(c), \ref{fig:27276_time_evolution}(c) and \ref{fig:27269_time_evolution}(c), also show that these MHD modes have much longer correlation times ($>0.3\:ms$) than the turbulent density patterns. The Mirnov signals show that plasmas develop chirping/fishbone modes with low toroidal and poloidal mode numbers \cite{gryaznevich_ppcf_2004} within the filtered frequency band after the shaded region in Figures \ref{fig:27278_time_evolution}(a), \ref{fig:27276_time_evolution}(a) and \ref{fig:27269_time_evolution}(a).  The effects of these MHD (global) modes on the CCTD method are investigated in \secref{sec:influence_global_mode}, where it is found that such activity can increase not only the absolute values of the bias errors but also the linear fitting errors on $\vbes$.  Thus, comparisons between $\vbes$ and $U_z\tan\alpha$ are difficult to make during the periods where the MHD activity is strong.
\newline\noindent
(2) During the periods of weak MHD activity, i.e., $0.11$-$0.15\:s$ for the L- and H-mode discharges, and $0.16$-$0.22\:s$ for the ITB discharge, it is clear that the apparent mean poloidal velocity of turbulent density patterns is dominated by the `rotating barber-pole' velocity, i.e., \eqref{eq:simple_v_bes_relation} holds, and the sum of all the terms of the order of $\rhostar$ or higher in \eqref{eq:definition_gamma} is indeed small.
\newline\indent
Note that the H-mode discharge (shot \#27276) goes into its H-mode at $\sim0.21\:s$ (thus, $\vbes=-U_z\tan\alpha$ is only true before the L-H transition, strictly speaking), which can be seen from the $D_\alpha$ intensity trace in \figref{basic_info} or from the BES cross-power spectrogram in \figref{27276_time_evolution}: the turbulence level drops at the start of the H-mode.  Any changes of $\vbes$ during the L-H transition cannot be discussed, because the CCTD method with the current data analysis scheme is not reliable at this time due to strong MHD activity.

\subsection{High-poloidal-beta discharge (shot \#27385): $\vbes \ne -U_z\tan\alpha$}\label{sec:exp_non_barberpole}
\myfig[3.5in]{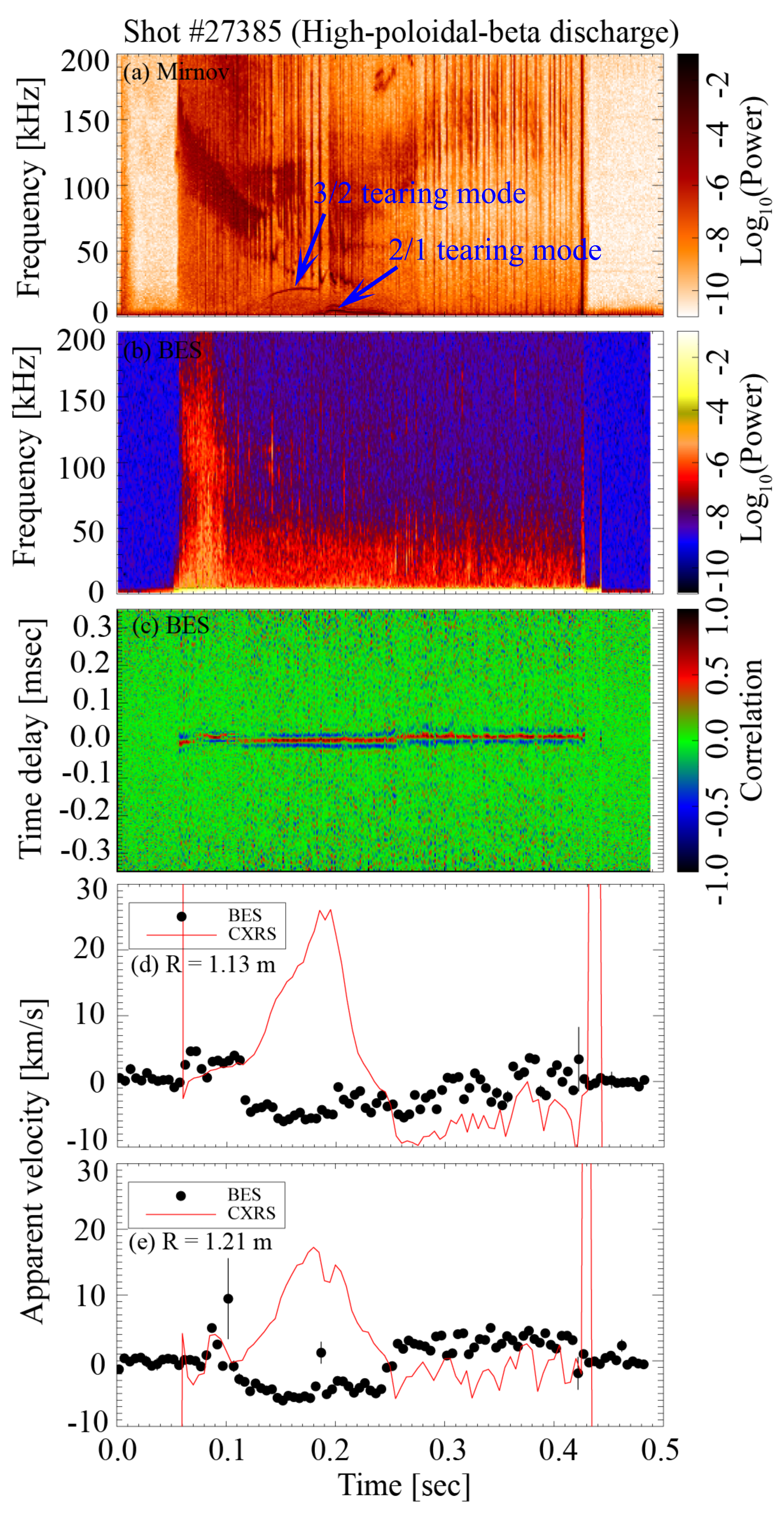}{Time evolution of shot \#27385 (High-poloidal-beta discharge)}{Same as \figref{27278_time_evolution} for shot \#27385 (high-poloidal-beta discharge).  Note that neither (b) the cross-power spectrogram nor (c) the temporal cross-correlation of the BES signal show any MHD activity.}
Shot \#27385 has a relatively higher poloidal beta (the ratio of the plasma pressure to the poloidal magnetic field energy density) than the three discharges discussed in \secref{sec:exp_barberpole} (see \figref{basic_info}).  Thus, it is more susceptible to tearing modes (i.e., formation of magnetic islands) \cite{haye_pop_2000, buttery_prl_2002}.  The cross-power spectrogram between the two toroidally separated outboard Mirnov coils displayed in \figref{27385_time_evolution}(a) shows a $m/n=3/2$ tearing mode on the $q=1.5$ flux surface starting at $\sim0.11\:s$; its frequency increases from $<10$ kHz to $\sim25$ kHz at $\sim0.19\:s$. Then, a $m/n=2/1$ mode (fundamental frequency $<10$ kHz) develops and locks to the wall resulting in complete braking of the toroidal rotation of plasmas at $\sim0.25\:s$.  Here, $m$ and $n$ denote the poloidal and toroidal mode numbers, respectively, and $q$ for the safety factor.  The $q=3/2 (2/1)$ surface moves from $R=0.85 (0.90)\:m$ to $1.24 (1.27)\:m$ during the initial $0.15\:s$ period and then stays at constant $R$.
\newline\indent
 The $2/1$ mode is not expected to be seen on the BES signal as it is bandpass-filtered from $20.0-100.0$ kHz, and no trace of the $3/2$ mode is visible in the BES signal.\footnote{Because the mode flattens the mean density profile within the island, shaking of flux surfaces does not induce density fluctuations in the BES signal.}  Consequently, the $\vbes$ determined by the CCTD method does not contain large error bars during the whole discharge.
\newline\indent
The time evolution of $-\vbes$ and $U_z\tan\alpha$ in \figref{27385_time_evolution}(d)-(e) at two different radial locations shows that the two velocities do not agree each other at all during the period when the $3/2$ and $2/1$ modes are present.  What we find, remarkably, is that while the plasma continues to rotate toroidally (as attested by the CXRS data), there is virtually no detectable corresponding motion of the density patterns.  In fact, they seem to exhibit a weak rotation in the opposite direction to the expected rotating barber-pole effect.  Formally, this means that the plasma effects in the right-hand-side of \eqref{eq:continuity_equation_BES} are not small and are able to cancel almost exactly the toroidal rotation, i.e., an effective velocity of the density patterns develops in the plasma frame that to lowest order is equal to minus the rotation velocity.  We do not currently have a theoretical explanation for this effect.  There is very little apparent difference between the turbulent density patterns in this discharge compared to others, except somewhat longer radial correlation lengths.

\section{Conclusions}\label{sec:eddy_motion_conc}
We have analysed 2D BES data from different types of discharges on MAST to determine the apparent mean poloidal velocities of the ion-scale density patterns using the cross-correlation time delay method.  The dominant cause of the apparent poloidal motion of the density patterns is experimentally identified to be due to the fact that field aligned patterns are advected by the background, dominantly toroidal, plasma rotational flow, i.e., the `rotating barber-pole' effect dominates the apparent mean motion of the density patterns in the lab frame.  This conclusion holds for the L-, H-mode and ITB discharges we have investigated.  An exception to this rule is found to be the investigated high-poloidal-beta discharge, where a large magnetic island is present, and the apparent velocity of the density patterns is very small, despite strong toroidal rotation.  Identifying the causes of this effect by investigating the behaviour of the turbulent density patterns quantitatively is left for future work.

\chapter{Experimental signatures of critically balanced turbulence in MAST}\label{ch:critical_balance}
\begin{center}
\textit{This chapter is largely taken from Ref. \cite{ghim_prl_2012}.}
\end{center}
Beam Emission Spectroscopy (BES) measurements of ion-scale density fluctuations in MAST are used to show that the turbulence correlation time, the drift time associated with ion temperature or density gradients, the particle (ion) streaming time along the magnetic field and the magnetic drift time are consistently comparable, suggesting a ``critically balanced'' turbulence determined by the local equilibrium. The resulting scalings of the poloidal and radial correlation lengths are derived and tested. The nonlinear time inferred from the density fluctuations is longer than the other times; its ratio to the correlation time  scales as $\nu_{*i}^{-0.8\pm0.1}$, where $\nu_{*i}=$ ion collision rate/streaming rate. This is consistent with turbulent decorrelation being controlled by a zonal component, invisible to the BES, with an amplitude exceeding the drift waves' by~$\sim \nu_{*i}^{-0.8}$.

\section{Introduction}
Microscale turbulence hindering energy confinement in magnetically confined hot plasmas is driven by gradients of equilibrium quantities such as temperature and density. These gradients give rise to instabilities that inject energy into plasma fluctuations (``drift waves'') at scales just above the ion Larmor scale. The most effective of these is believed to be the ion-temperature-gradient (ITG) instability \cite{cowley_pfb_1991, rudakov_doklady_1961, coppi_pof_1967}. A turbulent state ensues, giving rise to ``anomalous transport'' of energy \cite{horton_rmp_1999}. It is of interest, both for practical considerations of improving confinement and for the fundamental understanding of multiscale plasma dynamics, what the structure of this turbulence is and how its amplitude, scale(s) and resulting transport depend on the equilibrium parameters: ion and electron temperatures, density, angular velocity, magnetic geometry, etc. 
\newline\indent
Fluctuations in a magnetized toroidal plasma are subject to a number of distinct physical effects, which can be thought about in terms of various time scales such as the drift times associated with the temperature and density gradients, the particle streaming time along the magnetic field as it takes them around the torus toroidally and poloidally, the magnetic ($\grad B$ and curvature) drift times of particles moving across the field, the nonlinear time of the fluctuations being advected across the field by the fluctuating $\vct{E}\times\vct{B}$ velocity, the time between collisions, the shear time associated with plasma rotation. Some of these time scales and, consequently, the corresponding physics may be irrelevant, while others play a crucial role for the saturation of the linearly unstable fluctuations. There has been a growing understanding \cite{barnes_prl_2011_107}, driven largely by theory \cite{goldreich_apj_1995,cho_apj_2004,schekochihin_apjs_2009,nazarenko_jfm_2011},  observations \cite{horbury_prl_2008,podesta_apj_2009,wicks_mnras_2010} and simulations of magnetohydrodynamic \cite{cho_apj_2000,maron_apj_2001,chen_mnras_2011} and kinetic \cite{cho_apj_2004,tenbarge_pop_2012} plasma turbulence in space, that if a medium can support parallel (to the magnetic field) propagation of waves (and/or particles) and nonlinear interactions in the perpendicular direction, the turbulence in such a medium would normally be ``critically balanced,'' meaning that the characteristic time scales of propagation and nonlinear interaction would be comparable to each other and (therefore) to the correlation time of the fluctuations. This means that the turbulence is {\em not} weak and {\em not} two-dimensional, unless specially constrained to be so \cite{nazarenko_jfm_2011}. 
\newline\indent
Beam Emission Spectroscopy (BES) measurements of density fluctuations in tokamak plasmas \cite{field_rsi_2012, fonck_rsi_1990,fonck_prl_1993,mckee_rsi_1999,mckee_rsi_2003} have made it possible to probe ion-scale turbulence in these devices directly. In this chapter, we use such measurements in MAST, along with the local equilibrium parameters calculated by other diagnostics, to estimate and compare the characteristic time scales of the turbulent fluctuations in the energy-containing range. We obtain, for the first time, direct evidence that the correlation, drift and parallel streaming time scales are indeed comparable across a range of equilibrium parameters (cf. \cite{mckee_nf_2001, hennequin_ppcf_2004}) and that the magnetic drift time is part of this ``grand critical balance'' as well. We also find indirect evidence that the decorrelation rate of turbulence is controlled by a zonal component whose relative importance to the drift-wave-like fluctuations scales with the ion collisionality.  
\newline\indent
Before presenting this evidence and its implications (e.g., dependence of the  correlation lengths on equilibrium parameters), let us describe how it was obtained.

\section{Experimental data and its analysis}
During the 2011 campaign, density fluctuation data from the BES diagnostic \cite{field_rsi_2012} on MAST were collected in a variety of discharges (including L- and H-modes and internal transport barriers). Here we report the data from 39 neutral-beam heated ``double-null-diverted'' discharges, with no pellet injection and no resonant magnetic perturbations. The BES system on MAST collects photons from a 2D array of 8 radial $\times$ 4 vertical locations in the outboard midplane of the tokamak, with 2\:cm separation between the adjacent channels in either direction. The detected photon intensity (mean $+$ fluctuating, $I+\dI$) is used to infer, at each location, the density fluctuation level $\dn/n=\lp1/\betabes\rp\lp\dI/I\rp$ \cite{fonck_rsi_1990}, where $\betabes$ depends on the mean density $n$ and is estimated based on the Hutchinson model \cite{hutchinson_ppcf_2002} (dependence on the mean temperature is weak). As the BES array was moved radially for different discharges, our database contains cases with radial viewing positions $10$\:cm\:$<r<50$\:cm from the magnetic axis (the minor radius of the plasma is $\approx60$\:cm).
\newline\indent
Local equilibrium parameters are measured by standard diagnostics: mean electron densities $n_e$ and temperatures $T_e$ by the Thomson scattering system \cite{scannell_rsi_2010}, impurity ion (C$^{6+}$) mean temperatures (assumed to equal the bulk ion temperature $T_i$) and toroidal flow velocity $U_\phi$ by the Charge eXchange Recombination Spectroscopy (CXRS) system \cite{conway_rsi_2006}, local magnetic pitch angle $\alpha$ by the Motional Stark Effect (MSE) system \cite{debock_rsi_2008}, and further equilibrium magnetic field information is obtained from pressure- and MSE-constrained \texttt{EFIT} equilibria \cite{lao_nf_1985}.
\newline\indent
We filter the BES data to the frequency interval $[20,100]$ kHz and calculate the spatio-temporal correlation function
\begin{align}
\label{eq:turb_scale_corr_def}
\nonumber
&\corr\lp\Dx,\DZ,\Dt\rp = \\
&\frac{\lab \dI\lp x, Z, t \rp \dI\lp x+\Dx, Z+\DZ, t+\Dt\rp \rab}{\sqrt{\lab\dI^2\lp x, Z, t\rp\rab \lab\dI^2\lp x+\Dx, Z+\DZ,  t+\Dt\rp\rab}},
\end{align}
where $x$, $Z$ and $t$ are the radial, vertical and time coordinates, respectively, and $\Dx$, $\DZ$ and $\Dt$ are the corresponding channel separations and the time lag; $\lab \cdot \rab$ is the time average over 5\:ms periods. At $\Dx=\DZ=0$, the auto-covariances $\lab \dI\lp x, Z, t \rp \dI\lp x, Z, t+\Dt\rp \rab$ contain not only the physical signal but also photon and electronic noise. We remove this effect by applying LED light to the BES channels, obtaining 150 different DC levels of BES signal from $0$ to $1.5$\:V, calculating the noise auto-covariance $C_N\lp\Dt\rp$ at each DC level with the same band frequency filter of $[20,100]$ kHz, then finding $C_N\lp\Dt\rp$ whose DC level of the signal matches the DC level of the BES data from the MAST discharges, and subtracting it from the calculated auto-covariances (see \chref{ch:bes_principle} for more detailed description). From the correlation function \eqref{eq:turb_scale_corr_def}, we calculate the local characteristics of the density fluctuations.  The spatio-temporal correlation function and spatio-spatio correlation function are illustrated in \figref{spatio_temporal_cont} and \figref{spatio_spatio_cont}, respectively.
\mydoublesidefig{2.8in}{2.8in}{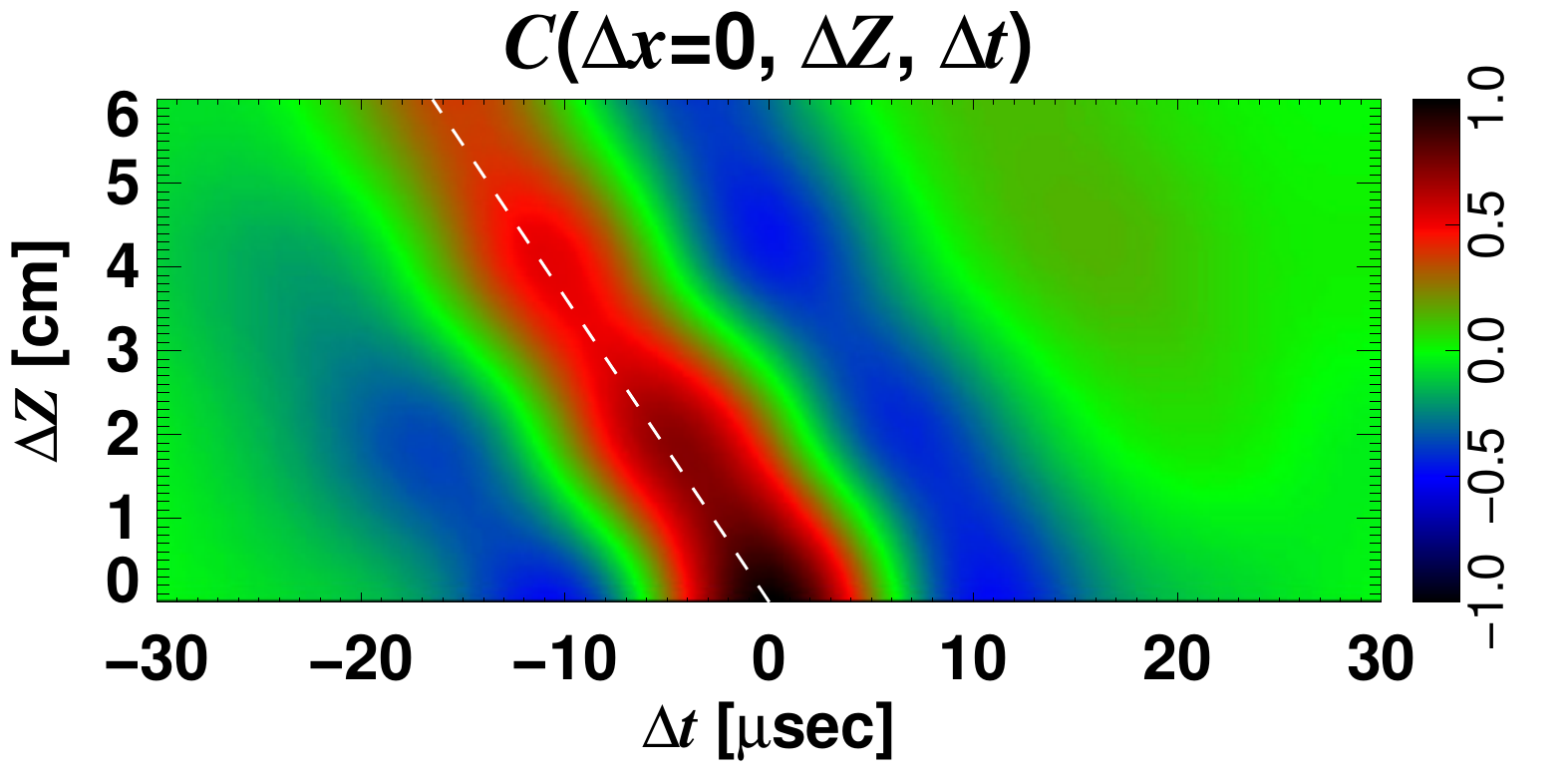}{Spatio-temporal correlation function of BES data}{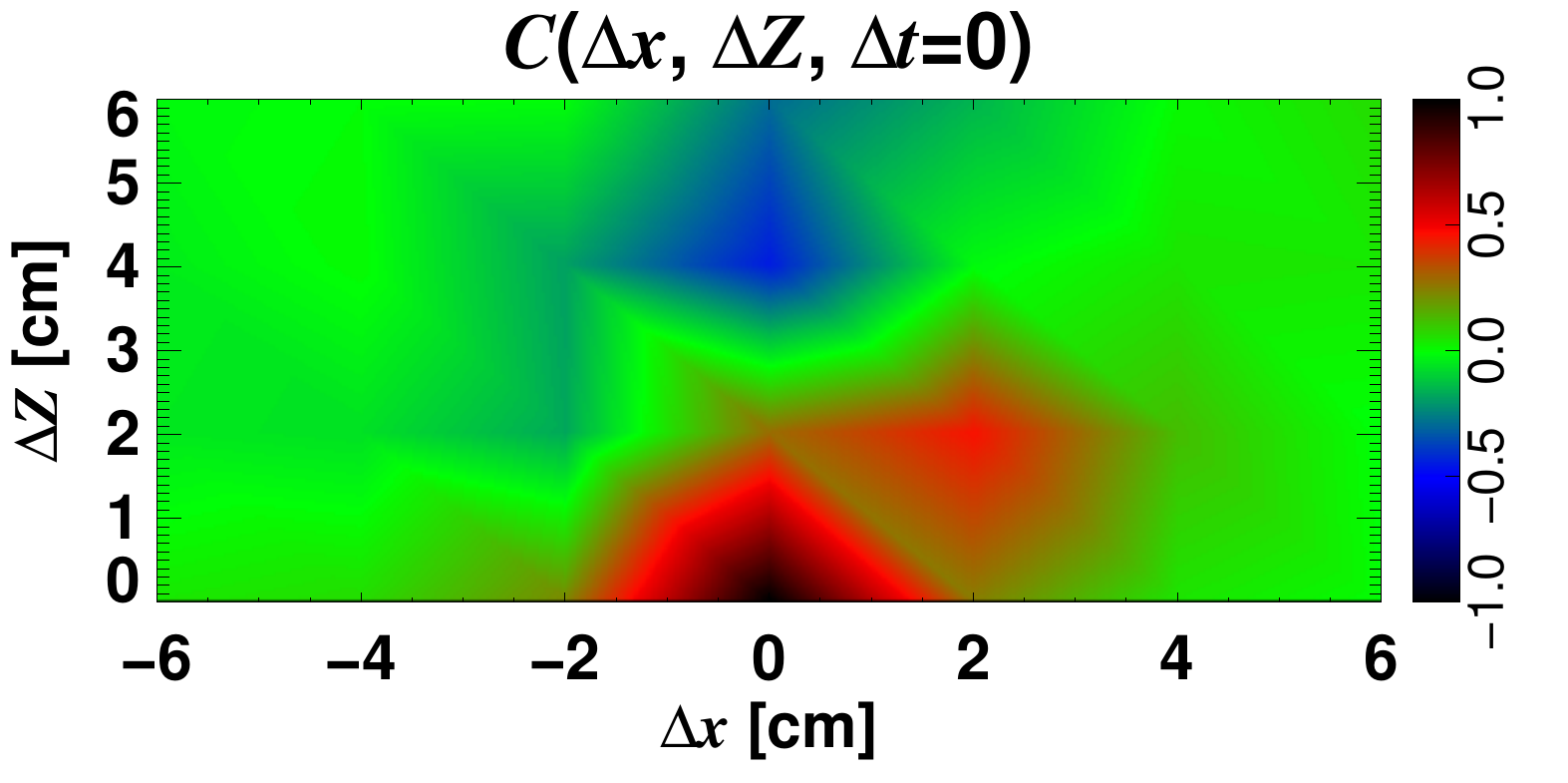}{Spatio-spatio correlation function of BES data}{An example of the correlation function in the poloidal-temporal plane, $\corr\lp\Dx=0,\DZ,\Dt\rp$. This data was taken at $r=30$\:cm, toroidal rotation speed was $U_\phi=10$\:km/s and magnetic pitch angle $\alpha=20^\circ$. The direction of maximum correlation is the direction of the magnetic field (dashed line).}{An example of the correlation function in the poloidal-radial plane, $\corr\lp\Dx, \DZ, \Dt=0\rp$.  This data was taken at the same location and the time as in \figref{spatio_temporal_cont}.}
\newline\indent
The fluctuation level at each radial location is obtained from the (noise-subtracted) auto-covariance function $\dn/n = \lp1/\betabes\rp\sqrt{\lab\dI^2(x,Z,t)\rab}/I$ at all 32 locations and then averaged over the four poloidally separated channels at the same radial location. 
\newline\indent
The correlation length $\ly$ in the direction parallel to the flux surface and perpendicular to the magnetic field is obtained from the vertical (poloidal) correlation length $\lZ$ via $\ly=\lZ\cos\alpha$, assuming that the parallel correlation length is sufficiently long: $\lpar\gg\ly\tan\alpha$. The correlation length $\lZ$ is estimated using four poloidal channels at each radial location (the top channel is the reference channel) by fitting $\corr\lp\Dx=0, \DZ, \Dt=0\rp$ to the function $f_Z\lp\DZ\rp = p_Z + \lp 1 -p_Z \rp\cos\lsb2\pi \DZ/\lZ\rsb\exp\lsb-\labs\DZ\rabs/\lZ\rsb$, where $p_Z$ is a fitting constant that serves to account for global structures such as coherent MHD modes (for which $\corr\lp \Dx=0, \DZ=\infty, \Dt=0\rp = p_Z \neq 0$). In choosing $f_Z\lp\DZ\rp$, we assumed wave-like fluctuations in the poloidal direction \cite{fonck_prl_1993} (drift-wave turbulence), with the wavelength and correlation length comparable to each other. It is not possible to distinguish meaningfully between the two with only four poloidal channels. Assuming wave-like structure is essential as in most cases, we find that $\corr\lp\Dx=0, \DZ, \Dt=0\rp$ goes negative and/or is non-monotonic over the vertical extent of the BES array.
\newline\indent
The radial correlation length $\lx$ is estimated using eight radial channels at each poloidal location (the fourth channel from the inward side is the reference channel). The correlation function $\corr\lp\Dx, \DZ=0, \Dt=0\rp$ is fitted to the function $f_x\lp\Dx\rp = p_x + \lp 1 -p_x \rp\exp\lsb-\labs\Dx\rabs/\lx\rsb$, where $p_x$ plays the same role as $p_Z$ did for $f_Z$. The values of $\lx$ from four poloidal locations are averaged, assuming that the radial correlations do not change significantly within the poloidal extent of the BES array. Because we have to use the entire array to estimate $\lx$, the number of data points for $\lx$ is 8 times smaller than for $\ly$.
\newline\indent
To estimate the correlation time $\tc$, we use the fact that the fluctuating density patterns are advected poloidally past the BES array with an apparent velocity $\vbes = U_\phi\tan \alpha$ due to the toroidal rotation velocity $U_\phi$ \cite{ghim_ppcf_2012} (discussed in \chref{ch:eddy_motion}). We fit $\corr\lp\Dx=0, \DZ, \Dt=\taupeakcc\lp\DZ\rp\rp$ taken at the time delay $\taupeakcc\lp\DZ\rp$ when the correlation function is maximum at a given $\DZ$ \cite{durst_rsi_1992}, to the function $f_\tau\lp\DZ\rp = \exp\lsb -\labs\taupeakcc\lp\DZ\rp\rabs/\tc\rsb$. The reliability of this method relies on the temporal decorrelation dominating over the parallel spatial decorrelation, viz., we require $\tc\ll\lpar\cos\alpha/U_\phi$. Anticipating the critical balance assumption $\tc\sim \lpar/\vti$ \cite{barnes_prl_2011_107}, where $\vti = \sqrt{2T_i/m_i}$ is the ion thermal speed, and denoting the Mach number $\mathrm{Ma}=U_\phi/\vti$, we estimate that the fractional error in $\tc$ is $\sim \mathrm{Ma}/\cos\alpha$, which was never more than 20\% in the MAST discharges we used.
\newline\indent
The four quantities $\dn/n$, $\ly$, $\lx$ and $\tc$ are calculated (see \secref{sec:get_turb_info} for more detailed descriptions) at $8$ radial locations (except $\lx$), every $5$\:ms for all 39 discharges. All the fits described above are obtained via the \texttt{mpfit} procedure \cite{markwardt_mpfit}. We consider a data point unreliable and remove it from the database if 
\newline
(i) $I<0.3$\:V (the signal-to-noise ratio (SNR) is too low, i.e., SNR$<140$ \cite{field_rsi_2012}),
\newline
(ii) the estimated correlation lengths are smaller than the distance between the channels, $\lx$ or $\ly < 2$\:cm (to be more precise, data points are valid only if $\lx$ and $\ly$ are larger than the size of the point-spread-functions (PSFs) shown in \figref{bes_psf}; however, most of estimated $\ly$ is greater than $10\:cm$, and $2\:cm$ is a reasonable approximation of the PSFs in radial direction), 
\newline
(iii) the assumption that plasma rotation is mostly toroidal is suspect, viz., $\labs\lp\vbes-U_\tor\tan\alpha\rp/\vbes\rabs \ge 0.2$ (see Ref. \cite{ghim_ppcf_2012} and \chref{ch:eddy_motion}), where $\vbes$ is calculated at each radial location using the cross-correlation time delay (CCTD) method \cite{durst_rsi_1992},
\newline
(iv) the estimated error in the calculation of  $\vbes$ is $>20\%$ (see \chref{ch:eddy_motion} for the possible sources of error in the $\vbes$),
\newline
(v) $p_Z$ or $p_x >0.5$,
\newline
The last two exclusion criteria pick out the cases when MHD modes are too strong; they are known to degrade the reliability of the BES data \cite{ghim_ppcf_2012}. The remaining database contains 448 points.

\section{Time scales}
\myfig[5.0in]{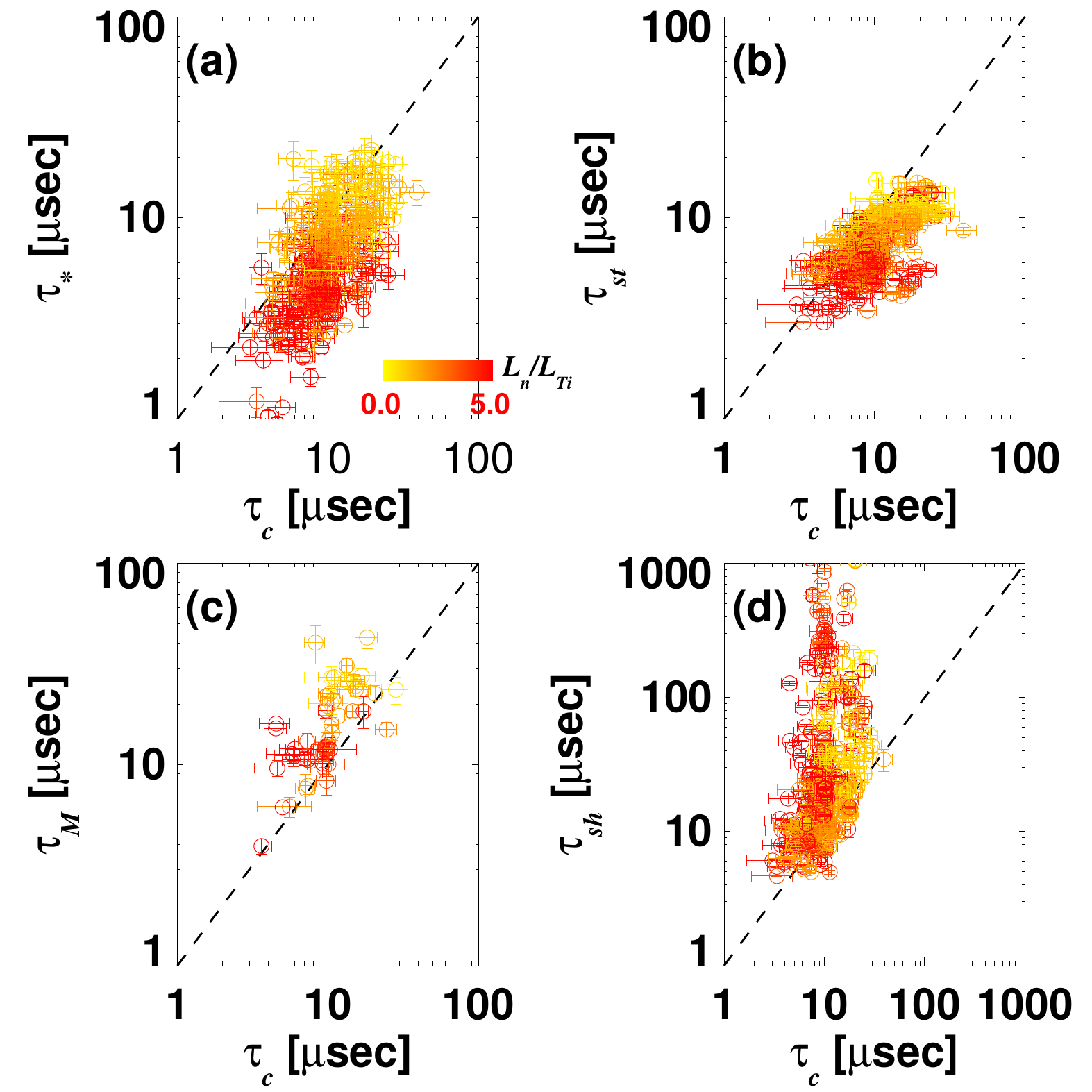}{Comparisons among various time scales}{(a) Drift time $\tstar=\lp\ly/\rhoi\rp\Lstar/\vti$ vs.\ correlation time $\tc$; (b) streaming time $\tst=\Lambda/\vti=\lp B/B_p\rp\pi r/\vti$ vs.\ $\tc$; (c) magnetic drift time $\tM = \lp\lx/\rhoi\rp R/\vti$ vs.\ $\tc$; (d) perpendicular velocity shear time $\tshear=\lsb\lp B_p/B\rp dU_\phi/dr]
\rsb^{-1}$ vs.\ $\tc$. In all cases, the colour of points represents $\eta_i=\Ln/\LTi$.}
\subsection{Correlation time vs. drift time}
The turbulence can be driven by radial gradients in the mean ion and electron temperatures $T_{i,e}$ and density $n$. Denoting $\LTie^{-1}=|\grad\ln T_{i, e}|$ and $\Ln^{-1}=|\grad\ln n|$, the associated time scales are the inverse drift frequencies: 
\begin{equation}
\tstarie^{-1} = \frac{\rhoie}{\ly}\frac{\vtie}{\LTie},\quad 
\tstarn^{-1} = \frac{\rhoi}{\ly}\frac{\vti}{\Ln}, 
\end{equation}
where $\rhoie=\vtie/\Omega_{i,e}$ are the ion ($i$) and electron ($e$) Larmor radii, $\vtie=\sqrt{2T_{i, e}/m_{i, e}}$ the thermal speeds and $\Omega_{i,e}=eB/m_{i,e}c$ the Larmor frequencies. To estimate the drift times, we need information about the local equilibrium ($T_{i,e}$, $\LTie$ $\Ln$, $B$) and the correlation length $\ly$, calculated from the poloidal BES correlations. 
\newline\indent
In \figref{all_vs_tau_c_eta}(a), we compare the drift times with the correlation time $\tc$ calculated from the spatio-temporal BES correlations. We find that $\tstar=(0.7\pm0.3)\tc$, where $\tstar=\min\{\tstari,\tstarn\}$ and the spread is calculated as the root mean square deviation from the mean value. The scaling holds over an order of magnitude in either time scale. Thus, the turbulence appears to be driven by the larger of the ion temperature or density gradient. However, for $\tc\lesssim10\:\mu s$, $\tstare\sim\tstari$ and for $\tc\gtrsim10\:\mu s$, $\tstarn\sim\tstari$, so we cannot rule out ion-scale electron drive (e.g., trapped electron modes \cite{kadomtsev_nf_1971} or microtearing \cite{roach_ppcf_2005,guttenfelder_pop_2012,doerk_pop_2012}). We find no clear correlation of $\tstare$ with $\tc$, or with any of the other time scales discussed below.

\subsection{Critical balance}
The standard argument behind the critical balance conjecture is causality \cite{nazarenko_jfm_2011}: two distant points on a field line cannot stay correlated if information cannot be exchanged between them over a turbulence correlation time. Assuming information travels at $\vti$, one gets $\lpar\sim\vti\tc$. This cannot be checked directly because there are no diagnostics capable of measuring $\lpar$ on MAST. \footnote{As noted above, our method for measuring $\tc$ would instead yield $\lpar/\vti$ if $\mathrm{Ma}>\cos\alpha$, but that would require much stronger rotation (the smallest value in our database is $\cos\alpha\approx0.76$).} Considering that the inboard side of the torus is a region of ``good'' (stabilizing) curvature, not much turbulence is expected there, so we assume that, at the energy injection scale, $\lpar\sim\Lambda$ \cite{barnes_prl_2011_107}, where the distance along the field line that takes a particle from the outer to the inner side of the torus is $\Lambda=\pi r B/B_p$ ($r$ is the minor radius at the BES position on the outer side and $B_p$ the poloidal component of the magnetic field).\footnote{In a conventional tokamak, $\Lambda\approx \pi q R$, where $q$ is the safety factor and $R$ major radius, but in a spherical tokamak, the local estimate we use is more appropriate.} Then critical balance means that $\tc$ should be comparable to 
\begin{equation}
\tst^{-1}=\frac{\vti}{\Lambda} = \frac{\vti}{\pi r}\frac{B_p}{B}\sim\frac{\vti}{\lpar},
\end{equation}  
the ion streaming time 
(the first two equalities are its definition, the last an assumption). Indeed, we find $\tst=(0.8\pm0.3)\tc$ (see \figref{all_vs_tau_c_eta}(b)). 
\newline\indent
The balance $\tst\sim\tstar$ implies that the poloidal correlation scale is $\ly/\rhoi\sim \Lambda/\Lstar$, where $\Lstar=\min\{\LTi,\Ln\}$ \cite{barnes_prl_2011_107}. This is tested in \figref{spatial_eta}(a), showing that while the two quantities are certainly of the same order, we do not have enough of a range of equilibrium parameters to state conclusively that this theoretically predicted scaling works.

\subsection{Magnetic drift time and radial correlation scale}
\myfig[5.0in]{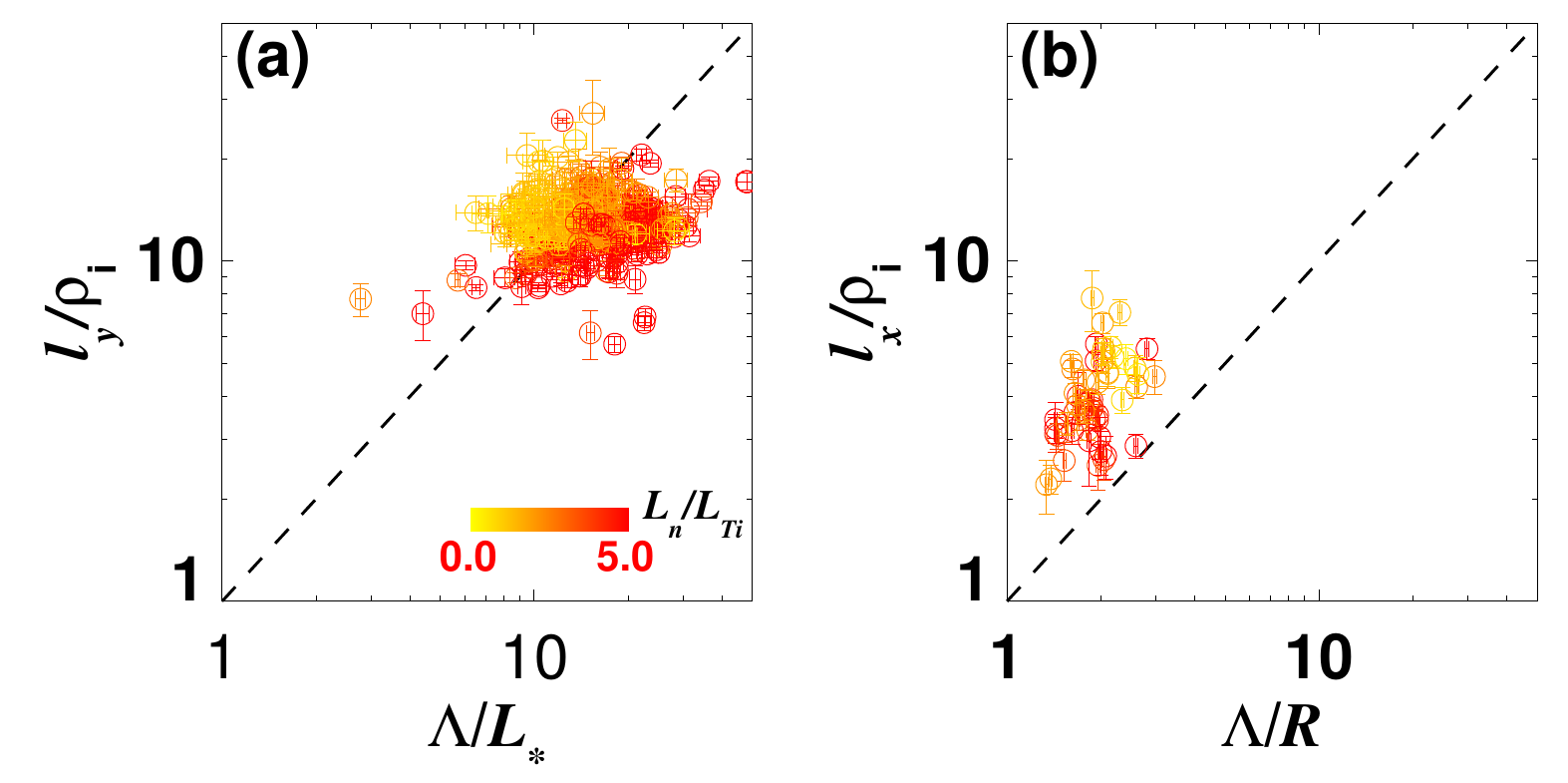}{Comparisons of measured and expected perpendicular correlation lengths}{(a) Poloidal correlation length $\ly/\rhoi$ vs.\ $\Lambda/\Lstar$; (b) Radial correlation length $\lx/\rhoi$ vs.\ $\Lambda/R$. Colour as in \figref{all_vs_tau_c_eta}.}
The time scale of the magnetic ($\grad B$ and curvature) drifts is
\begin{equation}
\tM^{-1}=\frac{\rhoi}{\lx}\frac{\vti}{R},
\end{equation}
where we have assumed that the scale length of the background magnetic field is $R$ (major radius at the viewing location) and $\lx<\ly$ (this will shortly prove correct). While magnetic drift physics may matter (in a torus, curvature contributes to the ITG drive \cite{horton_rmp_1999}), it does not have to affect scalings, as, for example, it would not in a slab and as it did not in the numerical simulations of \cite{barnes_prl_2011_107}. In contrast, \figref{all_vs_tau_c_eta}(c) shows that in the MAST discharges we have analyzed, $\tM$ is not negligible and scales with $\tc$, similarly to $\tstar$ and $\tst$. As $\tM$ contains $\lx$, there are 8 times fewer data points here than in previous two figures, as explained above. We find $\tM=(1.6\pm0.7)\tc$. Thus, a ``grand critical balance'' appears to hold in MAST, viz., $\tc\sim\tstar\sim\tst\sim\tM$.
\newline\indent
This suggests that the balance of all relevant timescales determines correlation scales of the turbulence in all three spatial directions. Indeed, balancing $\tM\sim\tst$, we find the radial correlation scale $\lx/\rhoi\sim \Lambda/R$, the scaling tested in \figref{spatial_eta}(b), with a degree of success. This means that the density fluctuations we are measuring in MAST are not isotropic in the perpendicular plane, but rather elongated in the poloidal direction $\ly/\lx\sim R/\Lstar$ ($\sim5$ in our data). Interestingly, this clashes with the reported approximate isotropy ($\lx\sim\ly$) both in Cyclone Base Case simulations \cite{barnes_prl_2011_107} and in measured DIII-D turbulence (where $\ly/\lx\sim1.4$ \cite{shafer_pop_2012} and $\lx$ does not appear to depend on $B_p$ \cite{rhodes_pop_2002}). Whether this is a difference between spherical and conventional tokamaks is not as yet clear.

\subsection{Nonlinear time}
Since we know the fluctuation amplitude, we can directly estimate the time scale associated with the advection of the fluctuations ($\vct{\delvper}\cdot\vct{\grad}\dn$) by the fluctuating $\vct{E}\times\vct{B}$ velocity $\delvper = c\vct{B}\times\vct{\grad}\varphi/B^2$. The electrostatic potential $\varphi$ is not directly measured, but can be estimated assuming Boltzmann response of the electrons: $\dn/n\approx e\varphi/T_e$. This estimate ignores trapped particles and, more importantly as we are about to argue, also does not apply to ion-scale zonal flows (poloidally and toroidally symmetric perturbations of $\varphi$ with $\dn=0$ \cite{diamond_ppcfreview_2005,fujisawa_nf_2009}). Thus, the non-zonal nonlinear time~is
\begin{equation}
\label{eq:tnl}
\lp\tnlnz\rp^{-1}=\frac{1}{\lx\ly}\frac{c\varphi}{B}=\frac{1}{\lx\ly}\frac{T_i}{eB/c}\frac{T_e}{T_i}\frac{e\varphi}{T_e}=\frac{\vti\rhoi}{\lx\ly}\frac{T_e}{T_i}\frac{\dn}{n}.
\end{equation}
\myfig[5.0in]{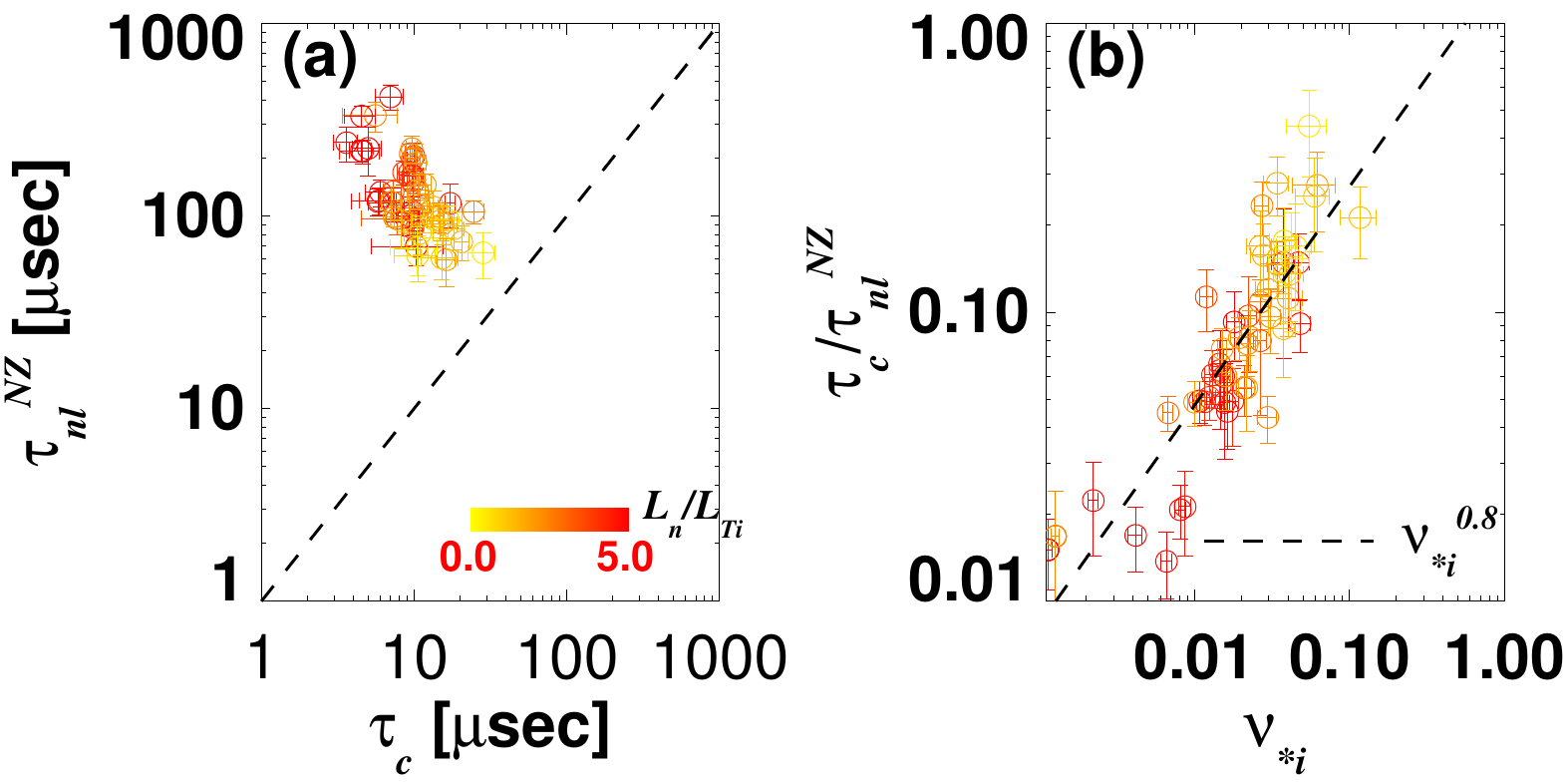}{Comparisons of nonlinear time associated with non-zonal component of turbulence with correlation time and ion-ion collisionality}{(a) The nonlinear time associated with density fluctuations, $\tnlnz$, vs.\ the correlation time $\tc$; (b) their ratio vs.\ normalized ion collision rate $\nu_{*i}=\nu_{ii}\tst$. Colour as in \figref{all_vs_tau_c_eta}.}
\figref{tau_nl_eta}(a) shows that $\tnlnz$ is always larger than $\tc$ (or the other time scales discussed above) and, furthermore, observed to have an inverse rather than direct correlation with it. Since turbulence clearly cannot be saturated by linear physics alone, this means that our estimate does not capture the correct nonlinear time. We conjecture that it is in fact the coupling to the zonal flows, invisible (directly) to BES (because their $\dn=0$)\footnote{Zonal flows can be detected using the BES system by looking at the velocity modulation of $\dn$ \cite{gupta_prl_2006} which requires more sophisticated statistical analysis. However, this does not change our estimation of $\tnlnz$.}, that dominates over the nonlinear interaction between the drift-wave-like fluctuations represented by $\tnlnz$ \cite{dimits_pop_2000, waltz_pop_1994, diamond_ppcfreview_2005, hammett_ppcf_1993, lin_science_1998, rogers_prl_2000, nakata_pop_2012, makwana_pop_2012}. It has long been suspected that the relative amplitude of the zonal flows compared to that of the drift waves depends on the ion collisionality \cite{diamond_ppcfreview_2005, hinton_ppcf_1999, xiao_pop_2007, ricci_prl_2006}. We can test this expectation by assuming that $\tc$ is the characteristic time associated with the coupling of the drift waves to the zonal flows and so depends on their amplitude. \figref{tau_nl_eta}(b) indeed shows a strong collisionality dependence: $\tc/\tnlnz \sim \nu_{*i}^{0.8\pm0.1}$, where $\nu_{*i}=\nu_{ii}\tst$ (the ion collision time itself, $\nu_{ii}^{-1}$, is at least an order of magnitude longer than the time scales that participate in the ``grand critical balance''). We note that a similar scaling is obtained for vs.\ $\nu_{ii}\tc$ and $\nu_{ii}\tstar$ or just straightforwardly for $(\tnlnz)^{-1}$ vs.\ $\nu_{ii}$. If $\tc^{-1}\sim (\vti\rhoi/\lx\ly)e\varphi^\mathrm{ZF}/T_i$, where $\varphi^\mathrm{ZF}$ is the amplitude of the zonal potential, this result implies that the ratio of zonal to non-zonal component of the turbulence is $\varphi^\mathrm{ZF}/\varphi^\mathrm{NZ}\sim \nu_{*i}^{-0.8\pm0.1}$. A scaling popular in theoretical models of zonal-flow-ITG turbulence is $\nu_{*i}^{-1/2}$ \cite{diamond_ppcfreview_2005}.
\newline\indent
We note that this situation is qualitatively distinct from what is seen in numerical simulations of ITG turbulence far from the threshold \cite{barnes_prl_2011_107}, where the drift-wave nonlinearity appears to dominate ($\tnlnz\sim\tc$). However, the turbulence in a real tokamak is likely to be close to marginal and so possibly in the state of reduced transport controlled by weakly-collisionally damped zonal flows \cite{fujisawa_nf_2009} and usually associated with the so-called ``Dimits upshift'' of the stiff-transport threshold \cite{dimits_pop_2000, rogers_prl_2000, lin_prl_1999, mikkelsen_prl_2008}. 

\section{Discussion and conclusions}
Our results support the notion that the statistics of turbulence are determined by the local equilibrium properties of the plasma. We find little correlation between the quantities reported above and the radial location\footnote{There is a slight bias in \figref{tau_nl_eta} for larger $\nu_{*i}$ to be found farther from the magnetic axis.} (note that we have limited our consideration to temporal and spatial scales and did not touch on the fluctuation amplitudes or transport properties, which do of course depend on radius). Our results also appeared insensitive to (i.e., not measurably correlated with) three other parameters that might in principle have proven important: $T_i/T_e$ (varied between $0.5$ and $2$), the magnetic shear $\hat s=d\ln q/d\ln r$ (varied between $-1$ and $5$) and the perpendicular component of the toroidal velocity shear $\tshear^{-1} = (B_p/B)dU_\phi/dr$. In much of our data, $\tshear \ge \tc,\tst$ (see \figref{all_vs_tau_c_eta}(d)), so it stands to reason that the statistics of the turbulence would not be dramatically affected; in the instances of $\tshear\sim\tst$, the effect of $\tshear$ could not be isolated. In general, we expect that a strong velocity shear would change $\lpar$ via a modified critical balance: if $\tshear<\tst$, then $\tc\sim\tshear\sim\lpar/\vti$, so $\lpar\sim\vti\tshear<\Lambda$. It would be interesting to investigate higher-rotation plasmas, as $\tshear^{-1}$, when sufficiently large, is expected to have a dramatic effect on transport \cite{highcock_prl_2012, barnes_prl_2011, highcock_prl_2010, roach_ppcf_2009, waltz_pop_1994, mantica_prl_2011, devries_nf_2009, parra_prl_2011}; even in our database, there is in fact some evidence that velocity shear might raise the critical temperature gradients \cite{ghim_prl2}, but we see no signature of this effect in the correlation properties of the turbulence.
\newline\indent
We have presented experimental results statistically consistent with a turbulent state in MAST set by the local equilibrium and in which the time scales of the linear drive, turbulence decorrelation, ion streaming and magnetic drifts are all similar and scale together as equilibrium parameters are varied. This ``grand critical balance'' implies a three-dimensionally anisotropic turbulence, with parallel, poloidal and radial correlation lengths  having different parameter dependences and $\lpar\gg\ly>\lx$. Our results also suggest the presence of a zonal component with an amplitude $\nu_{*i}^{-0.8\pm0.1}$ greater than the drift-wave density fluctuations. Furthermore, it provides a way to estimate the turbulent fluxes (at least an order of magnitude) solely based on equilibrium parameters as we can predict the spatial and temporal structures of turbulence and its fluctuation level.

\mypart{Better Tokamak Performance}

\chapter{Local dependence of ion temperature gradient on magnetic configuration, rotational shear and turbulent heat flux in MAST}\label{ch:larger_RLTi}
\begin{center}
\textit{This chapter is largely taken from Ref. \cite{ghim_prl2}.}
\end{center}

Experimental data from the Mega Amp Spherical Tokamak (MAST) is used to show that the inverse gradient scale length of the ion temperature $\RLTi$ (normalized to the major radius $R$) has its strongest {\em local} correlation with the rotational shear and the pitch angle of the magnetic field (or, equivalently, an inverse correlation with $q/\varepsilon$, the safety factor/the inverse aspect ratio). Furthermore, $\RLTi$ is found to be {\em inversely} correlated with the gyro-Bohm-normalized local turbulent heat flux estimated from the density fluctuation level measured using a 2D Beam Emission Spectroscopy (BES) diagnostic. These results can be explained in terms of the conjecture that the turbulent system adjusts to keep $\RLTi$ close to a certain critical value (marginal for the excitation of turbulence) determined by local equilibrium parameters (although not necessarily by linear stability).

\section{Introduction}\label{sec:rlti_intro}
A key physics challenge posed by magnetically confined plasmas in fusion devices is how the internal energy can be kept from being transported too fast from the core to the periphery. This problem is primarily one of turbulent transport, the temperature gradient between the edge and the core of a toroidal plasma supplying the source of free energy for the excitation of the turbulent fluctuations that then act to enhance the effective thermal conductivity and relax the gradient. It is expected that it is the {\em ion} temperature gradient (ITG) that gives rise to the most virulent instabilities (on ion Larmor scales) \cite{cowley_pfb_1991, rudakov_doklady_1961, coppi_pof_1967} and that this is then self-consistently limited by the resulting turbulence \cite{horton_rmp_1999}. If we view the edge ion temperature as fixed by the physics and engineering aspects of the tokamak design that will not concern us here \cite{snyder_nf_2011}, the key question is how to maximize the ion temperature gradient. We therefore wish to inquire, experimentally, on what this gradient depends and how. Motivated by the fact (or the conjecture) that the state of the ion-scale microturbulence is largely determined by the {\em local} (to a given flux surface) equilibrium conditions \cite{abel_rpp_2012, ghim_prl_2012, candy_pop_2004, candy_pop_2009, barnes_pop_2010} and in turn acts back to adjust them {\em locally}, we ask what local parameters are most strongly correlated with the corresponding value of $\RLTi$, the inverse radial gradient scale length of the ion temperature ($\LTi^{-1}=\labs\partial\ln T_i/\partial r\rabs$) normalized to the major radius $R$ of the torus.
\newline\indent
One may wonder how universal any such measured dependences are likely to be for situations with different global conditions, e.g., different neutral-beam-injection (NBI) heating powers. It has been recognized for some time that the turbulent heat flux tends to increase very strongly (much faster than linearly) with $\RLTi$ --- a phenomenon known as ``stiff'' transport \cite{kotschenreuther_pop_1995,  mantica_prl_2009, mantica_prl_2011, barnes_prl_2011_107, mantica_ppcf_2011}. If (or when) the transport is indeed stiff, any experimentally measured relationship between $\RLTi$ and other equilibrium parameters should be quite close to some critical manifold in the parameter space separating dominant turbulent transport from an essentially non-turbulent or weakly turbulent state (``the zero-turbulence manifold'' \cite{highcock_prl_2012}). This critical manifold would be independent of the power input and can be thought of as a local parameter dependence of the critical temperature gradient $\RLTic = f(q,\varepsilon,\hat s,\Utor',\RLn,\RLTe,\nu_{ii},T_i/T_e,\beta_i,\dots)$, where $q$ is the safety factor (number of toroidal revolutions per one poloidal revolution of the magnetic field around the torus on a given flux surface), $\varepsilon=r/R$ the inverse aspect ratio ($r$ is the minor radius of the flux surface), $\hat s=\partial\ln q/\partial\ln r$ the magnetic shear, $\Utor'=\partial\Utor/\partial r$ the radial shear of the mean toroidal rotation velocity $\Utor$, $\Ln$ and $\LTe$ the gradient scale lengths of the plasma density and electron temperature, $\nu_{ii}$ the ion collision frequency, $T_i/T_e$ the ion-to-electron temperature ratio, $\beta_i=8\pi nT_i/B^2$ the ion-to-magnetic pressure ratio and ``$\dots$'' stand for everything else (e.g., the many parameters required to fully describe the magnetic configuration).\footnote{We do not suggest {\em causality} between all these parameters and $\RLTi$ --- all local equilibrium characteristics, including $\RLTi$, jointly adjust to form the critical manifold.}  An important caveat is that the ``zero-turbulence'' threshold need not be the same as the threshold for the existence of linearly unstable eigenmodes. Two known examples when this is not the case are the so-called ``Dimits upshift'' of $\RLTi$ above the linear stability threshold \cite{dimits_pop_2000,rogers_prl_2000,mikkelsen_prl_2008} and the case of sufficiently large $\Utor'$ when the system may be linearly stable but strong transient excitations \cite{newton_ppcf_2010,schekochihin_ppcf_2012} lead to sustained subcritical turbulence \cite{barnes_prl_2011,highcock_prl_2010,highcock_pop_2011}. Thus, in general, there is a nonlinear threshold with some definite dependence on local equilibrium parameters. 
\newline\indent
Recent theoretical \cite{newton_ppcf_2010,schekochihin_ppcf_2012} and numerical \cite{highcock_prl_2012} investigations suggest that $q/\varepsilon$ and $\Utor'$ may be the most important such parameters, at least at low $\hat s$. Let us explain why this is. It is well known, both from experimental measurements \cite{ghim_ppcf_2012,field_ppcf_2009} and theory \cite{hinton_pf_1985,cowley_clr_1986,abel_rpp_2012}, that strong (finite-Mach) flows in a tokamak are predominantly toroidal (certainly when plasma is heated by tangential neutral beams, which produce a toroidal torque).\footnote{Any mean poloidal flow exceeding the diamagnetic velocity would be  damped by collisions. \cite{connor_ppcf_1987,catto_pf_1987}.} Therefore, unless the magnetic field is purely toroidal, any radial shear in the toroidal flow results in sheared flow in both the perpendicular ($\Uper' = \lp\Bpol/B\rp \Utor'$, $\Bpol$ is the poloidal field) and parallel ($\Upar' = \lp\Btor/B\rp \Utor'$, $\Btor$ is the toroidal field) directions. While perpendicular flow shear is known (theoretically \cite{barnes_prl_2011, highcock_prl_2010, roach_ppcf_2009, camenen_pop_2009, kinsey_pop_2005, dimits_nf_2001, waltz_pop_1994, dorland_ppcnfr_1994, casson_pop_2009} and experimentally \cite{mantica_prl_2009, mantica_prl_2011, burrell_pop_1999, burrell_pop_1997, mantica_ppcf_2011, schaffner_prl_2012}) to suppress turbulence and the associated transport, parallel flow shear can drive turbulence via the ``parallel-velocity-gradient'' (PVG) instability \cite{catto_pf_1973,newton_ppcf_2010,schekochihin_ppcf_2012}. The average ratio of these two shearing rates on a flux surface, $\Upar'/\Uper' = \Btor/\Bpol$, can be approximated by $q/\varepsilon$ and so the degree to which sheared equilibrium flow suppresses or drives turbulence is expected to depend on this parameter. Indeed, numerical studies of ITG- and PVG-driven turbulence have shown that the critical $\RLTic$ at any given value of $\Utor'$ increases with decreasing $q/\varepsilon$ (at least for low $\hat s$ \cite{highcock_prl_2012}); while at any given $q/\varepsilon$, $\RLTic$ increases with increasing $\Utor'$ provided the latter is not too large \cite{highcock_prl_2012, barnes_prl_2011, highcock_prl_2010, highcock_pop_2011} (as in most real tokamaks). 
\newline\indent
A comprehensive numerical parameter scan of the dependence of $\RLTic$ on all other potentially important local quantities ($\hat s$, $\RLn$, $\RLTe$, $\nu_{ii}$, $T_i/T_e$, $\beta_i$, etc.) is probably unaffordable in the near future, so faster progress can be made experimentally. In this chapter, our first goal is to establish, based on a relatively sizable dataset for MAST, what the most important parameters for the critical manifold are: we will show that, indeed, the local value of $\RLTi$ is most strongly correlated inversely with the local $q/\varepsilon$ and positively with the local rotational shear --- consistently with the result obtained in \cite{highcock_prl_2012}.
\newline\indent
Our second goal is to obtain an experimental signature that the measured $\RLTi$ is determined by --- or, more precisely, correlated with --- the local characteristics of the ion-scale turbulence, directly measured by the 2D beam emission spectroscopy (BES) diagnostic \cite{field_rsi_2012}. We will show that not only does a strong correlation between $\RLTi$ and an estimated turbulent heat flux level exist but its (at the first glance, counterintuitive) inverse nature is consistent with $\RLTi$ staying close to the critical threshold $\RLTic$ and hence with stiff transport.

\section{Equilibrium parameters}
A database was compiled of equilibrium quantities (and turbulence characteristics; see below) from 39 neutral-beam-heated discharges from the 2011 MAST experimental campaign. These discharges had a double-null diverted (DND) magnetic configuration, no pellet injection and no applied resonant magnetic perturbations. Mean electron density $n_e$ and temperature $T_e$ were measured with the Thomson scattering system \cite{scannell_rsi_2010}, mean impurity ion (C$^{6+}$) temperature $T_i$ and the toroidal flow velocity $\Utor$ with the Charge eXchange Recombination Spectroscopy (CXRS) system \cite{conway_rsi_2006} (we assumed that in these discharges the impurity and bulk ions have negligible differences in their temperature and flow velocities \cite{kim_pfb_1990}).  The local magnetic pitch angle ($\Bpol/\Btor$) was measured with the Motional Stark Effect (MSE) diagnostic \cite{debock_rsi_2008}; pressure- and MSE-constrained \texttt{EFIT} equilibria \cite{lao_nf_1985} were used to obtain the field strength $B$. All parameters were determined over $5\:ms$ intervals either by averaging if the diagnostic's temporal resolution was smaller or by interpolation if it was larger than $5\:ms$. Only data points from a limited range of minor radii $0.6<r/a<0.7$ ($r=a$ is the edge of the plasma) were used, in order to minimize any correlations between various quantities due to their profile dependence alone (thus, we did not attempt to {\em prove} locality here; see, however, \cite{ghim_prl_2012}). In total, 988 data points were available. 
\newline\indent
From this information, we constructed 7 local dimensionless parameters, which, motivated by theoretical models or common sense, we deemed {\em a priori} the most important ones (we also give the range of variation of each parameter): $\RLTi \in [0.08,20.3]$, $q/\varepsilon\in[4.0,16.3]$, $\hat s \in [1.2,6.0]$, $\gEbar \equiv \Uper'\tst=\pi r\Utor'/\vti \in [0.005,2.5]$, $\RLn \in [0.04,13.8]$, $\RLTe \in [1.43,22.7]$, $\nust\equiv \nu_{ii}\tst\in[0.003,0.12]$, $T_i/T_e \in [0.5,1.7]$. The ion collision rate $\nu_{ii}$ and the perpendicular velocity shear $\Uper'$ (which is used instead of $\Utor'$) were normalized, as in \cite{ghim_prl_2012}, to the ion parallel streaming time $\tst=\Lambda/\vti$, where $\vti=\sqrt{2T_i/m_i}$ and $\Lambda=\pi r B/\Bpol$ is the connection length (the approximate distance along the field line from the outboard to the inboard side of the torus, expected to determine the parallel correlation scale of the turbulence \cite{barnes_prl_2011_107}; if the flux surfaces had been circular, $\Lambda\approx \pi qR$). The local magnetic configuration is represented by only two parameters: $q/\varepsilon$ and $\hat s$. The choice of $q/\varepsilon$ was motivated by the physical considerations outlined in \secref{sec:rlti_intro} (since $\varepsilon$ varied very little in our database, we do not claim to distinguish any individual correlations of $\RLTi$ with $q$ and $\varepsilon$). It is left for further study whether other properties of the flux surfaces matter (e.g., Shafranov shift, triangularity, elongation, etc.; some of these may, in fact, affect the stiff-transport threshold \cite{mikkelsen_prl_2008, beer_pop_1997}). We have not included $n$, $T_i$, $T_e$, which are not normalizable by any natural local quantities; note that $\RLTi$ usually has a large but trivial correlation with $T_i$: larger temperature gradients lead to larger temperatures in the core. We also have not included $\beta_i=8\pi nT_i/B^2$ because, in the absence of large variation of $B$ in our dataset, $\beta_i$ is simply the normalized ion pressure and, similarly to $T_i$, has a large positive correlation with $\RLTi$ (it remains to be investigated whether the larger level of magnetic fluctuations at larger $\beta_i$ is large enough to have a nontrivial effect on turbulent transport \cite{guttenfelder_pop_2012, doerk_pop_2012, rechester_prl_1978, pueschel_pop_2010, nevins_prl_2011, hatch_prl_2012, abel_njp_2012}). 

\subsection{Correlation analysis}
We perform a Canonical Correlation Analysis (CCA) \cite{hotelling_bio_1936} with $\RLTi$ treated as the dependent variable and the other 7 local parameters itemized above as independent ones. This amounts to finding the maximum correlations between  $\ln(\RLTi)$ and linear combinations of logarithms of 1, 2, 3, \dots, or 7 
other parameters, leading to an effective statistical dependence 
\begin{equation}
\label{eq:cca_fit}
\frac{R}{\LTi} = \lp\frac{q}{\varepsilon}\rp^{\alpha_1}\gEbar^{\alpha_2}\nust^{\alpha_3} 
 \lp\frac{R}{\LTe}\rp^{\alpha_4}\hat s^{\alpha_5}\lp\frac{R}{\Ln}\rp^{\alpha_6}\lp\frac{T_i}{T_e}\rp^{\alpha_7}.
\end{equation} 
This is of course not valid if the dependence of $\RLTi$ on any of the parameters is non-monotonic. A non-monotonic dependence on $\gEbar$ is, in fact, expected, with $\RLTi$ first increasing, then decreasing at larger values of $\gEbar$ due to increased transport from the PVG-driven turbulence \cite{highcock_prl_2012, barnes_prl_2011, highcock_prl_2010, highcock_pop_2011}. However, the range of values of $\gEbar$ in our database does not extend to sufficiently high values for such a dependence to be observed (see \figref{q_eps_tau_st_tau_sh_R_LTi}). 
\newline\indent
\begin{table}[t]\caption[Canonical correlation analysis with equilibrium parameters]{Results of CCA performed assuming \eqref{eq:cca_fit}. Wherever $0$ appears, that means that the CCA was performed without including the corresponding parameter.} 
\label{table:cca_coeff}
\begin{tabular}{c | c c c c c c c } 
Canonical  & $q/\varepsilon$ & $\gEbar$ & $\nust$ & $\RLTe$ & $\hat s$ & $\RLn$ & $T_i/T_e$   \\
correlation & $\alpha_1$ & $\alpha_2 $ & $\alpha_3 $ & $\alpha_4 $ & $\alpha_5 $ & $\alpha_6$ & $\alpha_7$ \\
\hline
2.7\% & 0 & 0 & 0 & 0 & 0 & 0 &$-3.1$ \\
13.5\% & 0 & 0 & 0 & 0 & 0 & $-0.83$ & 0 \\
15.4\% & 0 & 0 & 0 & 0 & $-3.41$ & 0 & 0 \\
16.8\% & 0 & 0 & 0 & $-2.3$ & 0 & 0 & 0 \\
36\% & 0 & 0 & $-0.93$ & 0 & 0 & 0 & 0 \\
46\% & 0 & $0.94$ & 0 & 0 & 0 & 0 & 0 \\
61\% & $-1.69$ & 0 & 0 & 0 & 0 & 0 & 0 \\
\hline
62\% & $-1.50$ & 0 & $-0.21$ & 0 & 0 & 0 & 0 \\
66\% & $-1.30$ & $0.40$ & 0 & 0 & 0 & 0 & 0 \\
\hline
67\% & $-1.19$ & $0.38$ & $-0.15$ & 0 & 0 & 0 & 0 \\
\hline
69\% & $-1.12$ & $0.37$ & $-0.19$ & $-0.27$ & $0.21$ & $-0.09$ & $-0.51$ \\
\hline
\end{tabular}
\end{table}
The results are shown in \tableref{table:cca_coeff}, where the values of the canonical correlation (i.e., the correlation coefficient between the logarithms of $\RLTi$ and the right-hand of \eqref{eq:cca_fit}) are given together with the corresponding exponents $\alpha_1,\dots,\alpha_7$. We start by calculating the individual correlations of $\RLTi$ with each of the 7 parameters and then include pairs, triplets, etc., only if the correlation improves. We see that the strongest individual correlation of $\RLTi$ are with $q/\varepsilon$ (61\%) and $\gEbar$ (46\%). The overall fit is measurably improved (66\%) if both are included. Including further parameters does not make a significant difference; the third strongest (although not very strong) dependence is on $\nust$. 
\myfig[4.5in]{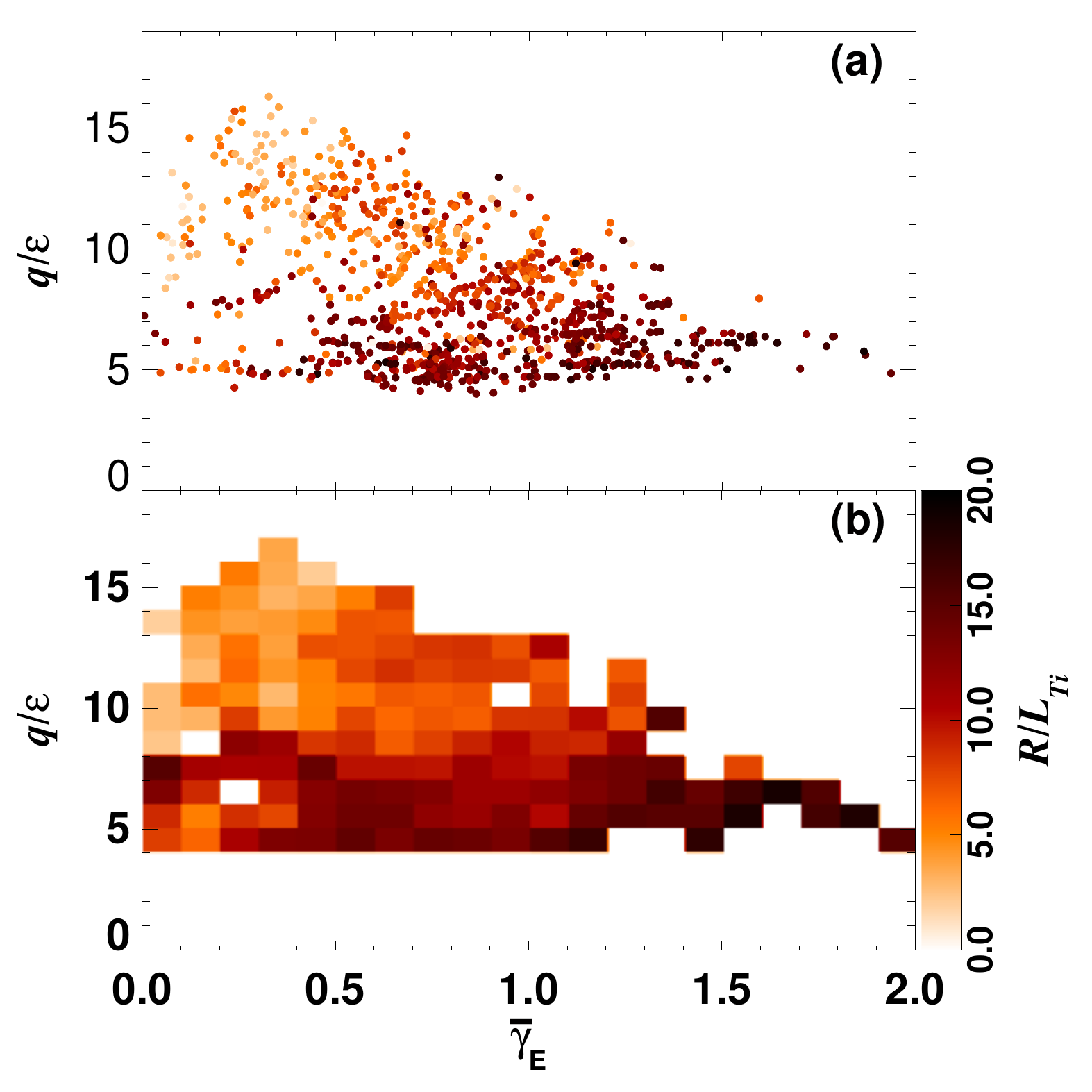}{Dependence of $\RLTi$ on $q/\varepsilon$ and $\gEbar$}{The dependence of $\RLTi$ (color) on $q/\varepsilon$ and $\gEbar = \pi r\Utor'/\vti$, showing (a) the individual data points and (b) mean values of $\RLTi$ within rectangular bins.}
The dependence of $\RLTi$ on\ $q/\varepsilon$ and $\gEbar$ is shown in \figref{q_eps_tau_st_tau_sh_R_LTi}, which generally increases with decreasing $q/\varepsilon$ and increasing $\gEbar$.\footnote{The broad scatter of data points in \figref{q_eps_tau_st_tau_sh_R_LTi}(a) suggests that the correlation between $q/\varepsilon$ and $\gEbar$ is weak; the lack of higher values of $\gEbar$ at large $q/\varepsilon$ is due to the fact that the flow shear is weak at earlier times in the discharges, when the central value of $q$ is high.} This is broadly consistent with the expectations based on intuitive physical reasoning (explained in \secref{sec:rlti_intro}) and on the numerical study of \cite{highcock_prl_2012}. We note that $\hat s$ was set to zero in the numerical study reported in \cite{highcock_prl_2012} which is different from our experimental cases. However, the values of $\hat s$ will not change the quantitative trend of observed $\RLTi$ as a function of $q/\varepsilon$ and $\gEbar$ as reported in \cite{barnes_prl_2011} with $\hat s=0.8$.
\newline\indent
The conclusion is that, at least on a very rough qualitative level, it is sensible to consider $\RLTi$ to be a function primarily of $q/\varepsilon$ and $\gEbar$. We note that equilibrium database shown in \figref{q_eps_tau_st_tau_sh_R_LTi} contains time periods when MHD modes are active unlike the BES database (see \figref{qi_plots}). As a result, it is possible that our database may include some outliers due to a fast change of heat transport during the MHD activities. Since the $q$ profile tends to change more slowly in tokamaks than other equilibrium profiles,\footnote{It can be proven that the functional dependence $q(\psi)$, where $\psi$ is the flux-surface label, only changes on the resistive timescale of the mean magnetic field \cite{abel_njp_2012}.} it may be useful to think of a critical curve $\RLTic(\gEbar)$ \cite{parra_prl_2011} parametrized by $q/\varepsilon$ \cite{highcock_prl_2012}, the latter quantity containing the essential information about the nature of the magnetic cage confining the plasma.

\subsection{Collisionality dependence}
Even though the $\nust$ dependence of $\RLTi$ is not as strong as $q/\varepsilon$ and $\gEbar$, some discernible inverse correlation between $\RLTi$ and $\nust$ might be expected because zonal flows, believed to suppress turbulence \cite{diamond_ppcfreview_2005, rogers_prl_2000}, should be more strongly damped at higher ion collisionality \cite{hinton_ppcf_1999, xiao_pop_2007, ricci_prl_2006, lin_prl_1999}. To isolate this dependence, we selected data points for approximately fixed $\gEbar\in[0.7,0.8]$ and $q/\varepsilon\in[5,6]$  (the largest number of data points could be found within these narrow ranges and no measurable correlation between $\RLTi$ and $q/\varepsilon$ or $\gEbar$ was present). The resulting \figref{R_LTi_others} confirms a degree of inverse correlation between $\RLTi$ and $\nust$. 
\myfig[4.5in]{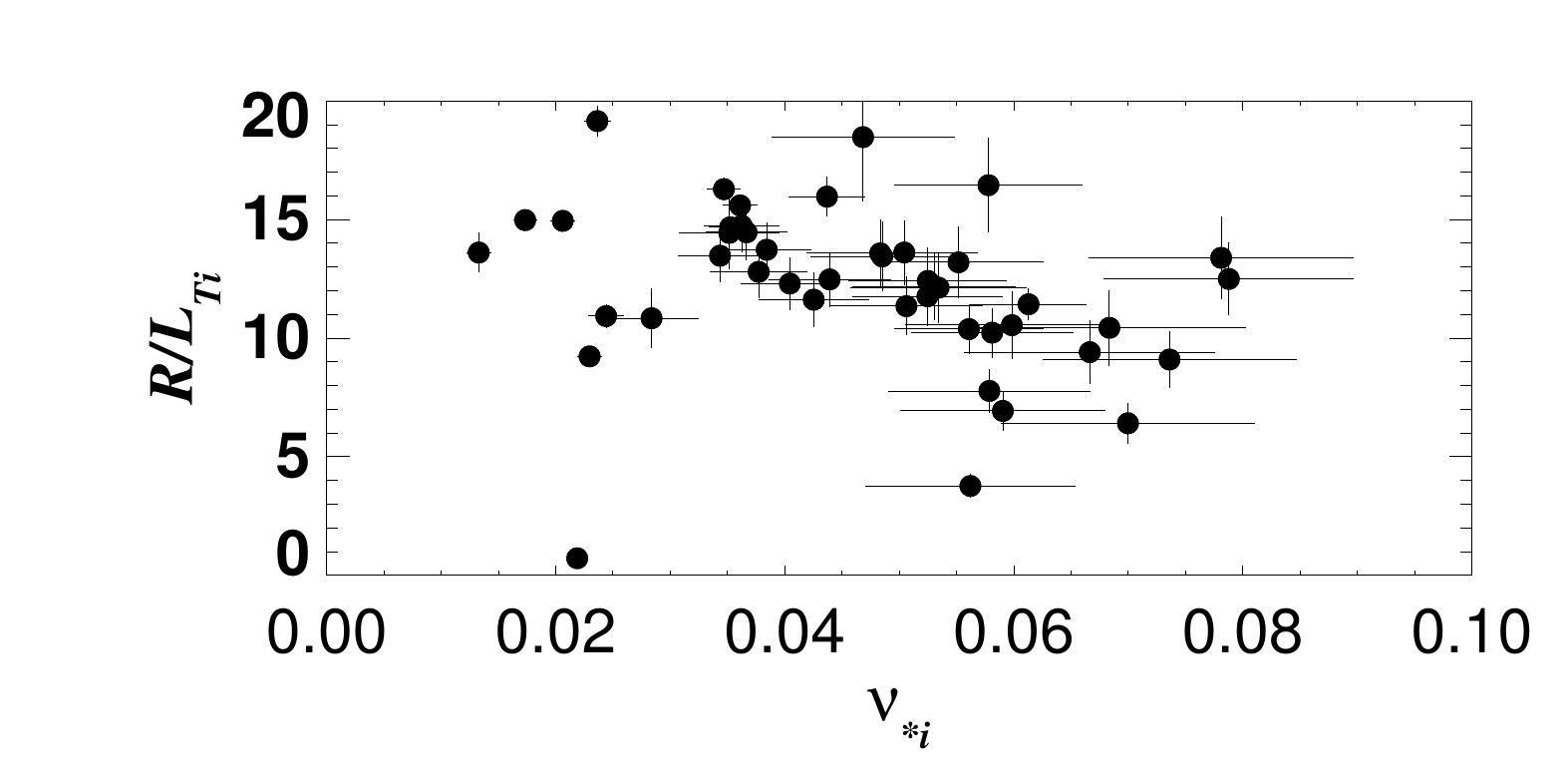}{Dependence of $\RLTi$ on $\nust$}{$\RLTi$ vs.~$\nust$ for fixed $\gEbar\in[0.7,0.8]$ and $q/\varepsilon\in[5,6]$.}

\section{Turbulent heat flux}
The turbulent ion heat flux through a given flux surface is (very approximately!) $Q_i\sim n T_i\chi_i/\LTi$, where the effective turbulent diffusivity is $\chi_i\sim \delta u^2\tau_c$, $\delta u\sim (c/B)\varphi/\ly$ is the (radial) fluctuating $\mathbf{E}\times\mathbf{B}$ velocity, $\tau_c$ its correlation time, $\ly$ its poloidal correlation scale and $\varphi$ the fluctuating electrostatic potential. The latter can be estimated from density fluctuations using the approximation of Boltzmann electrons: $e\varphi/T_e\approx\dn/n$ ($e$ is the proton charge, $n$ and $\dn$ the mean and fluctuating density, respectively). Both theory of ITG turbulence \cite{barnes_prl_2011_107} and the BES measurements in MAST \cite{ghim_prl_2012} suggest that $\tau_c\sim\tau_{*i}=\ly\LTi/\vti\rhoi$ (the drift time; $\rhoi$ is the ion Larmor radius). Collecting all this together, we estimate the gyro-Bohm-normalized turbulent ion heat flux:
\begin{equation}
\frac{Q_{i,\mathrm turb}}{Q_{\mathrm gB}}\sim 
\frac{\rhoi}{\ly}\lp\frac{R}{\rhoi}\rp^2\lp\frac{T_e}{T_i}\frac{\dn}{n}\rp^2\equiv\Qturb,
\label{eq:Q}
\end{equation}
where $Q_{\mathrm gB} = n T_i \vti \rhoi^2 / R^2$.
\newline\indent
Since ion-scale density fluctuations in MAST can be measured directly by the BES system, $\Qturb$ can be obtained independently of any transport reconstruction models such as TRANSP \cite{hawryluk_TRANSP_1980}. The method of determining $\dn/n$ and $\ly$ using the BES system on MAST (8 radial $\times$ 4 vertical channels with spatial resolution of $\approx 2$~cm \cite{field_rsi_2012}) is explained in detail in \cite{ghim_prl_2012}. This is done from the covariance and correlation functions of the photon intensity fluctuations, averaged over the same $5\:ms$ intervals for the same 39 discharges as the equilibrium quantities studied above, although not in all intervals there was good BES data. Restricted to the radial range $0.6<r/a<0.7$, the number of available data points for $\dn/n$ and $\ly$ was 102. 

\subsection{Inverse correlation between $\RLTi$ and $\Qturb$}
\myfig[4.0in]{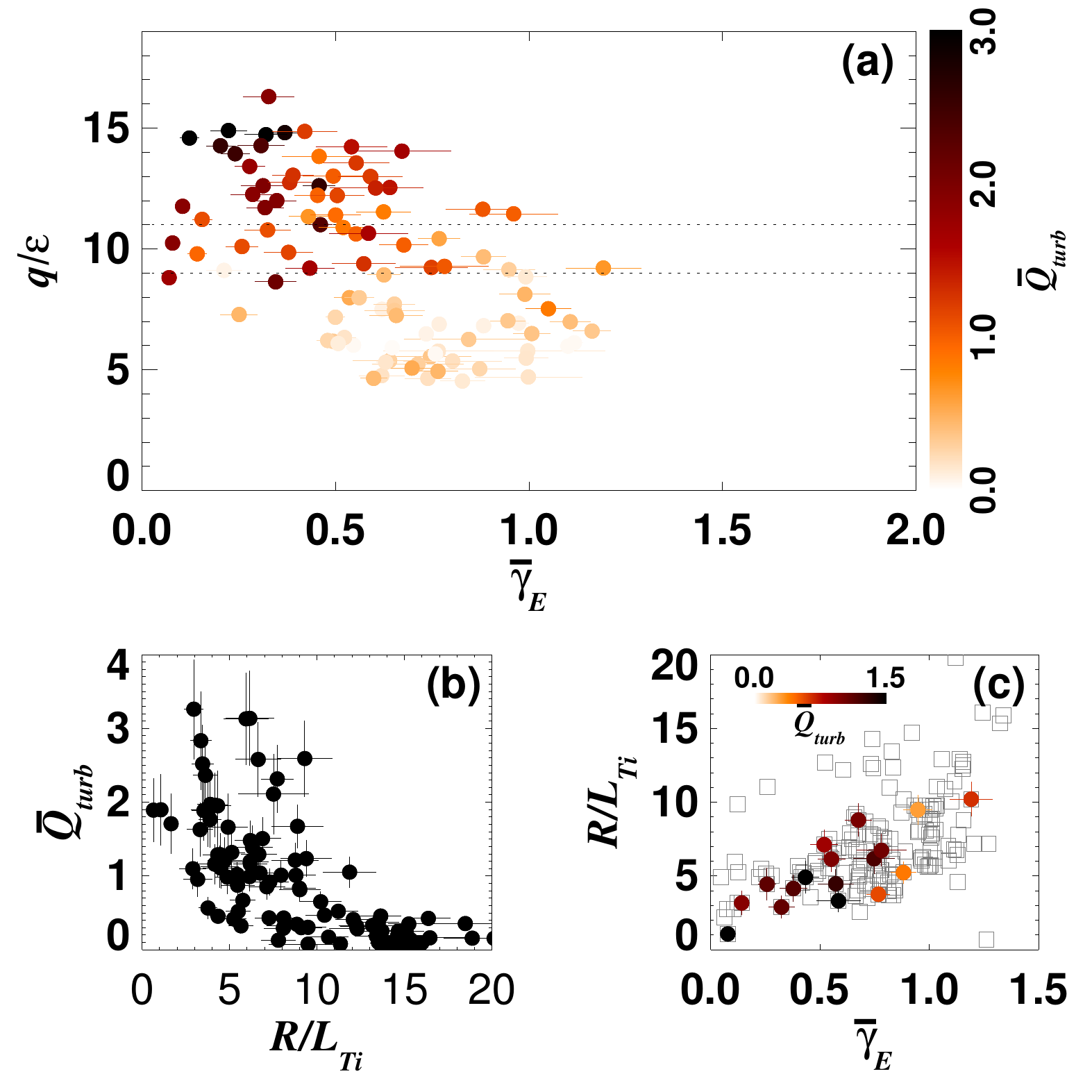}{Dependence of $\Qturb$ on $q/\varepsilon$, $\gEbar$ and $\RLTi$}{(a) $\Qturb$ (color) calculated from BES data according to \eqref{eq:Q} vs.\ $q/\varepsilon$ and $\gEbar$ (cf.\ \figref{q_eps_tau_st_tau_sh_R_LTi}); (b) $\Qturb$ vs.\ $\RLTi$(cf. \figref{fig_highcock_Q_vs_RLTi}); (c) $\Qturb$ (color) vs.\ $\RLTi$ and $\gEbar$ for a fixed range of $q/\varepsilon\in[9,11]$, indicated by the dotted horizontal lines in (a); open squares are data from \figref{q_eps_tau_st_tau_sh_R_LTi}(a) for which BES measurements were not available (cf. \figref{fig_parra_RLTi_vs_gE}).} 
\mydoublesidefig{2.5in}{2.5in}{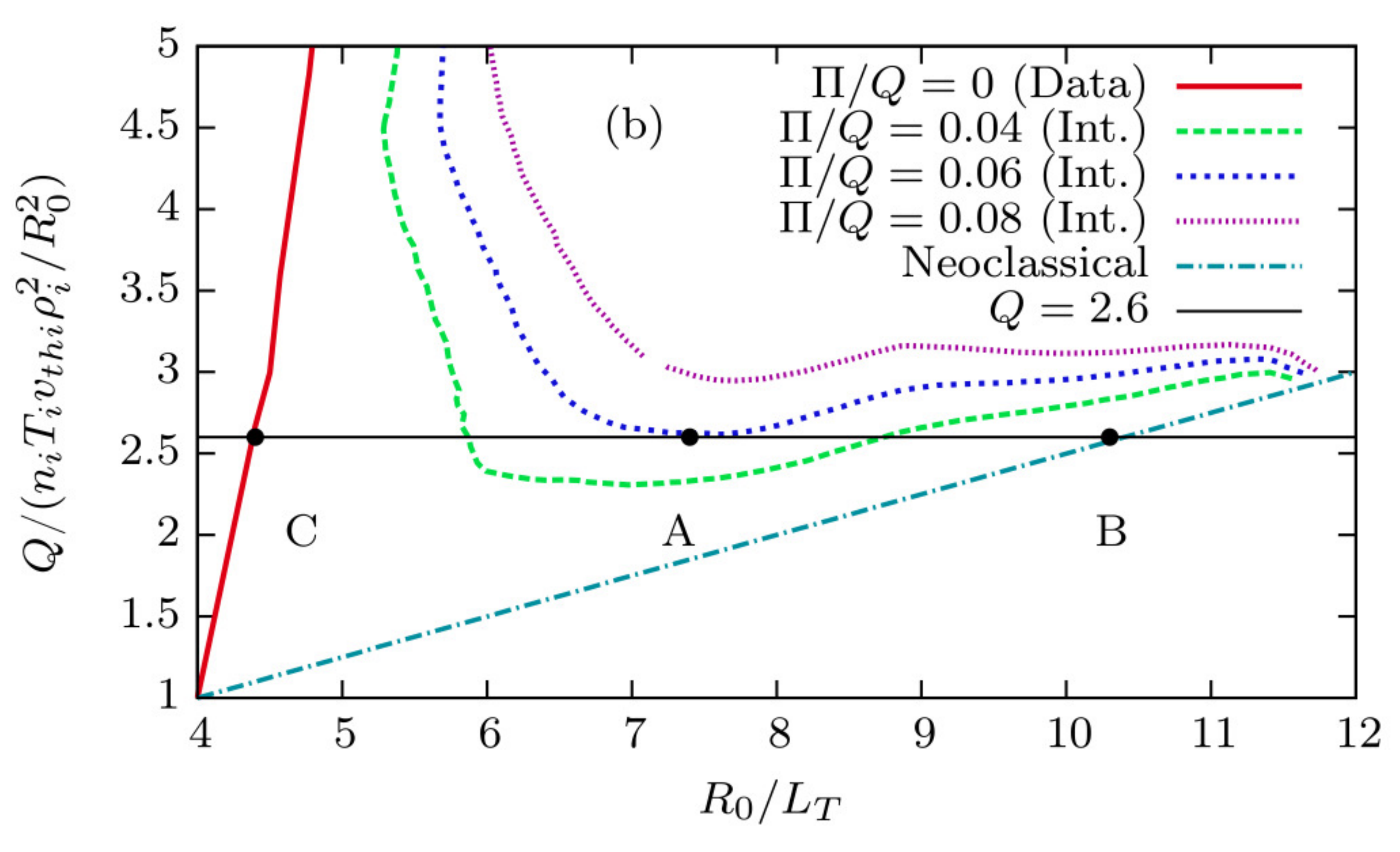}{Numerically generated $Q$ vs. $\RLTi$}{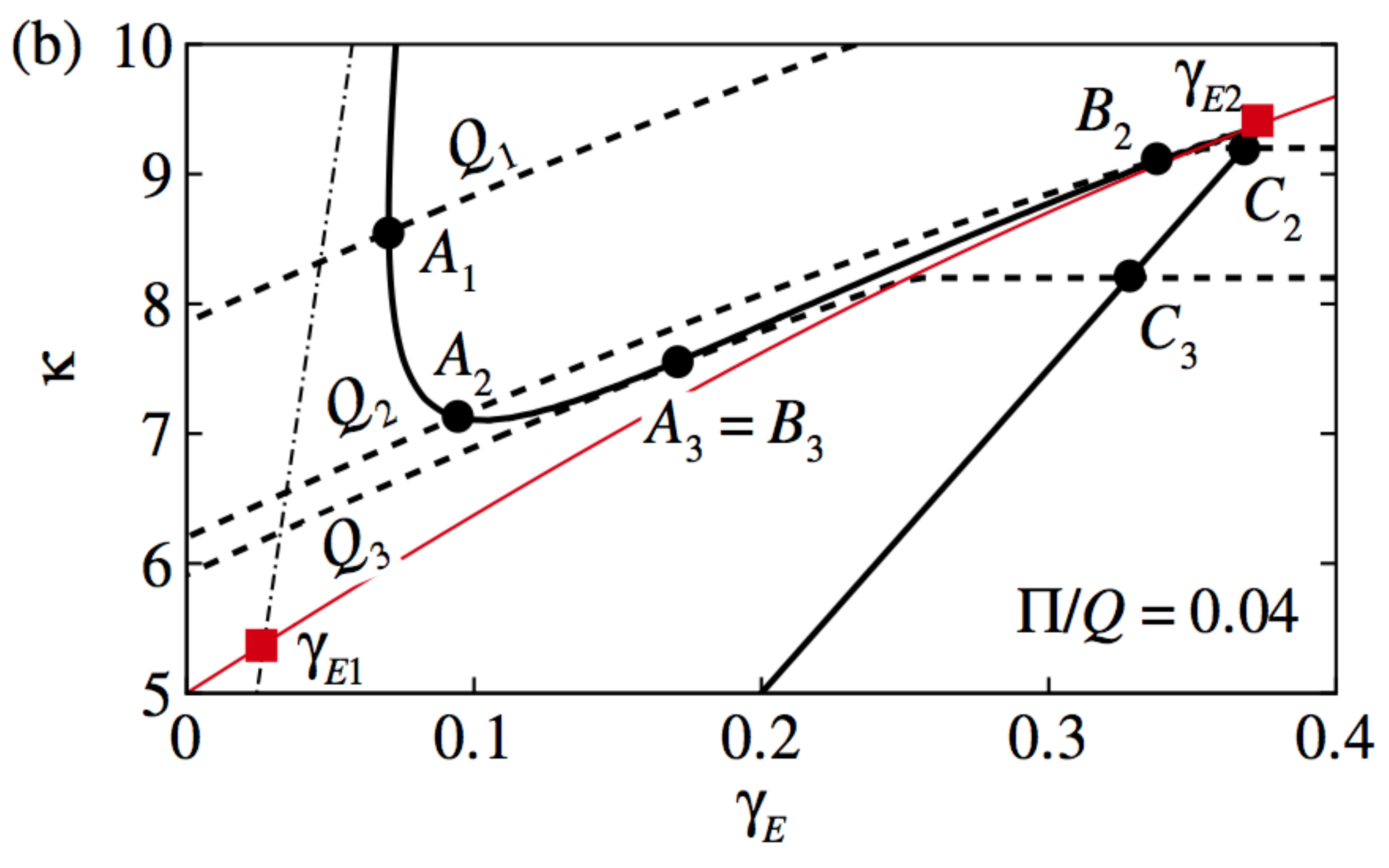}{Analytically calculated $\RLTi$ vs. $\gE$}{Numerically generated total heat flux $Q$ vs. $\RLTi$ showing that there exist regions where a smaller $Q$ corresponds to a larger $\RLTi$. Figure is taken from Fig.~4(b) of \cite{highcock_prl_2010}}{Analytically calculated $\kappa=\RLTi$ vs. $\gE$ with $Q_1 > Q_2 > Q_3$. Point $A_3$ has a larger $\RLTi$ than $A_2$ despite $Q_3 < Q_2$. The red line is the critical temperature gradient $\RLTic$ below which there is no turbulence. Figure is taken from Fig.~3(b) of \cite{parra_prl_2011}}
It is shown in \figref{qi_plots}(a) ($\Qturb$ vs.\ $q/\varepsilon$ and $\gEbar$; cf.\ \figref{q_eps_tau_st_tau_sh_R_LTi}) and \figref{qi_plots}(b) ($\Qturb$ vs.\ $\RLTi$) that smaller $\Qturb$ is observed where $\RLTi$ is large and vice versa. This is perhaps counterintuitive as one might expect that the larger $\RLTi$, the more turbulent the plasma and so the larger the turbulent heat flux. This would indeed have been the case had $\RLTi$ been externally fixed, as in local flux-tube simulations, where $\Qturb$ does increase with $\RLTi$ \cite{dimits_pop_2000,barnes_prl_2011_107}. In contrast, in the real plasma, a certain amount of power flows through a flux surface and, if transport is stiff, the temperature gradient (along with other equilibrium quantities) adjust accordingly, to stay close to the critical gradient defined by the manifold $\RLTic(\gEbar,q/\varepsilon)$. Indeed, in plasmas with both power and momentum injection, there is a regime with $\RLTi$ close to $\RLTic$ where the turbulent and the neo-classical (collisional) transport are comparable and in which a larger heat flux results in lower $\RLTi$. For a detailed explanation, we refer the reader to \cite{parra_prl_2011} (see also Fig.~4(b) of \cite{highcock_prl_2010} (or \figref{fig_highcock_Q_vs_RLTi}), which should be compared to our \figref{qi_plots}(b)). In brief, in the neoclassical regime, the momentum transport is much less efficient than the heat transport, while in the turbulent regime, they are comparable (the turbulent Prandtl number is order unity \cite{barnes_prl_2011, highcock_pop_2011, casson_pop_2009, meyer_nf_2009} while the neoclassical one is small \cite{hinton_pf_1985}), so, as a larger heat flux takes the system (slightly) farther from the marginal state, this leads to much more efficient momentum transport, hence smaller velocity shear $\gEbar$, hence a regime with less suppression of turbulence and smaller $\RLTic$ (see \figref{qi_plots}(c), where the correspondence between larger $\Qturb$, lower $\RLTi$ and lower $\gEbar$ is shown at approximately fixed $q/\varepsilon\in [9,11]$; cf.\ Fig.~3(b) of \cite{parra_prl_2011} (or \figref{fig_parra_RLTi_vs_gE})). We stress that all of this happens quite close to marginality and so our experimental observation of an inverse correlation between $\RLTi$ and $\Qturb$ provides circumstantial evidence that $\RLTi$ in MAST is indeed close to its critical value.\footnote{This situation is distinct from the experiments on transport stiffness on JET \cite{mantica_prl_2009, mantica_prl_2011, mantica_ppcf_2011} in that they provided vigorous extra heating power using localized ion-cyclotron-resonance heating (ICRH) to depart far from marginality.}

\section{Conclusion}
We have found that the normalized inverse ion-temperature-gradient scale length $\RLTi$ has its strongest local correlation with $q/\varepsilon$ and the shear in the equilibrium toroidal flow: $\RLTi$ increases with increasing shear, which is a well known effect, and with decreasing $q/\varepsilon$, which corresponds to an increasing ratio of the perpendicular to the parallel shearing rates. We note that a similar dependence of $\RLTi \lp q/\varepsilon, \gEbar\rp$ is also observed in JET \cite{fern_unpub_2012}, suggesting that the inverse correlation between $\RLTi$ and $q/\varepsilon$ is perhaps ubiquitous, as would be the case if $\RLTi$ were generally fixed at some locally determined critical value \cite{highcock_prl_2012}. Furthermore, we have found an {\em inverse} correlation between $\RLTi$ and the gyro-Bohm-normalized turbulent heat flux (estimated via direct measurements of density fluctuations) and argued that this is consistent with $\RLTi$ always remaining close to a critical manifold $\RLTic\lp q/\varepsilon, \gEbar\rp$ separating the turbulent and non-turbulent regimes \cite{highcock_prl_2010, highcock_pop_2011, parra_prl_2011} (stiff transport). It is thus plausible that we have essentially produced this critical manifold for the MAST discharges we investigated. Practically, our results suggest that $\RLTi$ can be increased by lowering the $q/\varepsilon$, which is relatively easier and less expensive than increasing the shearing rate in tokamak operations.\footnote{With tangential NBI heating, it is difficult to increase the toroidal Mach number --- and hence the equilibrium flow shear --- because of the fixed ratio of injected torque to power at a fixed injection energy.}

\mypart{Conclusions}
\chapter{Conclusions}\label{ch:conclusions}
\begin{flushright}
The poet should say what sorts of thing might happen, that is, the things possible according to likelihood or necessity.\\
-- Aristotles, Poetics Ch. 9\\
\end{flushright}
In this work, we have presented experimental results on the characteristics of plasma turbulence measured by the 2D beam emission spectroscopy (BES) system on the MAST spherical tokamak.
\newline\indent
First, we have described the principle of the 2D BES system used to measure ion-Larmor scale fluctuating densities up to a few $100$ kHz down to a few $0.1\:\%$ level.  We find from the generated point spread functions (PSFs) that the radial spatial smearing does not depend significantly on the observation locations; whereas the poloidal spatial smearing becomes greater as the observation locations are moved toward the edge of MAST due to increasing magnetic field pitch angles.  The PSFs at different major radii vary which complicates the deconvolution of the measured signal.  However, as the measured poloidal correlation lengths are found to be larger than the poloidal widths of the PSFs, the deconvolution of the signal would only slightly correct the measured correlation lengths. BES data with shorter radial correlation lengths than the radial widths of the PSFs are not used so that our results are not biased even if the deconvolution is not performed.  Then, detailed procedures of obtaining statistical properties of turbulence such as spatial and temporal correlation lengths and generating synthetic 2D BES data are discussed.
\newline\indent
The first physics result we have obtained from the measurements is on the subject of interpretation of mean motion of fluctuating density patterns.  Using the cross-correlation time delay method whose statistical reliability is extensively investigated with the synthetic 2D BES data, the apparent poloidal motion of fluctuating density patterns in the lab frame is found to arise because the longest correlation lengths of the patterns are not parallel to the dominant toroidal flow of the bulk plasma. This projection effect holds for the investigated L-, H-mode and internal transport barrier discharges.  An interesting exception to this rule is found for the investigated high-poloidal-beta discharge, where a large magnetic island exists.  Whether this exception is related to presence of the island is not known, nor do we successfully identify any feasible physical explanations for this phenomenon.  Further investigation is left for future work.
\newline\indent
With the understanding of mean motion of turbulence, spatial and temporal characteristics of it are compared with the local equilibrium quantities.  Such comparisons are performed in terms of various time scales each of which is related to a distinct physical effect. The measured turbulence is shown to be consistent with the idea of turbulence being critically balanced, that is, the correlation time of the turbulence, the drift time associated with the background ion-temperature or density gradients, the ion streaming time along the magnetic field line and the magnetic drift time are consistently comparable.  The balance between the drift time associated with the background gradient and the ion parallel streaming time provides how the perpendicular (with respect to the magnetic field within a flux surface) correlation length of turbulence scales with the ratio of the connection length to the gradient scale length; while the balance between the magnetic drift time and the ion parallel streaming time gives the scaling of the radial correlation length as a function of the ratio of the connection length to the size of a tokamak.  Furthermore, estimation of non-zonal component of nonlinear time of the turbulence seems to indicate that the measured turbulence is dominantly decorrelated by the scattering of the drift waves by the zonal flows rather than the nonlinear interactions between the drift-wave-like fluctuations themselves.  It is found that the ratio of zonal to non-zonal components of the turbulence scales inversely with the ion collisionality.
\newline\indent
Finally, we have experimentally shown that the normalised inverse ion-temperature-gradient scale length $\RLTi$ has the strongest local correlation with $q/\varepsilon$ (safety factor/inverse aspect ratio) and the shear in the equilibrium toroidal flow.  Furthermore, we argued that the counterintuitive observation of the inverse correlation between the gyro-Bohm-normalised turbulent ion heat flux and the $\RLTi$ is evidence that observed turbulence level is quite close to its critical value indicating that we have essentially produced the critical manifold in the local equilibrium parameter space for the MAST discharges. In practice, we can use this extra experimental knob of $q/\varepsilon$ to control $\RLTi$ which is much cheaper and easier than the `shear flow' knob. Thus, the $q/\varepsilon$ may well become a key parameter to ignite the plasmas in the ITER.

\appendix
\appendixpage

\chapter{Drift waves} \label{ch:drift_wave}
In this chapter, we consider three types of drift waves associated with gradients of density, ion temperature and parallel velocity.  First, we provide a pictorial description of a density-gradient driven drift wave for easier understanding of physical mechanism of the wave following ref. \cite{horton_rmp_1999}.  Then, we use fluid equations, being more quantitative, to describe ion-temperature-gradient (ITG) and parallel-velocity-gradient (PVG) driven drift waves following Cowley's lecture.\footnote{The lecture was given during the Culham Summer School in 2009.}
\newline\indent
In all cases, we assume that perturbations are electrostatic for the simplicity, and we use a cartesian coordinate system where $x$ is the inhomogeneous direction, i.e., equilibrium quantities such as density, temperatures and plasma flows are varying in the $x-$direction, the $y$-direction is taken to be periodic, and the $z$-direction is parallel to the homogeneous straight magnetic field lines, $\vct{B}=B\vct{\hat z}$.  Note that the assumption of $\vct{B}$-field not varying in the $x-$ and $y-$directions is valid as long as spatial variation of $\vct{B}$ in these two directions is much slower than that of perturbation. 

\section{Drift waves associated with density gradients}\label{sec:dens_driven_dr_wave}
Let us impose a small electrostatic potential perturbation $\varphi$ at $y=0$ as shown in \figref{dr_wave_inphase} with a finite spatial structure in all three spatial directions.\footnote{Finite parallel structure is critical for a drift wave to exist \cite{horton_rmp_1999, goldston_taylor_2000}}  Furthermore, consider a case where the dynamics of the perturbation is much slower than electron parallel streaming time with its thermal velocity, which is proven to be true later.
\newline\indent
In the case that electrons satisfy Boltzmann response, i.e., $\delta n/n=e\varphi/T_e$ where $n \lp\delta n\rp$ is the mean (perturbed) density, $e$ the proton charge and $T_e$ the electron temperature, there is neither net transport of particles in the $x-$direction nor unstable modes, i.e., the mode is purely oscillatory \cite{goldston_taylor_2000}.  This is illustrated with \figref{dr_wave_inphase}.  The potential structure with a local maximum at $y=0$, i.e., $\varphi_1>\varphi_2>0$, creates a radial electric field which produces $\vct{E}\times\vct{B}$ drift velocity $\vct{v_E}$ in the clockwise direction.  The particle flux towards positive $x-$direction (at the top yellow box \circled{2}) is $\Gamma_>=\lsb\lp n_>+\delta n_2\rp - \lp n_<+\delta n_2\rp\rsb v_E$; while towards negative $x-$direction (at the bottom yellow box \circled{1}) is $\Gamma_<=\lsb-\lp n_>+\delta n_2\rp + \lp n_<+\delta n_2\rp\rsb v_E$.  Note that $v_E$ at these two locations is purely in the $x-$direction.  Indeed, we have $\Gamma_>+\Gamma_<=0$.
\mydoublesidefig{2.5in}{2.5in}{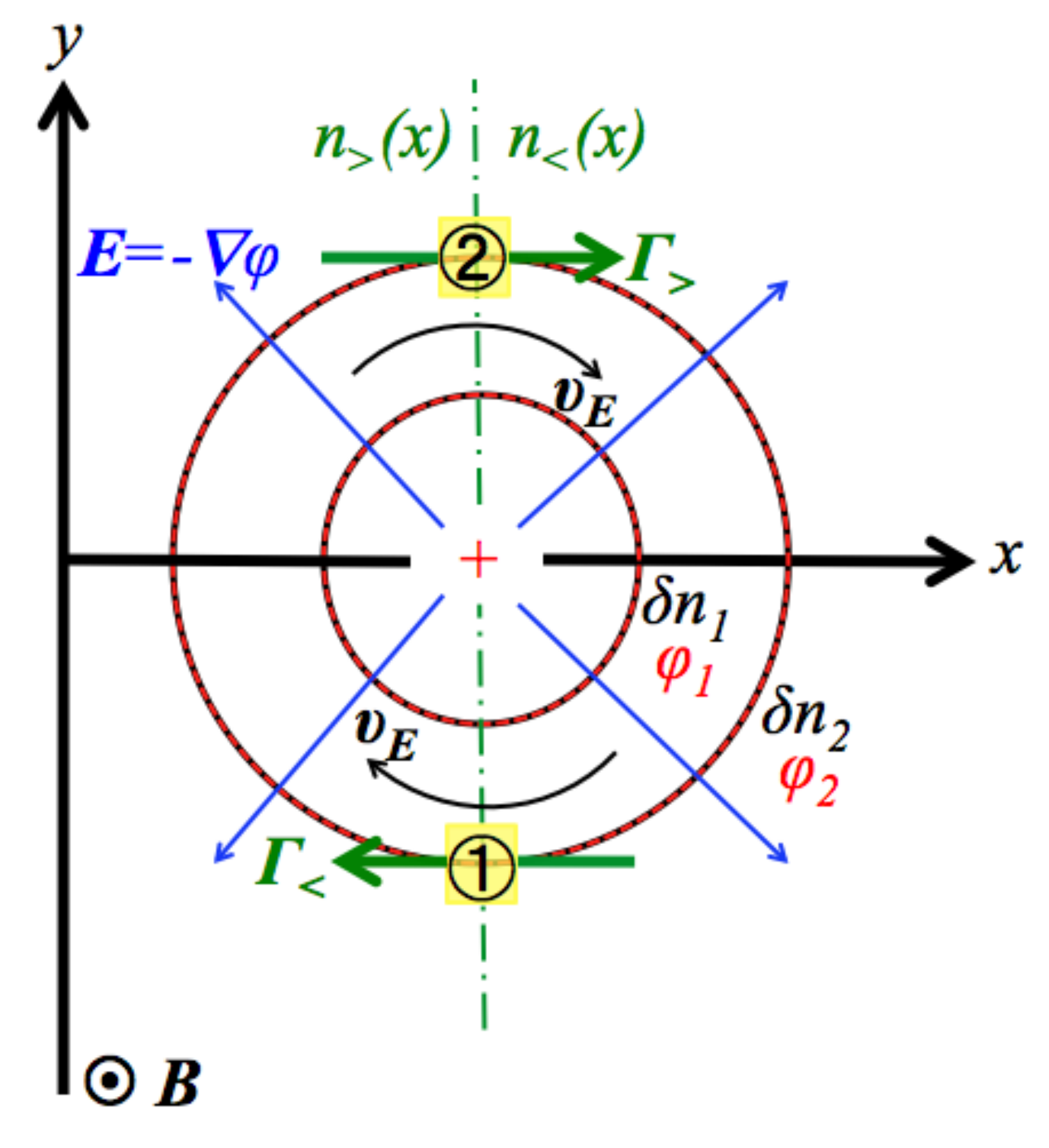}{Schematic of a drift wave (in-phase)}{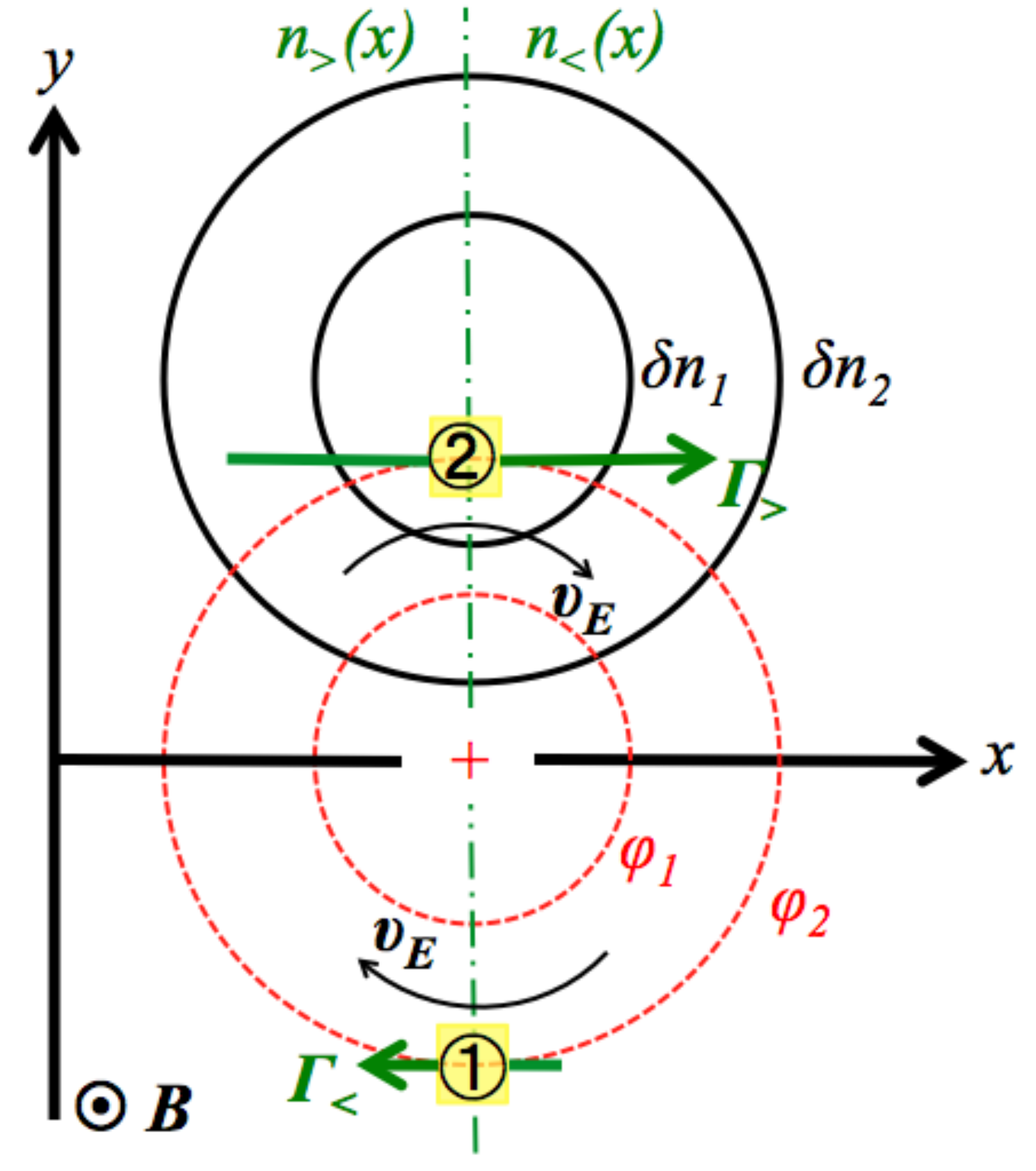}{Schematic of a drift wave (out-phase)}{A schematic of density-gradient driven drift waves when perturbations in density (black solid) and potential (red dashed) are in phase.  There is no net outward particle flux in the $x-$direction while the structures are propagating in the $+y-$direction (electron diamagnetic direction).  The mode is purely oscillatory, i.e., not unstable.}{A schematic of density-gradient driven drift waves when perturbations in density (black solid) and potential (red dashed) are not in phase.  If the density perturbation leads (lags) the potential, then there is net particle flux down (up) the density gradient, and the growth rate of the mode is positive (negative).}
\newline\indent
On the other hand, the location of the maximum positive potential is propagating in the $y-$direction. Consider a point at $y=\delta y$ and constant $x$ denoted as \circled{2} in \figref{dr_wave_inphase}.  The continuity equation $\partial n/\partial t + \grad\cdot\lp n\vct{v} \rp=0$ at this location can be written as:
\begin{equation}\label{eq:dr_cont}
\frac{\partial n}{\partial t}=-v_x\frac{\partial n}{\partial x}=-\frac{c}{\delta y}\frac{\varphi}{B}\frac{\partial n}{\partial x},
\end{equation}   
where we have replaced $v_x$ with $v_E$ at $y=\delta y$ in the last step.  The original perturbed maximum density at $y=0$ is $\delta n = ne\varphi/T_e$, and we can calculate how long it takes to reach this density at the location \circled{2} using \eqref{eq:dr_cont}:
\begin{eqnarray}\label{eq:vde_deriv}
\delta n & = & -\frac{c}{\delta y}\frac{\varphi}{B}\frac{\partial n}{\partial x}\delta t = n\frac{e\varphi}{T_e},  \nonumber \\
\frac{\delta y}{\delta t} & = & -\frac{cT_e}{eB}\frac{\partial}{\partial x}\ln x=v_{de}, 
\end{eqnarray}
which shows that the potential structure propagates in $+y-$direction (electron diamagnetic direction) with the speed of $v_{de}$. It was mentioned that the drift wave dynamics are much slower compared to the electron parallel streaming time.  This can be validated using \eqref{eq:vde_deriv}: $v_{de}=\vte\rhoe/\Ln\ll\vte$ as $\rhoe\ll\Ln$ where $\Ln^{-1}=\labs\grad\ln n\rabs$ and $\rhoe$ the electron Larmor radius.
\newline\indent
If the motion of electrons in the parallel direction is impeded by collisions, it creates a phase shift between the density and potential structures as shown in \figref{dr_wave_outphase}.  In this case, there exists net outward particle transport: the $\vct{E}\times\vct{B}$ velocities at the locations of \circled{1} and \circled{2} are the same as they lie on a constant contour line of the potential, but the number of particles at \circled{2} is larger, i.e., more perturbed density, than that at \circled{1}.  Thus, as the structure propagates in the $+y-$direction, the initial perturbation is reinforced, i.e., the mode is unstable.  Note that if density perturbation lags the potential perturbation, then the mode decays.

\section{Drift waves associated with ITG and PVG}\label{sec:temp_vel_driven_dr_wave}
In this section, we describe the ion-temperature-gradient (ITG) and parallel-velocity-gradient (PVG) driven modes using the fluid equations of ions where electrons are assumed to be in the Boltzmann equilibrium, i.e., we consider ion dynamics here.  Furthermore, we take the case of a flat density profile (a finite density gradient is considered in \appendixref{sec:dens_driven_dr_wave}) with $\labs\grad\ln T_i \rabs=\LT^{-1}$ and $\labs\grad\ln U_z \rabs=\LUz^{-1}$ where $T_i$ is the ion temperature and $U_z$ the incompressible parallel ($\vct{\hat z}$) plasma flow velocity.\footnote{Note that $U_z$ is used to denote the 'toroidal' plasma flow velocity in \chref{ch:eddy_motion}.}  Again, we consider electrostatic perturbations such that $\delta\vct{E}=-\grad\varphi$ for which the $\vct{E}\times\vct{B}$ drift velocity $\vct{v_E}=-c\lp\grad\varphi\times\hat z\rp/B$.  Note that $\vct{v_E}$ is a perturbed quantity and divergence free by definition. 
\newline\indent
The linearized ion continuity equation is
\begin{equation}\label{eq:ITG_dens}
\frac{\partial\delta n_i}{\partial t}+n_i\grad_\parallel\delta U_z + \vct{v_E}\cdot\grad n_i + U_z\grad_\parallel\delta n_i = 0,
\end{equation}
where we have dropped $n_i\grad\cdot\vct{v_E}$ (divergence free), $\delta n_i\grad_\parallel U_z$ (incompressible) and $\delta U_z\grad_\parallel n_i$ (mean density has no gradient in the parallel direction) terms.  Here, the prefix $\delta$ denotes the perturbed quantity.
\newline\indent
The linearized evolution equation of the parallel momentum\footnote{Perpendicular momentum equation is used to get the $\vct{E}\times\vct{B}$ drift velocity $\vct{v_E}$.} is 
\begin{equation}\label{eq:ITG_momentum}
n_i m_i \lp\frac{\partial\delta U_z}{\partial t} + \vct{v_E}\cdot\grad U_z + U_z\grad_\parallel\delta U_z \rp = -\grad_\parallel\delta p_i - e n_i \grad_\parallel\varphi,
\end{equation}
where we dropped $\delta U_z\grad_\parallel U_z$ term.
\newline\indent
The linearized evolution equation of the ion pressure is 
\begin{equation}\label{eq:ITG_pressure}
\frac{\partial\delta p_i}{\partial t}+U_z\grad_\parallel\delta p_i + \vct{v_E}\cdot\grad p_i = -\Gamma p_i\grad_\parallel\delta U_z,
\end{equation}
where we have dropped $\delta U_z\grad_\parallel p_i$ (no variation of equilibrium pressure in the parallel direction) and $-\Gamma\delta p_i\grad_\parallel U_z$.  Here, $\Gamma$ is the specific heat ratio.
\newline\indent
The unknown variables in the system are $\varphi$, $\delta n_i$, $\delta U_z$ and $\delta p_i$.  The fourth equation is constructed by invoking the Boltzmann electron with quasi-neutrality plasma condition:
\begin{equation}\label{eq:ITG_closure}
\delta n_i \approx \delta n_e \approx n_i\frac{e\varphi}{T_i},
\end{equation}
where we assume $T_i=T_e$. This closes the set of equations.
\newline\indent
In our system, a perturbed quantity $\xi$ whose amplitude is taken to be small enough can be Fourier decomposed in the $y-$ and $z-$directions, but such a decomposition cannot be done in $x-$direction as the equilibrium quantities are varying in this direction: $\xi=\hat\xi\lp x, t\rp\exp\lp-i\omega t+ik_yy+ik_zz\rp$ where $\hat\xi$ is the eigenfunction describing the amplitude of the wave-like structure $\xi$.  Thus, using the following ansatz,
\begin{eqnarray}\label{eq:ITG_ansatz}
\varphi & = & \hat\varphi\exp\lp -i\omega t+ik_yy+k_zz\rp, \nonumber \\
\delta n_i & = & \hat{\delta n_i}\exp\lp -i\omega t+ik_yy+k_zz\rp, \nonumber \\
\delta U_z & = & \hat{\delta U_z}\exp\lp -i\omega t+ik_yy+k_zz\rp, \nonumber \\
\delta p_i & = & \hat{\delta p_i}\exp\lp -i\omega t+ik_yy+k_zz\rp,
\end{eqnarray}
we look for a condition for a solution of the linearized equations to exist, which provides a dispersion relation, with assumptions of $k_y \gg k_z$ (perturbed quantities are elongated along the magnetic field line) and $\omega \gg k_z U_z$, i.e., we make sure that $k_z$ is small enough such that these assumptions are valid.  Note that the $\vct{E}\times\vct{B}$ drift velocity is (we drop the $\exp$ term), then,
\begin{equation}
\vct{v_E}=-ik_y c\frac{\hat\varphi}{B}\hat x.
\end{equation}
\newline\indent
Rewriting Eqs. (\ref{eq:ITG_dens})-(\ref{eq:ITG_closure}) using \eqref{eq:ITG_ansatz}, we have
\begin{eqnarray}\label{eq:ITG_set_eq}
-\omega\hat{\delta n_i} + k_z n_i \delta\hat U_z & \approx & 0, \nonumber \\
n_i m_i \lp-\omega\hat{\delta U_z} - k_y c\frac{\hat\varphi}{B}\grad U_z \rp & \approx & -k_z\hat{\delta p_i} - k_z e n_i \hat\varphi, \nonumber \\
-\omega\hat{\delta p_i} - k_y c\frac{\hat\varphi}{B}n_i\grad T_i & \approx & -\Gamma k_z n_i T_i \hat{\delta U_z}, \nonumber \\
\hat{\delta n_i} & \approx & n_i \frac{e\hat\varphi}{T_i}.
\end{eqnarray}
The dispersion relation we find is
\begin{equation}\label{eq:ITG_dispersion}
\omega^2 = k_z^2 C_s^2 - k_z C_s \omega_{*U} + k_z^2 C_s^2 \frac{\omega_{*T}}{\omega},  
\end{equation}
where
\begin{eqnarray}\label{eq:ITG_freq}
\omega_{*U} & = & \frac{c k_y T_i}{eB}\frac{\labs\grad U_z\rabs}{C_s} \sim k_y\rhoi\frac{\vti}{\LUz}, \nonumber \\
\omega_{*T} & = & \frac{c k_y T_i}{eB}\frac{\labs\grad T_i\rabs}{\lp 1+\Gamma\rp T_i} \sim k_y\rhoi\frac{\vti}{\LTi}, \nonumber \\
C_s^2 & = & \lp 1+\Gamma\rp\frac{T_i}{m_i}.
\end{eqnarray}
Using the discriminant $\Delta$ of the cubic function \eqref{eq:ITG_dispersion} in $\omega$, we can find a condition for which an unstable mode exists, i.e., $\Delta < 0$ such that one root has a positive imaginary part:
\begin{eqnarray}\label{eq:ITG_criterion}
\Delta  = 4\lp k_z^2 C_s^2 - k_z C_s \omega_{*U} \rp^3 - 27\lp k_z^2 C_s^2 \omega_{*T}\rp^2 < 0, \nonumber \\
\therefore k_z^2 C_s^2 <  k_z C_s \omega_{*U} + \frac{3}{2}\lp\frac{1}{2}k_z^2 C_s^2 \omega_{*T}\rp^{2/3}.
\end{eqnarray}
It was mentioned in \appendixref{sec:dens_driven_dr_wave} that the density-gradient driven drift wave becomes unstable when the density perturbation leads the potential perturbation, otherwise the wave is purely oscillatory (if they are in phase) or damped (if density lags potential).  The instability criterion given in \eqref{eq:ITG_criterion} can be explained in the exact same way.  Imaginary part of $\omega$, denoted as $\omega_I$, contains the phase information for \eqref{eq:ITG_set_eq}.  Thus, if $\Delta \ge 0$, we have $\omega_I=0$ meaning that all the perturbed quantities are in phase, i.e., purely oscillatory.  On the other hand, the $\Delta < 0$ condition (\eqref{eq:ITG_criterion}) ensures that there exists a pair of complex conjugate roots\footnote{This is alway true for a cubic equation with real coefficients.} so that there exist phase shifts; one of them is growing and the other decaying.  Note that perturbations in the density and the potential are always in phase here via the Boltzmann relation, but the perturbations in the pressure and the potential has a finite phase shift if $\Delta < 0$.
\newline\indent
The dispersion relation \eqref{eq:ITG_dispersion} contains three frequencies related to distinct physical effects: $k_z C_s$ (the sound wave), $\omega_{*U}$ (convection of sheared flow via $\vct{v_E})$ and $\omega_{*T}$ (convection of temperature via $\vct{v_E}$).  As in any system, a pressure perturbation launches a sound wave in our system, but only in the parallel direction due to strong magnetic field.  Then, we may expect that the system becomes unstable if the sound wave cannot keep up the driver of the pressure perturbation ($\omega_{*U}$ and/or $\omega_{*T}$), and this is exactly what \eqref{eq:ITG_criterion} dictates.  These two drivers contribute to destabilize the sound wave together.  \figref{ITG_PVG_stability} illustrates a region where unstable modes exist as functions of the sound frequency normalized driving frequencies.
\myfig[4.5in]{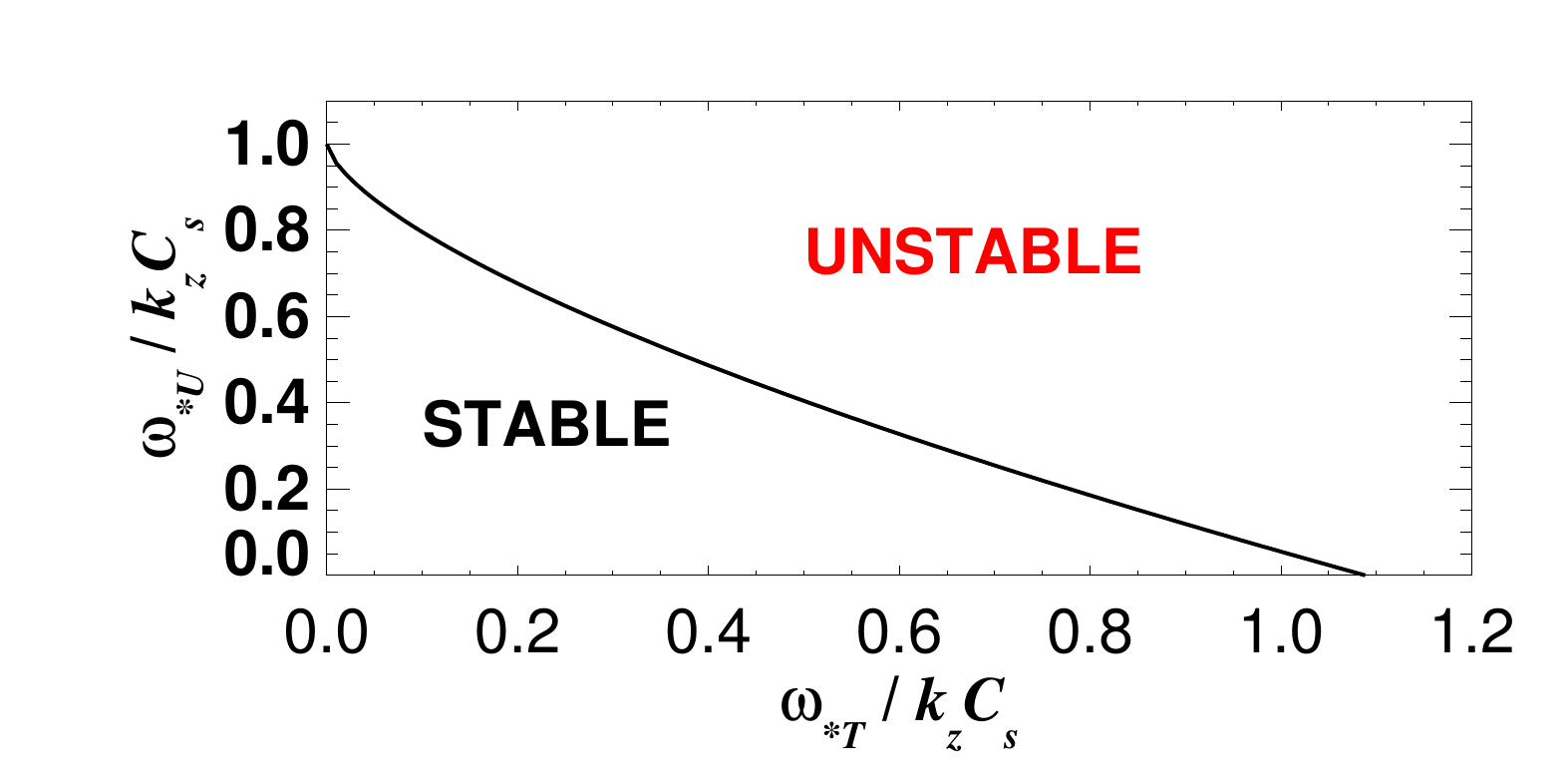}{Stability region of ITG and PVG}{A graphical representation of the ITG ($x-$axis) and PVG ($y-$axis) instability criterion using \eqref{eq:ITG_criterion}.}
\newline\indent
We finish this section with a few remarks: (i) both $\omega_{*U}$ and $\omega_{*T}$ require a finite $k_y$ (see \eqref{eq:ITG_freq}), but \eqref{eq:ITG_criterion} is not valid for large $k_y$ as the finite larmor radius effect\footnote{If scales of perturbed potential is smaller than the ion Larmor radius, then the gyro-motion effectively averages out the perturbed potential.} starts to play a role in which case a proper gyrokinetic equations must be used; (ii) shearing rate of the mean perpendicular plasma flow can change the instability criterion which was not included in this section; (iii) the smaller the $k_z$, the easier to satisfy the instability criterion, and the smallest possible $k_z$ in a tokamak is the inverse of the connection length.

\chapter{Examples of a correlation function and a power spectrum} \label{ch:ex_corr_power}
In this chapter, we describe, step by step, how we generate a correlation function and a power spectrum in this work from the raw BES data.

\myfig[5.5in]{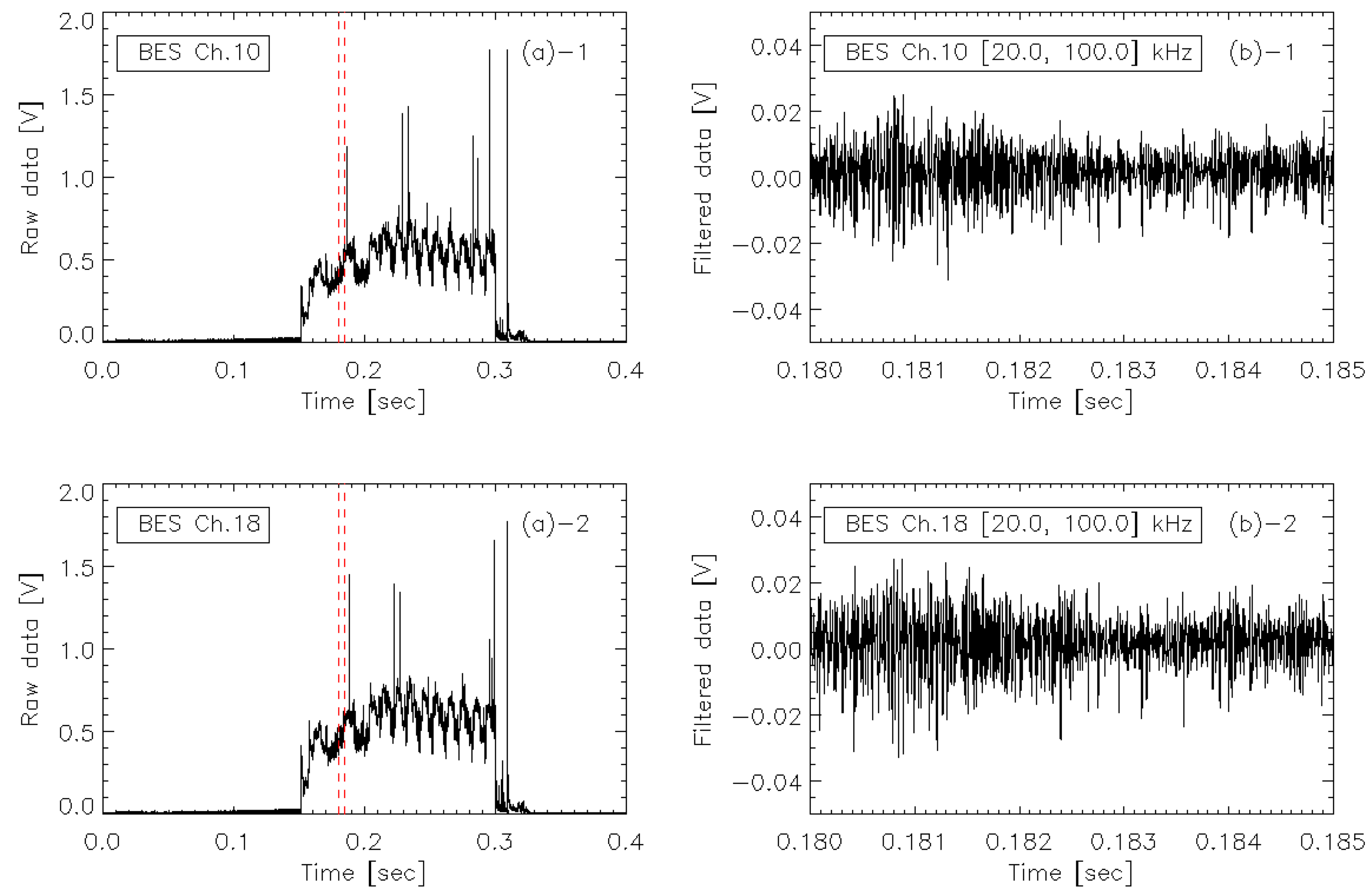}{Examples of raw BES data}{(a)-1 and (a)-2 show raw BES data of Ch. 10 and Ch. 18 from shot \#27267; (b)-1 and (b)-2 show frequency filtered ($[20, 100]$ kHz) BES data of (a) from $0.180$ to $0.185\:s$, i.e., red vertical dashes in (a). Note that Ch. 10 and Ch. 18 has a poloidal separation distance of $2.0\:cm$ with no radial separation.}

\section{Generating a correlation function}
\noindent
(1) Get the raw BES data and frequency-filter it from $20$ to $100$ kHz as shown in \figref{time_trace}(a) and (b).
\newline\noindent
(2) Select the time range where correlation functions are to be generated.  In \figref{time_trace}(b), we selected time range of $0.180-0.185\:s$.
\newline\noindent
(3) Divide the selected time range into subwindows.  In this example, we divided the time range into $50$ subwindows with $100\:\mu s$ duration for each subwindow.
(4) Using \eqref{eq:modified_cc_def}, generate a correlation function for each subwindow.  In our example, we, then, have $50$ correlation functions as shown in \figref{corr_freq}(a) with black lines. Note that we have cross-correlation functions using two poloidally separated channels, i.e., Ch. 10 and Ch. 18 whose separation distance is $2.0\:cm$.
\newline\noindent
(5) Average the $50$ correlation functions to obtain the averaged correlation function shown as red circles in \figref{corr_freq}(a).
\newline\noindent
(6) Uncertainty of the averaged correlation function is estimated as the standard deviation divided by the square root of number of subwindows \cite{bendat_wiley_2010}, i.e., $\sqrt{50}$ in our example.

\myfig[5.5in]{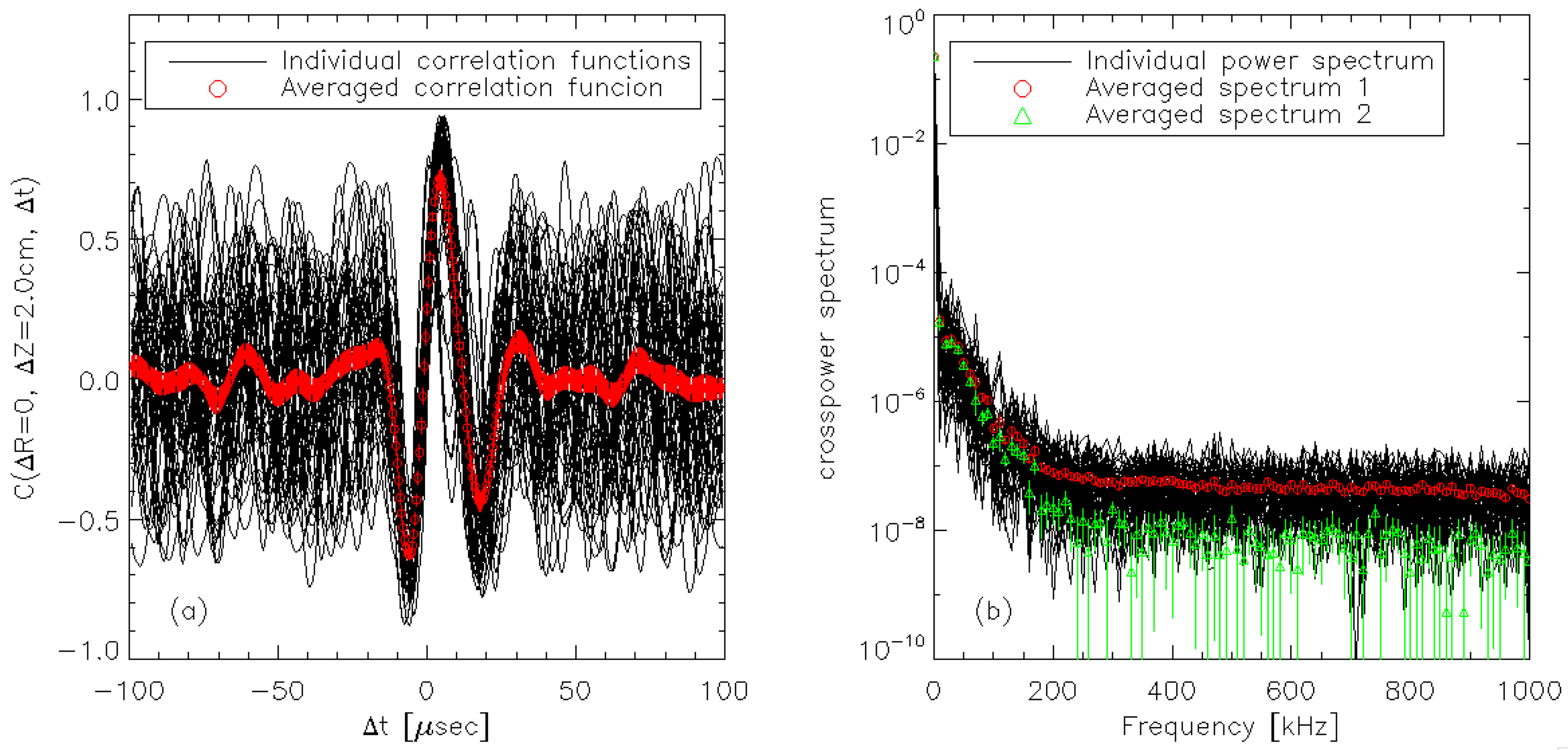}{Examples of correlation functions and power spectra}{(a) Cross-correlation functions and (b) cross-power spectra between Ch. 10 and Ch. 18 shown in \figref{time_trace}(b).}

\section{Generating a power spectrum}
\noindent
(1) Get the raw BES data as shown in \figref{time_trace}(a).
\newline\noindent
(2) Select the time range where power spectra are to be generated.  In \figref{time_trace}(b), we selected time range of $0.180-0.185\:s$.
\newline\noindent
(3) Divide the selected time range into subwindows.  In this example, we divided the time range into $50$ subwindows with $100\:\mu s$ duration for each subwindow.
\newline\noindent
(4) Generate a cross-spectrum for each subwindow as $S^{fg}_i\equiv FT\lcb f_i\lp t\rp \rcb FT^*\lcb g_i\lp t \rp \rcb$, where $FT\lcb\rcb$ and $FT^*\lcb\rcb$ are the Fourier-transform operator and its complex conjugate, respectively.  Here, the subscript $i$ denotes the $i$th subwindow, and $f\lp t\rp$ and $g\lp t\rp$ for \textit{raw} BES data from $0.180$ to $0.185\:s$ in our example.  Black lines in \figref{corr_freq}(b) shows the magnitude of $50$ cross-spectra, i.e., $|S^{fg}_i |$, where $|S|$ returns the magnitude of the $S$.
\newline\noindent
(5) Average the $50\:|S^{fg}_i |$ to get the averaged power spectrum shown as red circles in \figref{corr_freq}(b). (Differences between red circles and green triangles are explained later.)
\newline\noindent
(6) Uncertainty of the averaged power spectrum is estimated as the standard deviation divided by the square root of number of subwindows \cite{bendat_wiley_2010}.
\newline\noindent
We note that red circles in \figref{corr_freq}(b) are calculated as the average of the $|S^{fg}_i |$ while green triangles are calculated as the magnitude of the averaged $S^{fg}_i$.  These two estimates give different results because taking the magnitude of complex number is not a linear operator. Following Bendat and Piersol \cite{bendat_wiley_2010}, we actually take the magnitude of the averaged $S^{fg}_i$ in this work, i.e., green triangles, as the power spectrum of the signal. The reason we show $|S^{fg}_i |$ (black lines) and the averaged $|S^{fg}_i |$ (red circles) is because of the difficulty of plotting $S^{fg}_i$ for it being complex numbers.  Note that the uncertainty for the magnitude of the averaged $S^{fg}_i$ (green triangles) is estimated by first dividing the standard deviations of real and imaginary parts of the $S^{fg}_i$ by the square root of the number of subwindows, then by propagating these uncertainties to get the corresponding uncertainty.

\backmatter

\chapter{Definitions}

\section{Fields}
\mydotfill{$\vct{B}$}{Magnetic field}
\mydotfill{$\Btor$}{Toroidal component of the magnetic field}
\mydotfill{$\Bpol$, $B_p$}{Poloidal component of the magnetic field}
\mydotfill{$\vct{E}$}{Electric field}
\mydotfill{$\vct{J}$}{Current density}
\mydotfill{$\varphi$}{Perturbed electrostatic potential}

\section{Lengths}
\mydotfill{$\rhoi$}{Ion Larmor radius $=\vti/\Omegai$}
\mydotfill{$\rhoe$}{Electron Larmor radius $=\vte/\Omegae$}
\mydotfill{$\rhostar$}{$\rhoi/a$}
\mydotfill{$\Ln$}{Density gradient scale length $=\labs\grad\ln n\rabs^{-1}$}
\mydotfill{$\LT$}{Temperature gradient scale length $=\labs\grad\ln T\rabs^{-1}$}
\mydotfill{$\LTi$}{Ion temperature gradient scale length}
\mydotfill{$\LTe$}{Electron temperature gradient scale length}
\mydotfill{$\Lstar$}{Minimum of $\LTi$ and $\Ln$}
\mydotfill{$\LU$}{Flow velocity gradient scale length $=\labs\grad\ln U\rabs^{-1}$}
\mydotfill{$\lambda_x$}{Radial correlation length of Gaussian ``eddies'' in Chapters \ref{ch:synthetic_bes} and \ref{ch:eddy_motion}}
\mydotfill{$\lambda_y$}{Poloidal correlation length of Gaussian ``eddies'' in Chapters \ref{ch:synthetic_bes} and \ref{ch:eddy_motion}}
\mydotfill{$\kper$}{Perpendicular wavenumber of plasma turbulence}
\mydotfill{$\kpar$}{Parallel wavenumber of plasma turbulence}
\mydotfill{$\lper$}{Perpendicular correlation length of plasma turbulence}
\mydotfill{$\lpar$}{Parallel correlation length of plasma turbulence}
\mydotfill{$\lx$, $\lR$}{Radial correlation length of plasma turbulence}
\mydotfill{$\ly$}{Perpendicular (on a flux surface) correlation length of plasma turbulence}
\mydotfill{$\lZ$}{Poloidal (vertical) correlation length of plasma turbulence}
\mydotfill{$\kx$}{$2\pi/\lx$}
\mydotfill{$\ky$}{$2\pi/\ly$}
\mydotfill{$\Lambda$}{The connection length $=\pi r B/B_p$}

\section{Times (frequencies)}
\mydotfill{$\tau_E$}{Energy confinement time}
\mydotfill{$\Omegai$}{Ion gyro-frequency $=eB/m_i c$}
\mydotfill{$\Omegae$}{Electron gyro-frequency $=eB/m_e c$}
\mydotfill{$\tau_{life}$}{Lifetime of Gaussian ``eddies'' in the moving frame in Chapters \ref{ch:synthetic_bes} and \ref{ch:eddy_motion}}
\mydotfill{$\taupeakcc$}{Time delay at which the cross-correlation function reaches its maximum}
\mydotfill{$\tc$}{Correlation time of plasma turbulence}
\mydotfill{$\tnl$}{Nonlinear time of plasma turbulence}
\mydotfill{$\tnlnz$}{Nonlinear time associated with the non-zonal component of plasma turbulence}
\mydotfill{$\tst$}{Particle (ion) streaming time along the parallel direction $=\Lambda/\vti$}
\mydotfill{$\tshear$}{Perpendicular velocity shear time $=\lsb\lp B_p/B\rp d\Utor/dr\rsb^{-1}$}
\mydotfill{$\tM$}{Magnetic drift time $\lp\lx/\rhoi\rp R/\vti$}
\mydotfill{$\tstari$}{Drift time associated with ion-temperature gradient $\lp\ly/\rhoi\rp\LTi/\vti$}
\mydotfill{$\tstare$}{Drift time associated with electron-temperature gradient $\lp\ly/\rhoe\rp\LTe/\vte$}
\mydotfill{$\tstarn$}{Drift time associated with density gradient $=\lp\ly/\rhoi\rp\Ln/\vti$}
\mydotfill{$\tstar$}{Minimum of $\tstari$ and $\tstarn$}
\mydotfill{$\nu_{ii}$}{Ion collision frequency}
\mydotfill{$\nu_{*i}$}{Normalised ion collision frequency $=\nu_{ii}\tst$}
\mydotfill{$\omega_{*U}$}{Frequency of the drift wave associated with the parallel flow shear}
\mydotfill{$\omega_{*T}$}{Frequency of the drift wave associated with the temperature gradient}
\mydotfill{$\gE$}{Perpendicular flow shear $=\lp\Bpol/B\rp\partial\Utor/\partial r$}
\mydotfill{$\gEbar$}{Normalised perpendicular flow shear $=\gE\tst$}

\section{Velocities}
\mydotfill{$c$}{The speed of light}
\mydotfill{$\vti$}{Ion thermal velocity $=\sqrt{2T_i/m_i}$}
\mydotfill{$\vte$}{Electron thermal velocity $=\sqrt{2T_e/m_e}$}
\mydotfill{$C_s$}{Sound speed $=\lp 1+\Gamma\rp T_i/m_i$}
\mydotfill{$\vct{U}$}{Plasma flow velocity}
\mydotfill{$\Upar$}{Parallel component of plasma flow velocity}
\mydotfill{$\Uper$}{Perpendicular component of plasma flow velocity}
\mydotfill{$\Utor$}{Toroidal component of plasma flow velocity}
\mydotfill{$U_z$}{Toroidal component of plasma flow velocity in \chref{ch:eddy_motion}}
\mydotfill{$U_y$}{Poloidal component of plasma flow velocity in \chref{ch:eddy_motion}}
\mydotfill{$\vct{v_E}$, $\delvper$}{Fluctuating $\vct{E}\times\vct{B}$ drift velocity}
\mydotfill{$\vD$}{Magnetic drift velocity}
\mydotfill{$\vbes$}{Apparent poloidal velocity of measured fluctuating density patterns}
\mydotfill{$v_y$}{Apparent poloidal velocity of Gaussian ``eddies'' in Chapters \ref{ch:synthetic_bes} and \ref{ch:eddy_motion}}
\mydotfill{$v_y^{RMS}$}{RMS value of fluctuating $v_y$ of Gaussian ``eddies'' in Chapters \ref{ch:synthetic_bes} and \ref{ch:eddy_motion}}

\section{Other}
\mydotfill{$n$}{Plasma density}
\mydotfill{$\dn$}{Fluctuating plasma density}
\mydotfill{$n_i$}{Ion density}
\mydotfill{$n_e$}{Electron density}
\mydotfill{$T$}{Temperature in $eV$}
\mydotfill{$T_i$}{Ion temperature in $eV$}
\mydotfill{$T_e$}{Electron temperature in $eV$}
\mydotfill{$p$}{Plasma pressure}
\mydotfill{$m_i$}{Ion mass}
\mydotfill{$m_e$}{Electron mass}
\mydotfill{$\beta$}{The ratio of the plasma pressure to the magnetic field energy density}
\mydotfill{$e$}{Proton charge}
\mydotfill{$I$}{Mean part of the detected photon intensity by the 2D BES}
\mydotfill{$\dI$}{Fluctuating part of the detected photon intensity by the 2D BES}
\mydotfill{$I_{MHD}^{RMS}$}{RMS value of MHD modes in synthetic data in Chapters \ref{ch:synthetic_bes} and \ref{ch:eddy_motion}}
\mydotfill{$\Gamma$}{Specific heat ratio}
\mydotfill{$\Qturb$}{Gyro-Bohm-normalised turbulent ion heat flux}

\section{Geometrical quantities}
\mydotfill{$\psi$}{Flux surface label}
\mydotfill{$R$}{Distance from the centre of a torus (major radius)}
\mydotfill{$r$}{Distance from the magnetic axis of a flux surface (minor radius)}
\mydotfill{$Z$}{Height from the midplane of flux surfaces}
\mydotfill{$a_\psi$}{Half diameter of a flux surface $\psi$ at the midplane}
\mydotfill{$a$}{$a_\psi$ of the LCFS, a measure of total plasma size}
\mydotfill{$R_0$}{$R$ at the point where $a_\psi\rightarrow 0$}
\mydotfill{$\varepsilon_0$}{Inverse aspect ratio of a tokamak, i.e., $\varepsilon_0=a/R_0$}
\mydotfill{$\varepsilon$}{Local inverse aspect ratio, i.e., $\varepsilon=r/R$}
\mydotfill{$q$}{The safe factor $\approx\frac{r}{R}\frac{\Btor}{\Bpol}$ for large aspect ratio tokamak}
\mydotfill{$\alpha$}{Pitch angle of local magnetic field, $\tan\alpha=\frac{\Bpol}{\Btor}$}

\section{Acronyms}
\mydotfill{BES}{Beam Emission Spectroscopy}
\mydotfill{CCFE}{Culham Centre for Fusion Energy}
\mydotfill{CCTD}{Cross-correlation Time Delay}
\mydotfill{CUDA}{Compute Unified Device Architecture}
\mydotfill{CXRS}{Charge eXchange Recombination Spectroscopy}
\mydotfill{ETG}{Electron Temperature Gradient}
\mydotfill{GPU}{Graphical Processing Unit}
\mydotfill{ITB}{Internal Transport Barrier}
\mydotfill{ITG}{Ion Temperature Gradient}
\mydotfill{LCFS}{Last Closed Flux Surface}
\mydotfill{LoS}{Line-of-Sight}
\mydotfill{MAST}{Mega Amp Spherical Tokamak}
\mydotfill{MHD}{Magnetohydrodynamic}
\mydotfill{MSE}{Motional Stark Effect}
\mydotfill{NBI}{Neutral Beam Injection}
\mydotfill{PSF}{Point-Spread-Function}
\mydotfill{PVG}{Parallel Velocity Gradient}
\mydotfill{RMS}{Root-Mean-Square}
\mydotfill{START}{Small Tight Aspect Ration Tokamak}
\mydotfill{TEM}{Trapped Electron Mode}

\clearpage

{
\raggedright 

\phantomsection
\addcontentsline{toc}{chapter}{Bibliography}
\nobibintoc
\bibliographystyle{unsrt}
\bibliography{reference}
}

\end{document}